\documentclass[5p,twocolumn]{elsarticle}
\usepackage{graphicx,latexsym}
\usepackage{amssymb,amsmath,bm}
\usepackage{subfigure}
\usepackage{braket}

\usepackage{array}
\usepackage{float}
\usepackage{caption}

\newcommand{\angstrom}{\text{\normalfont\AA}}
\usepackage{hyperref}
\hypersetup{
    pdfnewwindow=true,       % links in new window
    colorlinks=true,         % false: boxed links; true: colored links
    linkcolor=blue,          % color of internal links
    citecolor=blue,          % color of links to bibliography
    filecolor=magenta,       % color of file links
    urlcolor=black           % color of external links
}

\usepackage[normalem]{ulem}

\def\sec#1{Sec.\ \ref{#1}}

\def\fig#1{Fig.\ \ref{#1}}

\journal{}

\begin{document}

\begin{frontmatter}

%-----------------------------------------------------------------

\title{Effects of bonded and non-bonded B/N codoping of graphene on its
    stability,\break interaction energy, electronic structure, and power factor}

\author[a1,a2]{Nzar Rauf Abdullah}
\ead{nzar.r.abdullah@gmail.com}
\address[a1]{Division of Computational Nanoscience, Physics Department, College of Science, 
             University of Sulaimani, Sulaimani 46001, Kurdistan Region, Iraq}
\address[a2]{Computer Engineering Department, College of Engineering, Komar University of Science and Technology, Sulaimani 46001, Kurdistan Region, Iraq}

\author[a1]{Hunar Omar Rashid}

\author[a3]{Mohammad T. Kareem}
\address[a3]{Chemistry Department, College of Science, 
             University of Sulaimani, Sulaimani 46001, Kurdistan Region, Iraq}
             
\author[a4]{Chi-Shung Tang}
\address[a4]{Department of Mechanical Engineering,
  National United University, 1, Lienda, Miaoli 36003, Taiwan}

\author[a5]{Andrei Manolescu}
\address[a5]{Reykjavik University, School of Science and Engineering,
              Menntavegur 1, IS-101 Reykjavik, Iceland}

\author[a6]{Vidar Gudmundsson}   
 \address[a6]{Science Institute, University of Iceland,
        Dunhaga 3, IS-107 Reykjavik, Iceland}

%
%----------------------------------------------------------------

\begin{abstract}
We model boron and nitrogen doped/codoped monolayer graphene to study its stability, interaction energy, 
electronic and thermal properties using density functional theory. It is found that a doped graphene sheet with 
non-bonded B or N atoms induces an attractive interaction and thus opens up 
the bandgap. Consequently, the power factor is enhanced. Additionally, bonded B or N atoms in doped
graphene generate a repulsive interaction leading to a diminished bandgap, and thus a decreased power factor. 
We emphasis that enhancement of the power factor is not very sensitive to the concentration of the boron and 
nitrogen atoms, but it is more sensitive to the positions of the B or N atoms in ortho, meta, and para 
positions of the hexagonal structure of graphene. 
In the B and N codoped graphene, the non-bonded dopant atoms have a weak attractive interaction 
and interaction leading to a small bandgap, while bonded doping atoms cause a strong attractive interaction and a large bandgap. 
As a result, the power factor of the graphene with non-bonded doping atoms is reduced while it is 
enhanced for graphene with bonded doping atoms.

\end{abstract}


\end{frontmatter}

\section{Introduction}

The 2D graphene structure consisting of carbon atoms arranged in a honeycomb structure with an sp$^2$ 
hybridization has been of interest for future nanotechnology \cite{Ren2014}, and 
it is the most promising material for implementing the next generation of electronic devices \cite{Kang2018}. 
Fascinating properties of graphene such as ballistic transport at room temperature, 
high mobility, high electronic conductivity, quantum Hall effect, electrically controllable spin transportand \cite{ZHANG20192957, ABDULLAH2018102}, and mechanical strength 
make graphene a potential candidate for an enormous breadth of applications \cite{Syama2019,C4TA01047G} in 
electronic \cite{Zhang2005}, optical \cite{doi:10.1021/nl102824h, ABDULLAH2016280, ABDULLAH20181432, ABDULLAH2018223}, and 
thermoelectric \cite{Shahil2012, Abdullah_2018} devices.

Pristine graphene itself is usually not considered a good thermoelectric material for application 
of thermal energy conversion because it has a small Seebeck coefficient or power factor, and a high 
conductivity due to the vanishing gap, facilitating opposite and canceling contributions of electrons and holes. 
This behavior leads to a very weak thermoelectric figure of merit which in turn 
produces a very limited performance or efficiency of pristine graphene \cite{doi:10.1002/smll.201303701}.
Consequently, it affects the application of graphene based materials for super capacitors, batteries, 
and fuel cells.

Scientists have been looking for semiconducting bandgap behavior in graphene systems 
which would be important for thermal devices.
It can be obtained by the means of opening up a bandgap that could lead to 
efficient devices \cite{Dvorak2013, RASHID2019102625}.
An appropriate nanostructuring and bandgap engineering of graphene can increase the Seebeck coefficient 
or the power factor and decrease to an extreme degree the lattice thermal conductance without dramatically 
reducing the electronic conductance.
Basically, the opening of a bandgap in graphene is achieved by a symmetry breaking of the hexagonal structure 
of graphene \cite{Sahu2017}.
Therefore, several methods have been proposed to break the high symmetry of monolayer and bilayer graphene, such as strain \cite{PhysRevB.78.075435,PhysRevB.80.167401}, 
magnetic field \cite{Jessen2019}, oxidation \cite{C6RA04782C}, 
application of a transverse electric field \cite{MCCANN2007110,NEMNES2018175}, chemical modification \cite{C2TC00570K, C0JM02922J}, and 
doping processes \cite{C2NR11728B,C3CS60401B,NEMNES20199}.

In the doping process of graphene, boron (B) and nitrogen (N) atoms have been used because the sizes of 
these two atoms are close to the carbon (C) atom \cite{Fei2017,C5TA10599D}. 
This affects the bonding of the B and N with the carbon atoms.
Equilibrium molecular dynamics simulations based on the Green-Kubo method have been utilized to investigate the
thermal conductivity and the heat transfer of N-doped graphene showing that the thermal conductivity of N-doped graphene is less sensitive to temperature than in pure graphene \cite{GOHARSHADI201574}.
Additionally, it has been shown that the N atoms at the edge of
a triangle doping defect can increase the thermal conductivity of graphene nanoribbons, 
but with increasing N-doping concentrations the thermal conductivity decreases 
sharply \cite{YANG20191306}. The site-dependent effects of a substitutional N or B atom on quantum transport 
have been investigated and shown that Coulomb interaction drops the transmission features \cite{PI2015196}.
Furthermore, the power factor has been found to be high in 
various B and N codoped graphene nanostructures which
makes them appropriate for energy conversion \cite{Duan14272}.

We believe that information on the effects of bonding between the B and N atoms in doped/codoped graphene 
on its electronic and thermal characteristics is still lacking. We thus theoretically study the electronic 
and thermal properties of a substitutionally B and N doped graphene using a self-consistent field 
approximation via the Kohn-Sham density functional theory (KS-DFT). We examine the effect of bonding and 
non-bonding of B and N atoms in doped graphene on its electronic and thermal properties.
In \sec{Sec:Model} the structure of graphene nanosheet is briefly overviewed. In \sec{Sec:Results} 
the main achieved results are analyzed. In \sec{Sec:Conclusion} the conclusions of the results is presented.

\section{Model and Computational Details}\label{Sec:Model}

A 2D graphene nanosheet in the $xy$-plane with a $2\times2\times1$ super-cell is modeled.
First-principles calculations are utilized to study the structural stability, and the electronic 
structure of the system. The calculations are carried out within the framework of 
a DFT \cite{PhysRev.140.A1133} in which the PAW potentials and the Perdew-Burke-Ernzerhof 
exchange and correlation (XC) functionals of the generalized gradient approximation are 
used \cite{PhysRevLett.77.3865}. 
In the DFT, the wavefunctions and the ground state electronic density of the crystal are calculated 
by a self-consistent (SCF) procedure for a coupled set of single-electron equations. 
The total energy of the system is calculated using the SCF Kohn-Sham equations. 
The DFT calculations with a plane-wave projector-augmented wave
method are implemented in the Quantum Espresso (QE) package \cite{Giannozzi_2009}. 
The plane wave cut-off energy is assumed to be $1088$ eV and the $k$-mesh points 
in the unit cell are selected to be 
$8\times8\times1$ with $\Gamma$ centered $k$-points. We assume a fully relaxed structure when 
the calculated force is less than $10^{-4}$ eV/${\angstrom}$.
In addition, for calculating non self-consistently (NSCF) the density of states (DOS) 
we assume a $30\times30\times1$ mesh of $k$-points.

Boltzmann theory is implemented in the BoltzTraP package \cite{Madsen2006} and is used for calculating 
the thermal properties of the system such as power factor (PF) 
\begin{equation}
 PF = S^2 \times \sigma.
\end{equation}
where $S$ is the Seebeck coefficient and $\sigma$ is the electrical conductivity of the system \cite{Madsen2006}.

\section{Results}\label{Sec:Results} 

\begin{figure}[htb]
\centering
\includegraphics[width=0.15\textwidth]{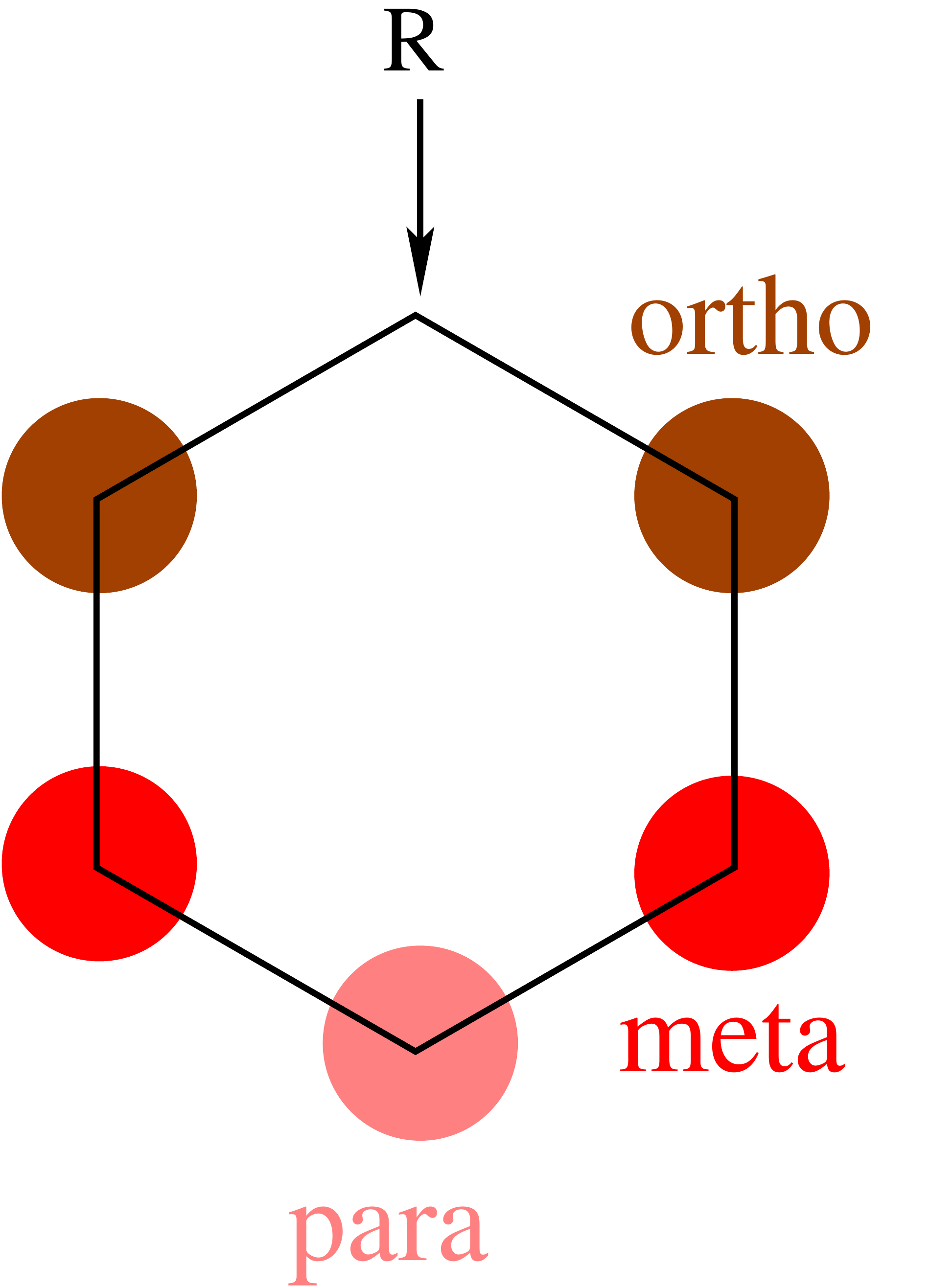}
 \caption{Schematic diagram presenting the ortho (brown), the meta (red), and the para (pink) positions 
 of atoms in the honeycomb structure where R is defined as a reference point.}
\label{fig01}
\end{figure}

In this section, we present the main results of our calculations of pristine and doped graphene nanosheets. 
We focus on the effects of doping ratio and positions on the stability, interaction energy, electronic properties, and thermal 
properties of the system. It is well known that graphene has a hexagonal crystal structure as is 
schematically illustrated in \fig{fig01}. 
The carbon atoms fill up the three main
positions consisting of the ortho-, the meta- and the para-positions with respect to R that 
is the reference point.

In the calculations, we manipulate the positions of doped atoms. 
We can thus determine the most stable doped graphene structure and investigate their 
electronic and thermal properties. The Fermi level
is fixed to be at the zero energy for all the structures investigated here.

\subsection{Boron- or Nitrogen-doped graphene structures}

In this section, we assume only B or N atoms are substitutionally doped in graphene.
For this purpose three concentration ratio of doped atoms are considered which are 
$12.5\%$, $25\%$, and $37.5\%$. 

\begin{figure}[H]
\begin{table}[H]
  \captionsetup{labelformat=empty}
\noindent
\begin{tabular}[]{ >{\centering\arraybackslash}m{0cm} >{\centering\arraybackslash}m{2.4cm}>{\centering\arraybackslash}m{2.4cm} >{\centering\arraybackslash}m{2.4cm} >{\centering\arraybackslash}m{2.4cm} }
	& a & b & c \\ 
	1 &
	\includegraphics[width=0.15\textwidth]{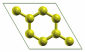} & 
	\includegraphics[width=0.15\textwidth]{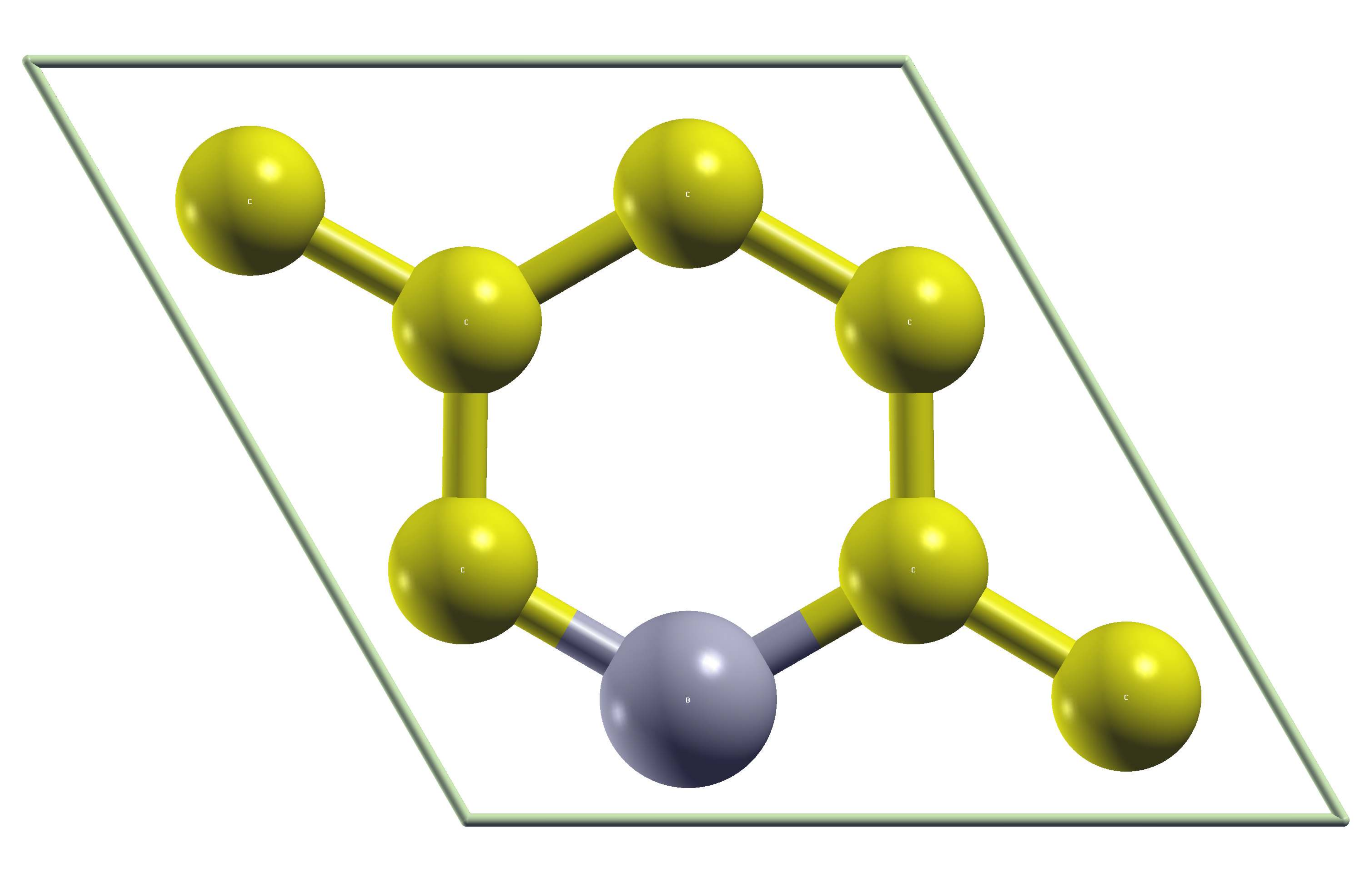} & 
	\includegraphics[width=0.15\textwidth]{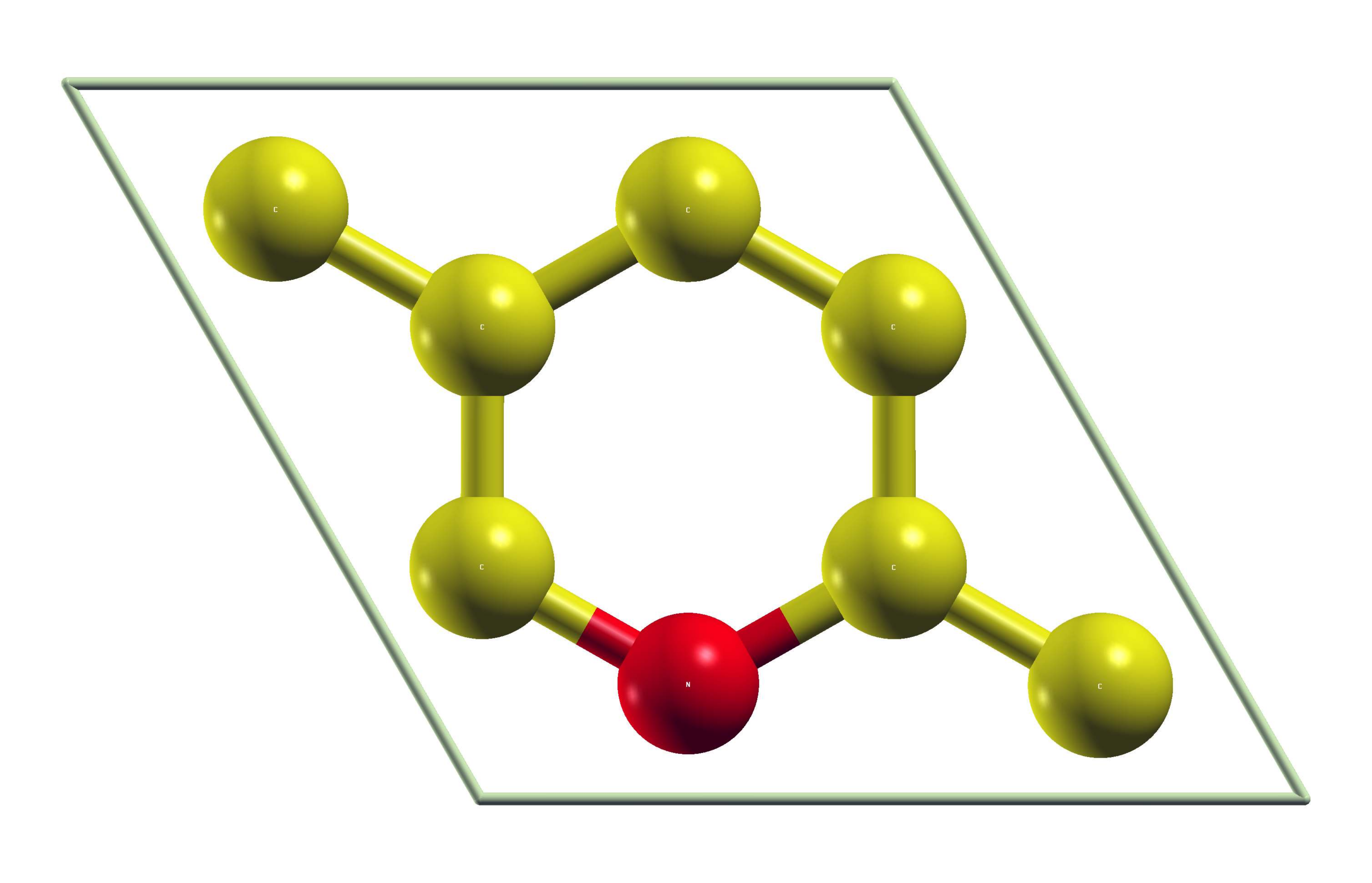} \\
	2 &
	\includegraphics[width=0.15\textwidth]{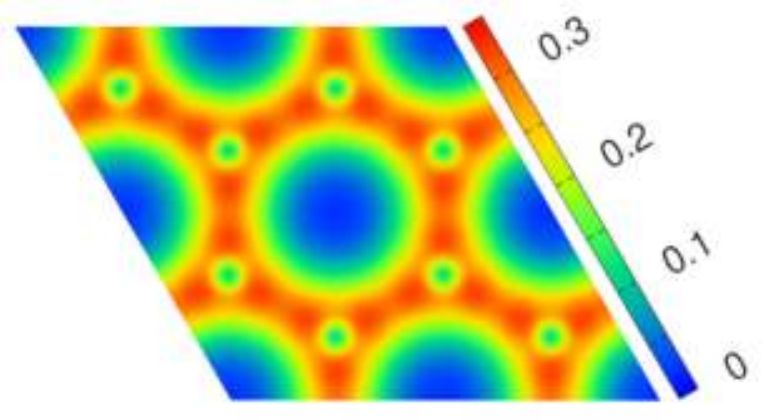}  &
	\includegraphics[width=0.15\textwidth]{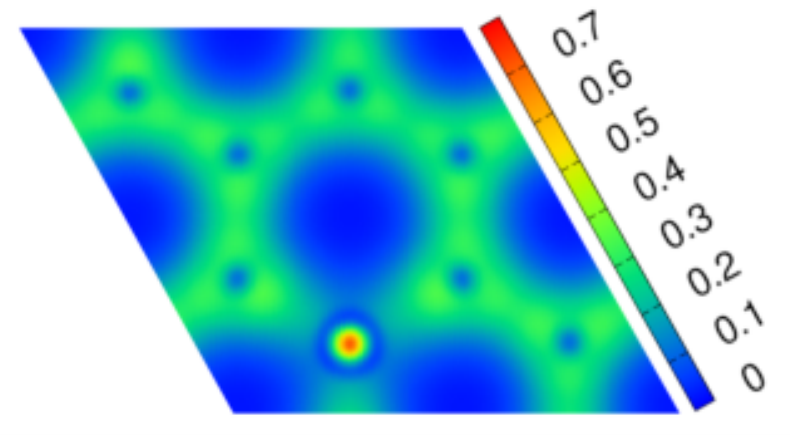} &
	\includegraphics[width=0.15\textwidth]{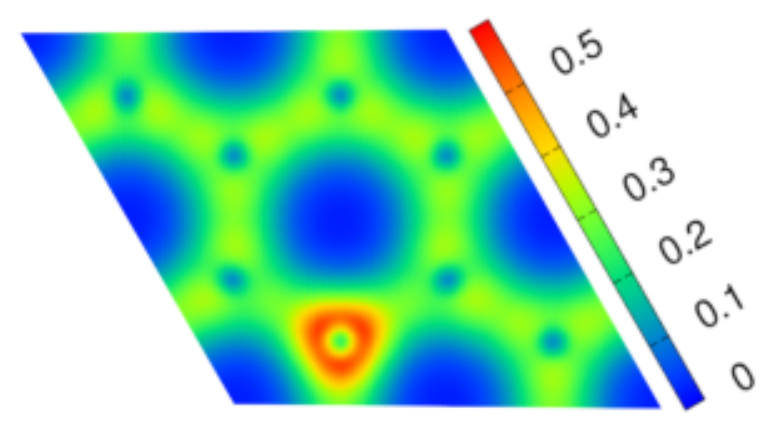} \\ 
	3 &
	\includegraphics[width=0.15\textwidth]{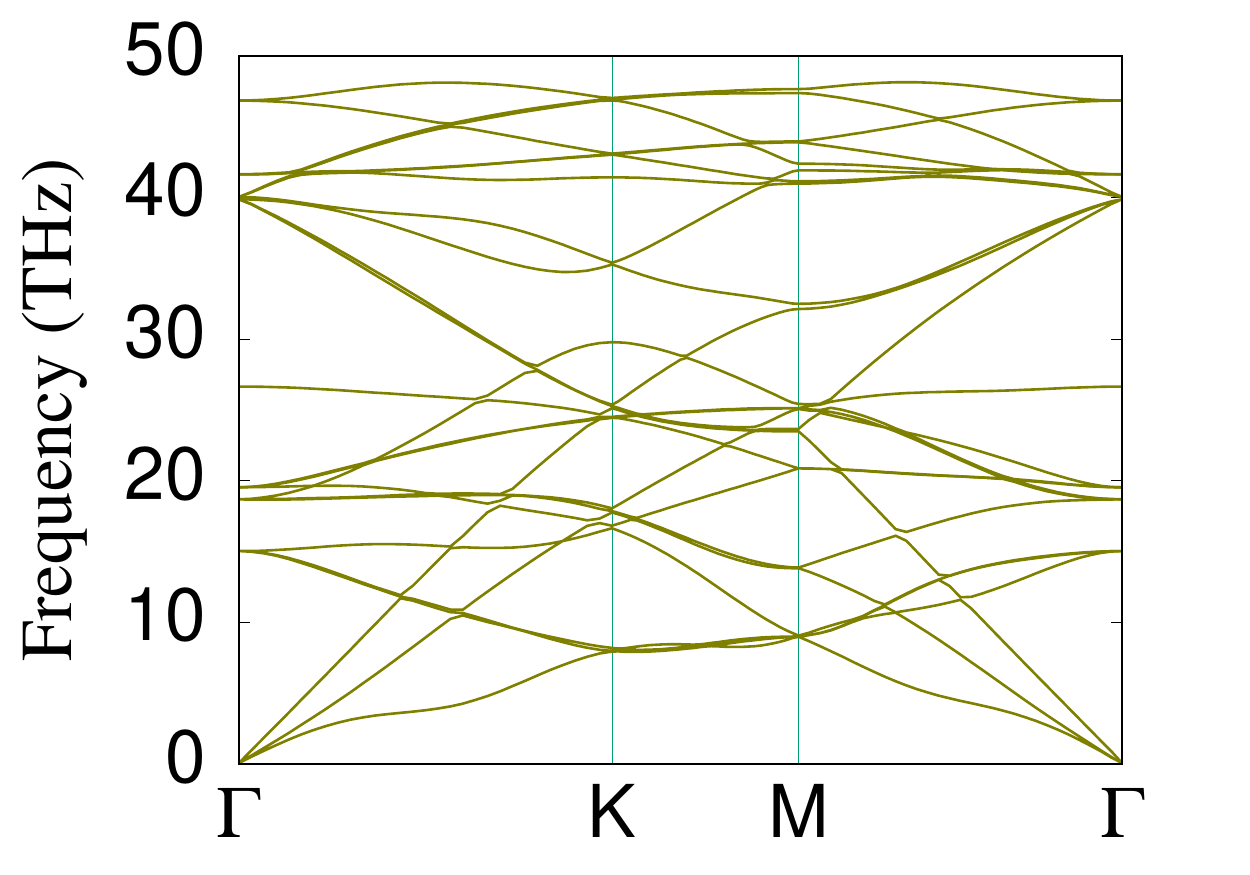} & 
	\includegraphics[width=0.15\textwidth]{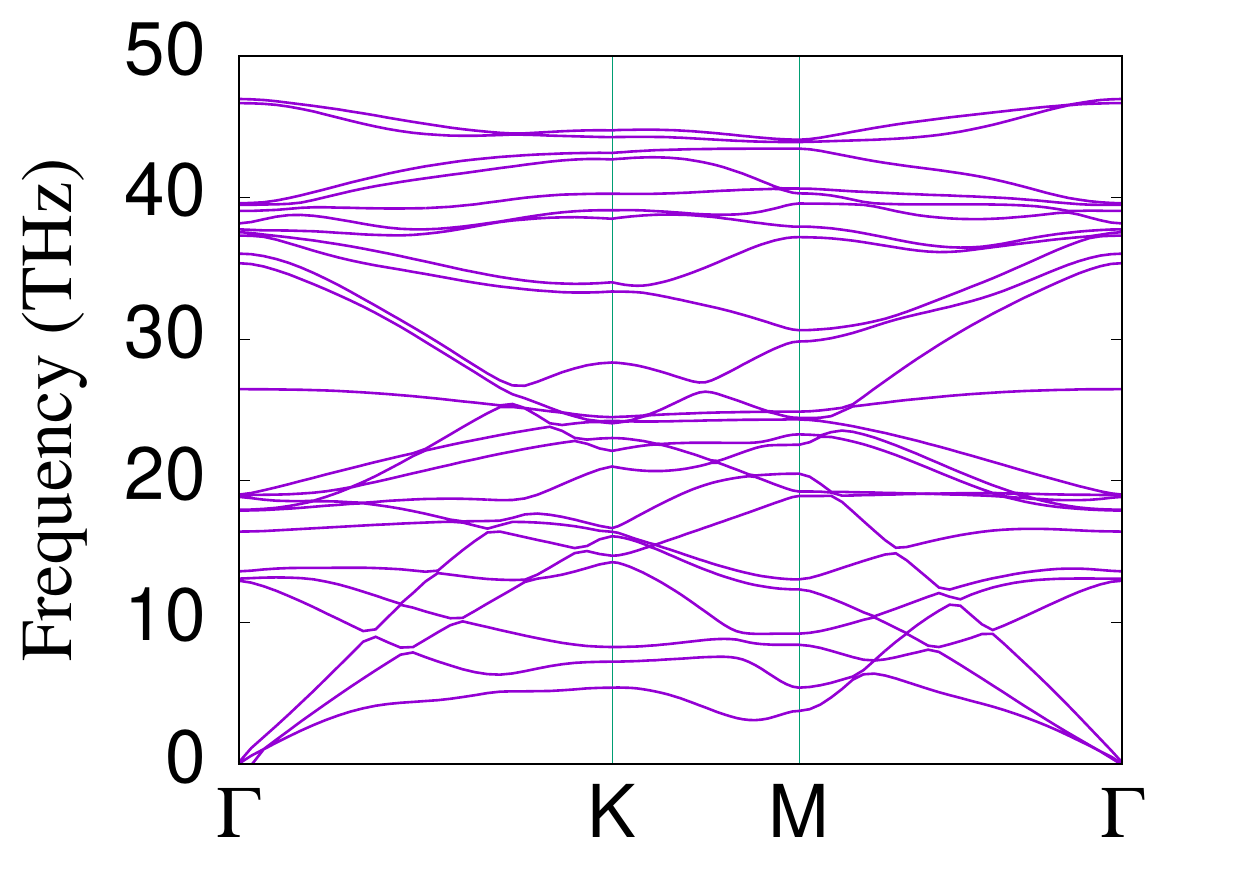} &
	\includegraphics[width=0.15\textwidth]{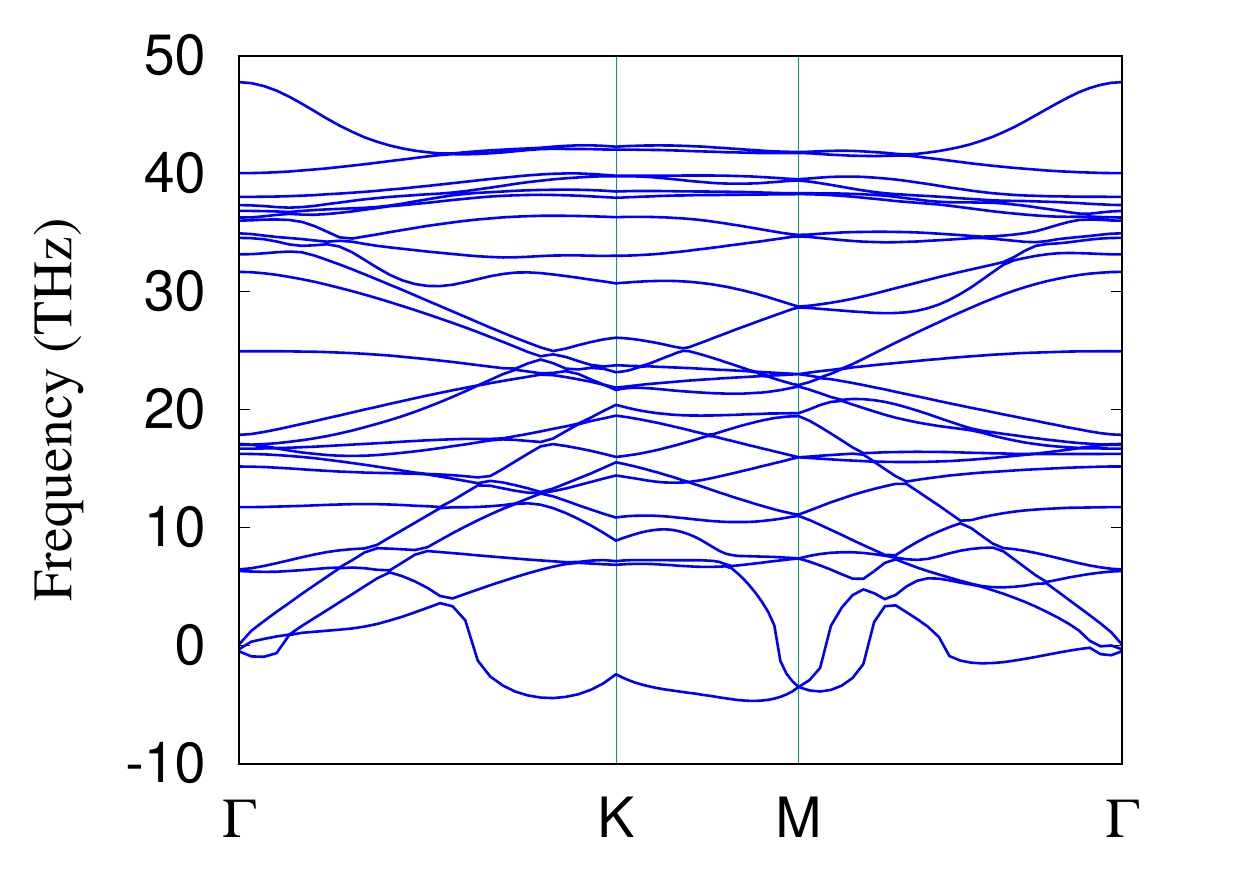} \\
	4 &
	\includegraphics[width=0.15\textwidth]{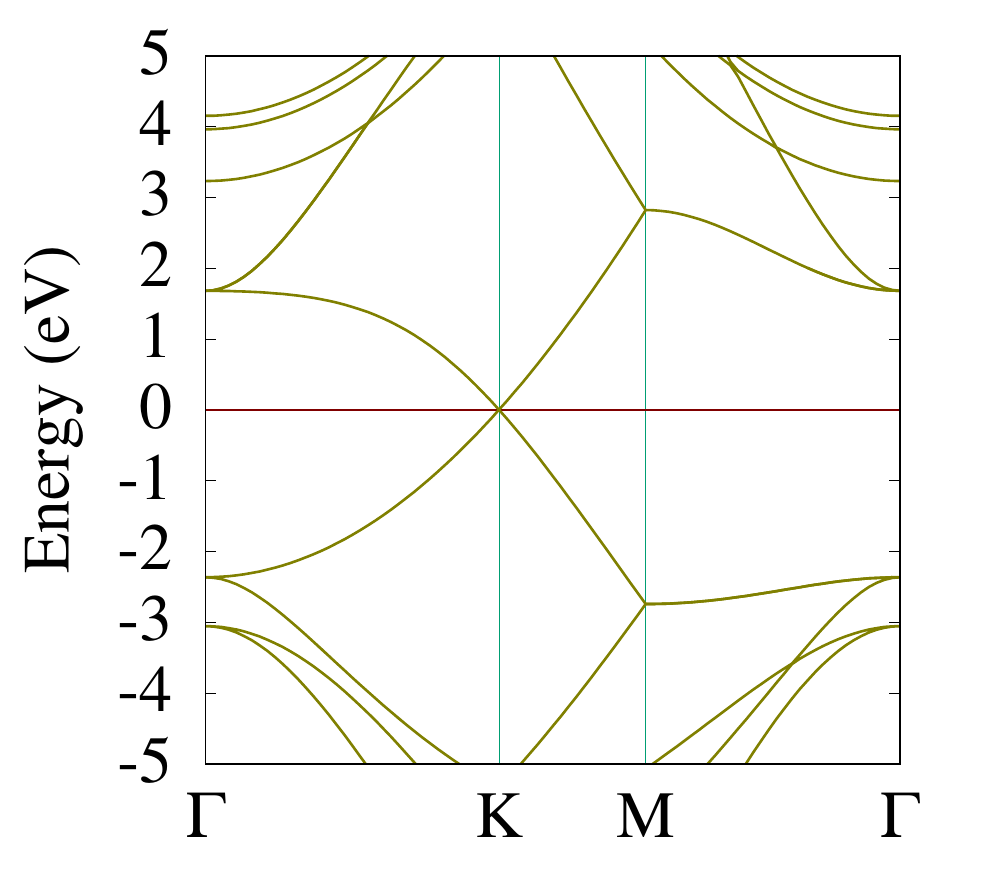} & 
	\includegraphics[width=0.15\textwidth]{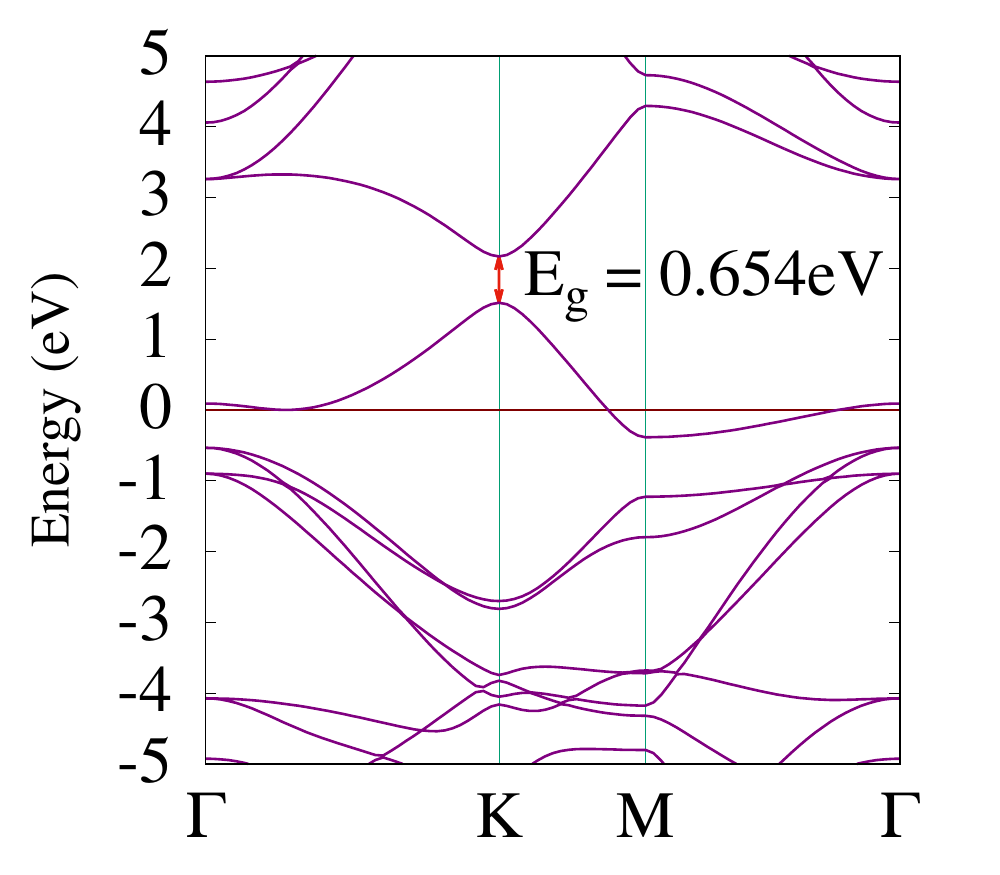} &
	\includegraphics[width=0.15\textwidth]{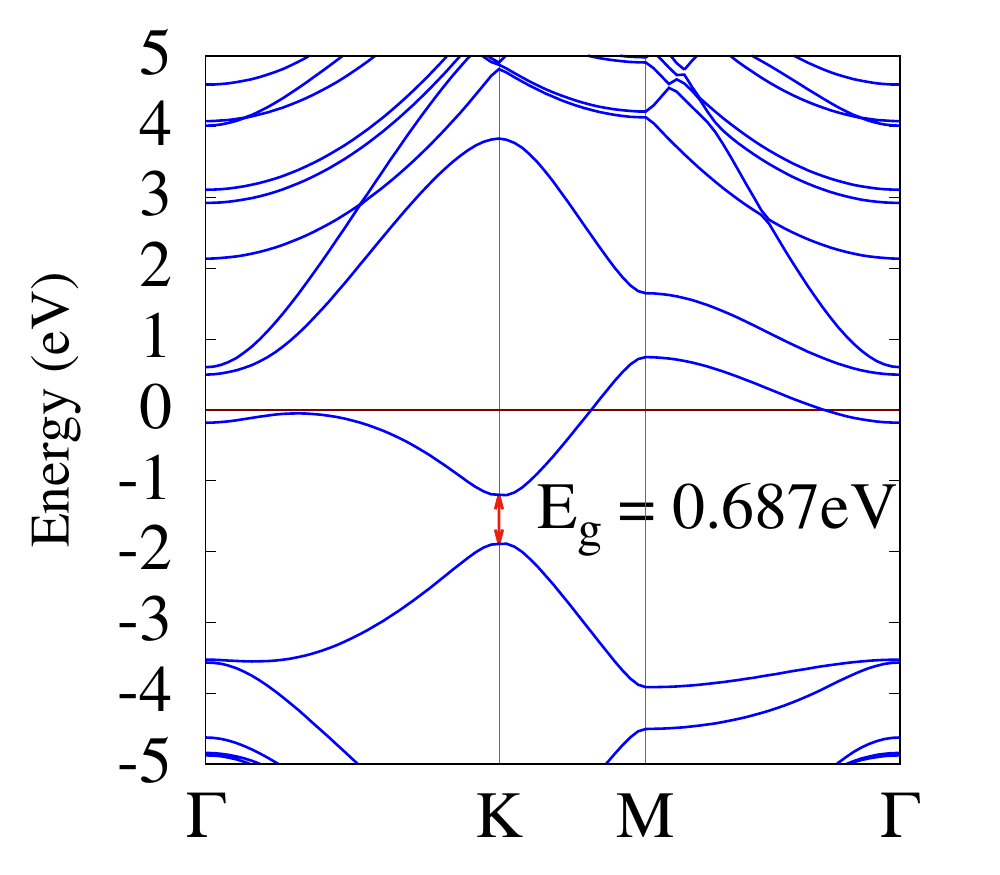} \\
	5 &
	\includegraphics[width=0.15\textwidth]{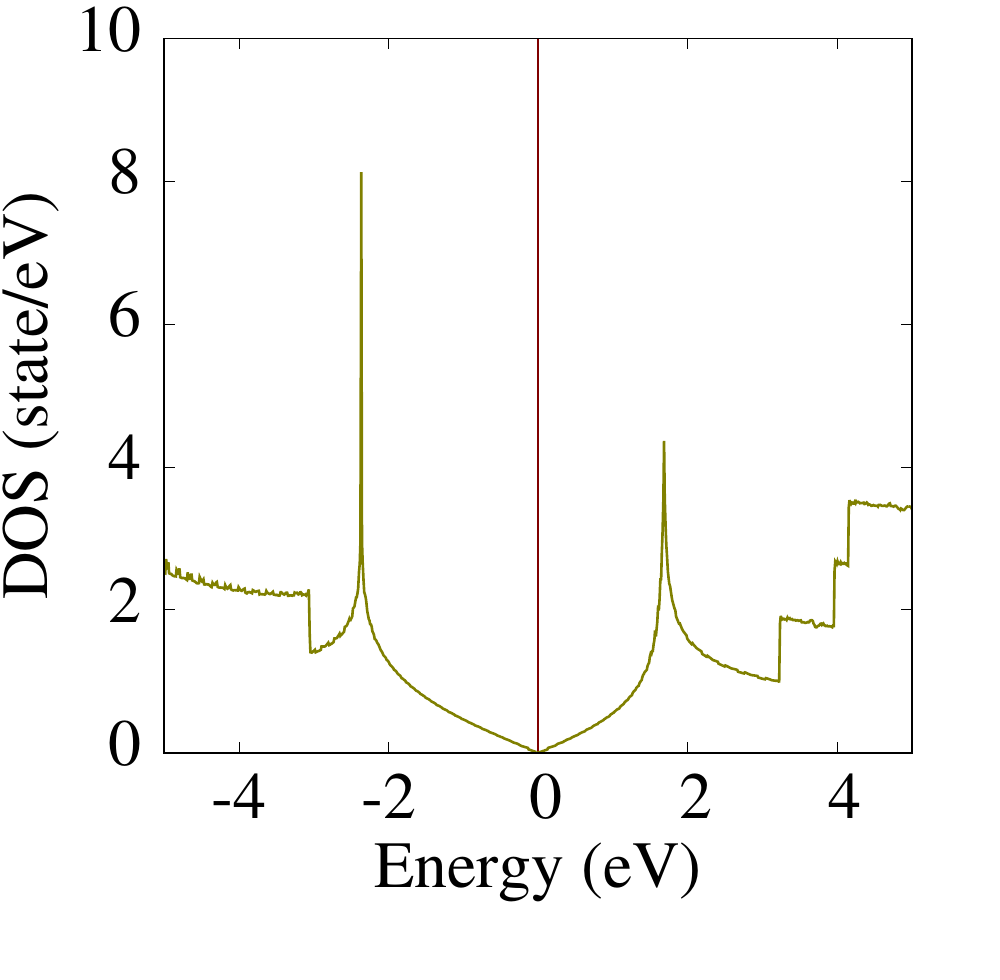} &
	\includegraphics[width=0.15\textwidth]{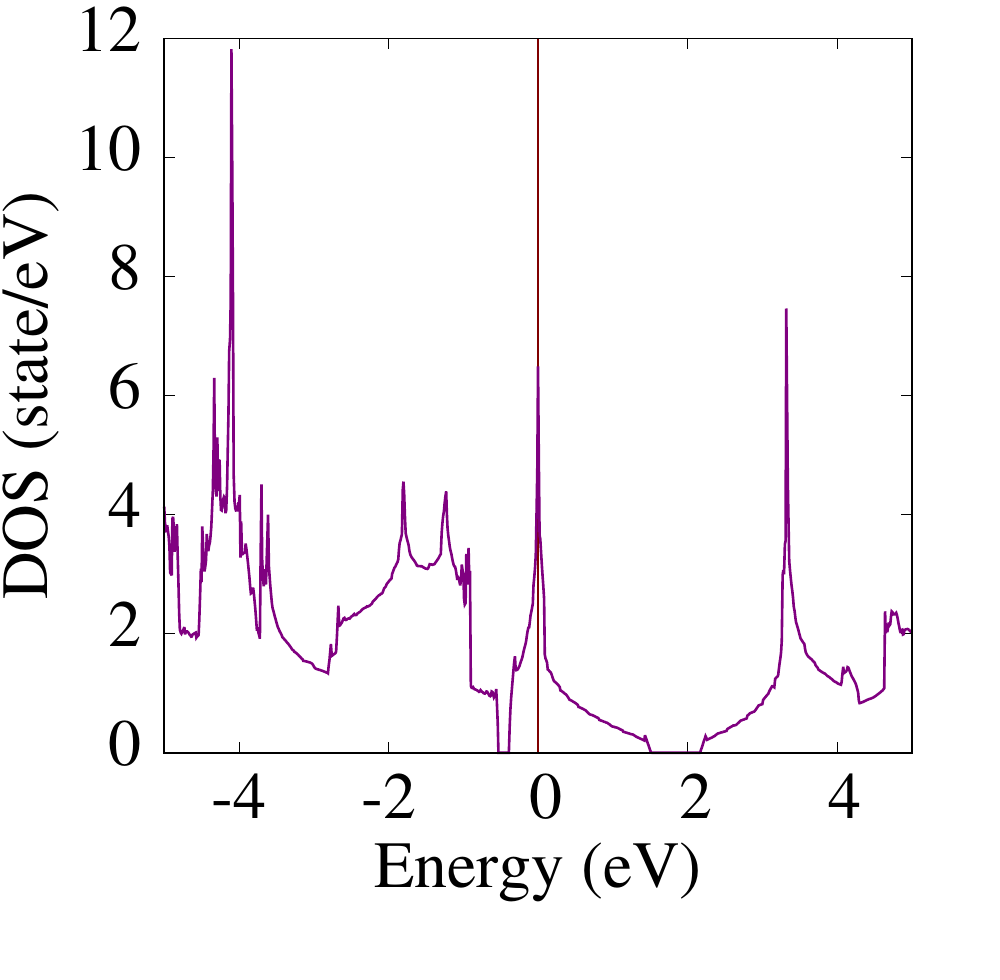} & 
	\includegraphics[width=0.15\textwidth]{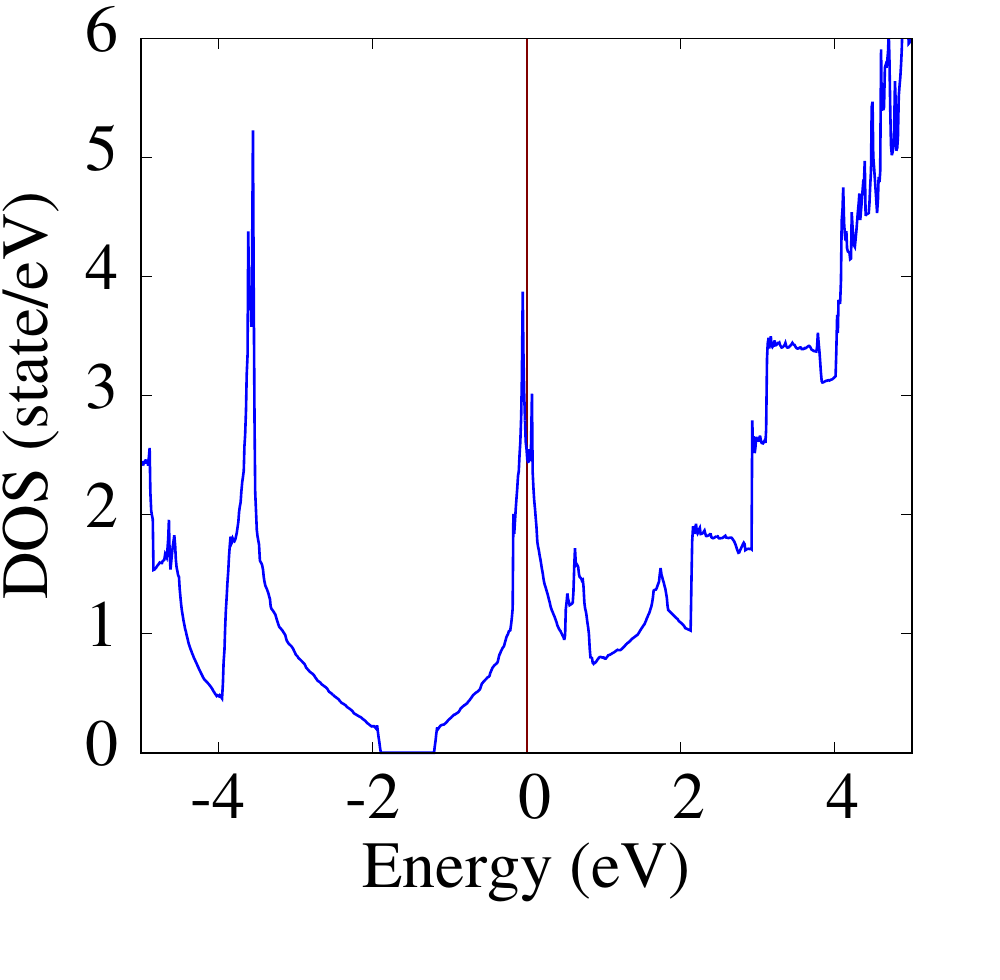} \\
	6 &
	\includegraphics[width=0.15\textwidth]{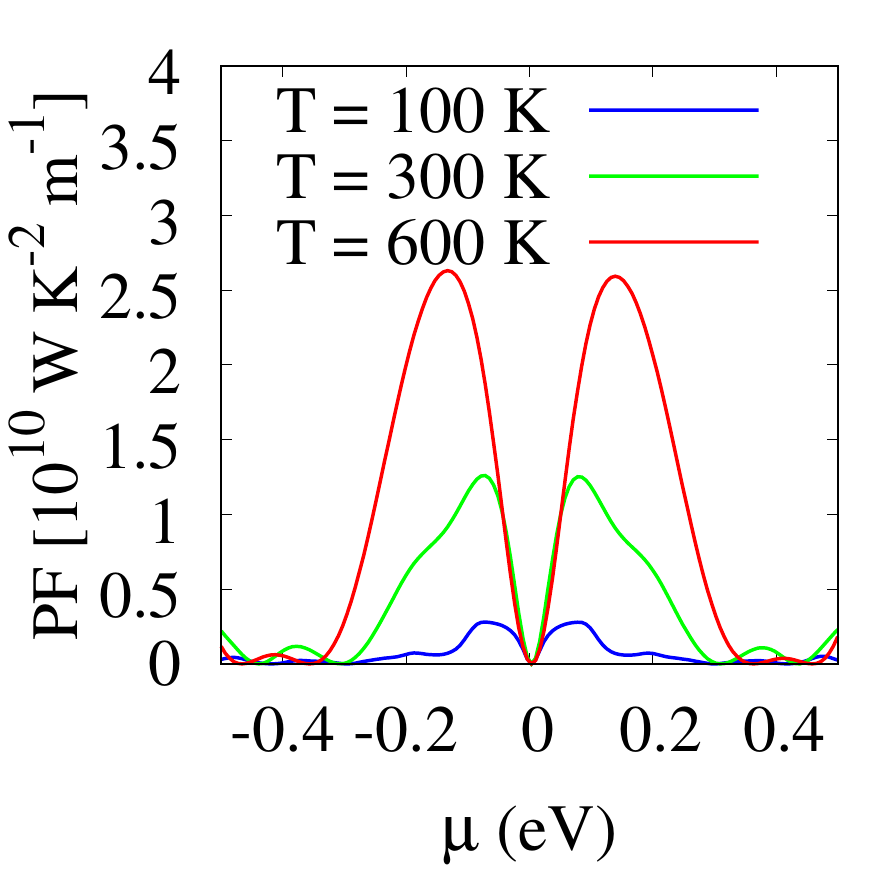} &
	\includegraphics[width=0.15\textwidth]{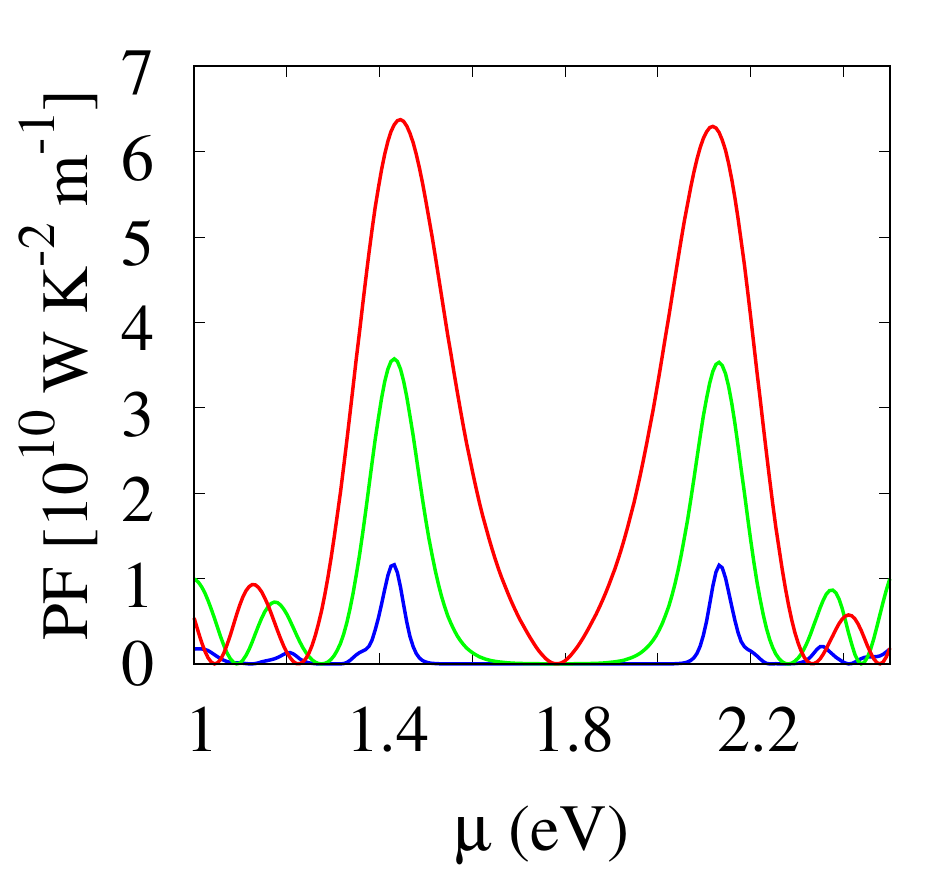} & 
	\includegraphics[width=0.15\textwidth]{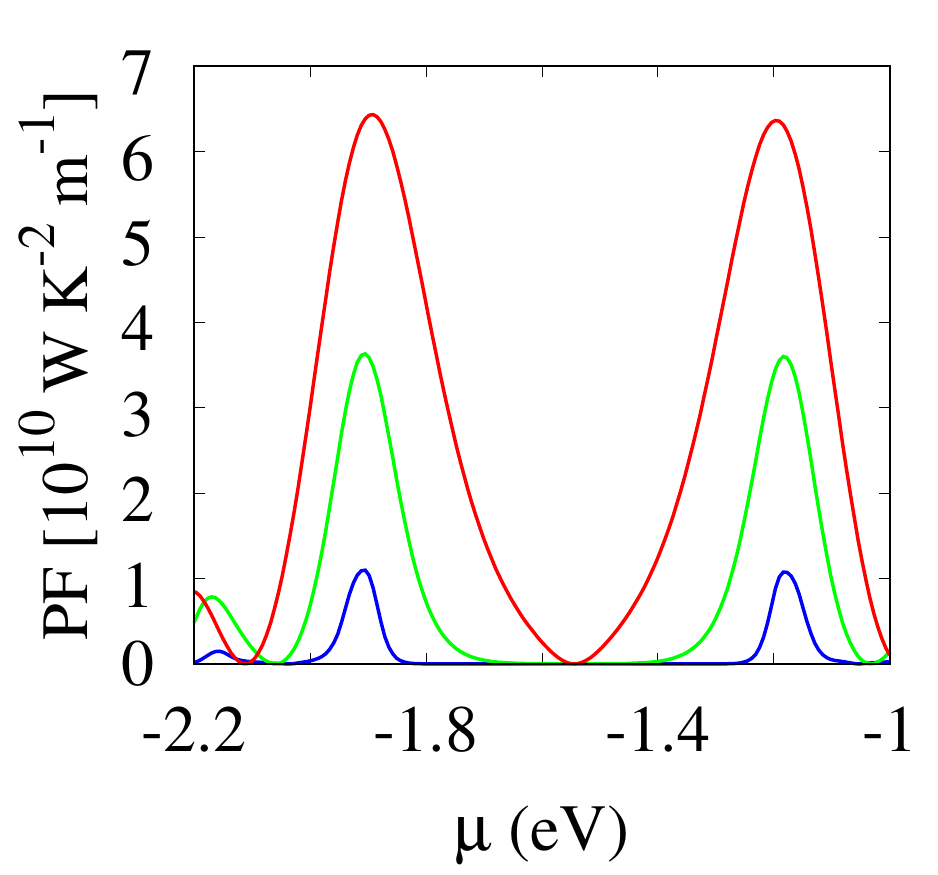} 
	\end{tabular} 
\end{table}
 \caption{First row shows the super-cell of pristine graphene (a1), B-doped graphene (b1), and N-doped graphene (c1).
 The bond length of the C-C atoms of pristine graphene is found to be $1.42$ ${\angstrom}$, 
 and the lattice constant becomes $a = 2.46$ ${\angstrom}$, and the C-B and C-N bond length are $1.494$ ${\angstrom}$ and 
  $1.405$ ${\angstrom}$, respectively.
 The corresponding electron density distribution, phonon dispersion, and the band structures are presented in the second, third, and fourth rows.
 In the fifth and sixth rows the corresponding density of state and the power factor 
 for three different values of temperatures, 100 (blue), 300 (green), and 600 K (red) are demonstrated.}
\label{fig02}
\end{figure}

Before presenting the effects of B or N doped atoms on the physical properties of the graphene, 
we demonstrate the results for pristine graphene in the leftmost column of \fig{fig02}(a). 
The super-cell consisting of $2\times2\times1$ is shown in (a1) where 
the bond length of the C-C atoms is observed to be $1.42$ ${\angstrom}$, and the lattice constant 
becomes $a = 2.46$ ${\angstrom}$, 
these values are in good agreement with the literature \cite{Avouris2010}. 
The electron density distribution is uniform for pristine graphene, with most of the electrons 
gathered between the C-C bonds, but rarely appearing in the center
of the hexagons, as it is shown in (a2) \cite{doi:10.1021/acs.jpcc.6b00136,Rauf_Abdullah_2016}. 
It is an indication of a strong covalent bond between the carbon atoms.
At this point, the $\pi$ electrons are inert and cannot contribute to the transport properties.
To ensure the dynamic stability, one may check the phonon dispersion that is shown in 
the third row of \fig{fig02}(a3). The positivity of the phonon dispersion of pure graphene indicates 
that the structure is stable \cite{MORTAZAVI2019733}.  	
The  acoustic modes, for the long waves around the $\Gamma$-points, are isotropic 
in the $xy$-plane as is appropriate for the symmetry of graphene \cite{FALKOVSKY20085189}.

In a pristine graphene sheet, where both C atoms
in the unit cell are equivalent, the $\pi$ and $\pi^*$ bands are
degenerate at K, leading to a linear crossing of the two bands.
The gap-less behavior of pristine graphene can be seen in the band structure (a4) where the Fermi level is at zero energy, and in turn, the
corresponding zero value of the DOS is found in (a5). Furthermore, the power factor of the 
pristine graphene versus the chemical potential 
is presented in the sixth row of \fig{fig02}(a) for three different values of temperature, 100 (blue), 300 (green), and 600 K (red).
The power factor is enhanced with increasing temperature since 
the Seebeck coefficient is directly proportional with temperature, $S \sim T$ \cite{C8TA03545H}.
The power factor is found to be $\backsimeq 2.5 \times 10^{10}$ WK$^{-2}$m$^{-1}$ for 600 K around the Fermi energy. 
The zero value of power factor at the Fermi energy is expected because the Seebeck coefficient 
is zero at the Fermi energy.

We start with the concentration ratio of $12.5\%$ where a single B- or N-atom is doped into a para-position of
the super-cell of the graphene nanosheet shown in \fig{fig02} for B-doped graphene (BG) in column b and 
N-doped graphene (NG) in column c. 
Although the B and the N atoms are expected to behave like p-type (or acceptor) and n-type (or donor) dopants, respectively,  
we observe a local depletion of the  electron distribution around the B atom, and a local excess  around the N atom,  
as shown in the second row of \fig{fig02}. These results are consistent with other calculations \cite{Wu_2010}.
Although, the atomic radius decreases and the electronegativity increases in the sequence of B $>$ C $>$ N, 
the average bond lengths of C-B is $1.494$ ${\angstrom}$ which is larger than that of C-N, 
$1.405$ ${\angstrom}$, respectively.

In the doped structure, it is important to see the most energetic stable structure.
To check the energetic stability of doped graphene we need 
to calculate the formation energy ($E_f$) via
\begin{equation}
      E_f = E_{\rm T} - N_{\rm C} \, \mu_{\rm C} - \sum_{i} N_i \, \mu_i.
\end{equation}
Herein, $E_{\rm T}$ is the total energy of the doped graphene system, $N_{\rm C}$, and $\mu_{\rm C}$ 
refer to the number and chemical potential of the carbon atoms, respectively, and $N_i$ and $\mu_i$ are 
the number and chemical potential of the doped atoms in the graphene nanosheet, respectively.
From the formation energy equation, we find that the formation energy of BG is $-27.309$ eV which 
is larger than that of NG, $-30.936$ eV.
The smaller formation energy, the more energetic stable the structure should be observed to be. 
As a result, the NG is more energetic stable than the BG system \cite{PhysRevLett.90.046103,C4NR03247K}.
We should remember that the formation energy is obtained from total energy calculations 
and does not take into account dynamic terms. The phonon dispersion delivers information about 
the dynamic stability, and the phonon dispersion of NG (see \fig{fig02}(c3)) demonstrates
a structure less stable dynamically, when compared to pure and B-doped graphene because of a negativity 
in the out-of acoustic plane mode (ZA). 
In addition, the acoustic modes are anisotropic in the $xy$ plane indicating a symmetry breaking of the 
two different sites by the B- and N-atoms.

The corresponding band structures are shown in the fourth row of \fig{fig02}, 
where the valence bands cross the Fermi energy and open up the bandgap in the BG (b3)
while the conduction bands cross the Fermi energy opening 
the bandgap in the NG (c3) with almost the same bandgap, $E_g \backsimeq 0.65$ eV. 
The crossing of the Fermi energy in both structures is expected because the B(N) atom has an electron less(more) than the C atom, respectively. A downward- or an upward-shift of the Fermi energy is 
thus seen in their band structures.
These changes in the band structure are related to the symmetry breaking of the 
graphene structure due to the B- or N-doped atom.
This can be seen from the wave functions of the $\pi$ and $\pi^*$ bands which are predominantly composed of atomic orbitals with an orientation perpendicular to the $xy$-plane, i.e.\ $p_z$ orbitals.
It is known that the $\pi$-band wave function is predominantly located at the N atom and the $\pi^*$-band wave function at the B atom; this is because of the higher electronegativity of an N atom. Therefore, it is the strong difference in electronegativity between B or N and C atoms that leads to the band gap \cite{RevModPhys.82.1843}.

The upward- and downward-shift of the band structure and the opening of a bandgap lead to the zero 
value of the DOS around $2$ eV and $-2$ eV in both the BG and NG, respectively, as is displayed 
in the fifth row. The crossing of valence bands in BG and conduction bands in NG 
increases the power factor of the system to almost $2.5$ times larger than that of the pristine graphene [see \fig{fig02}(b6)(c6)].
This enhancement refers to the increase of the Seebeck coefficient in both systems.
We note that both B and N doped graphene have almost the same power factor because both structures have almost the same bandgap and the valence and conduction bands 
are shifted by the same amount of energy.

\begin{figure}[htb]
	\centering
	\includegraphics[width=0.23\textwidth]{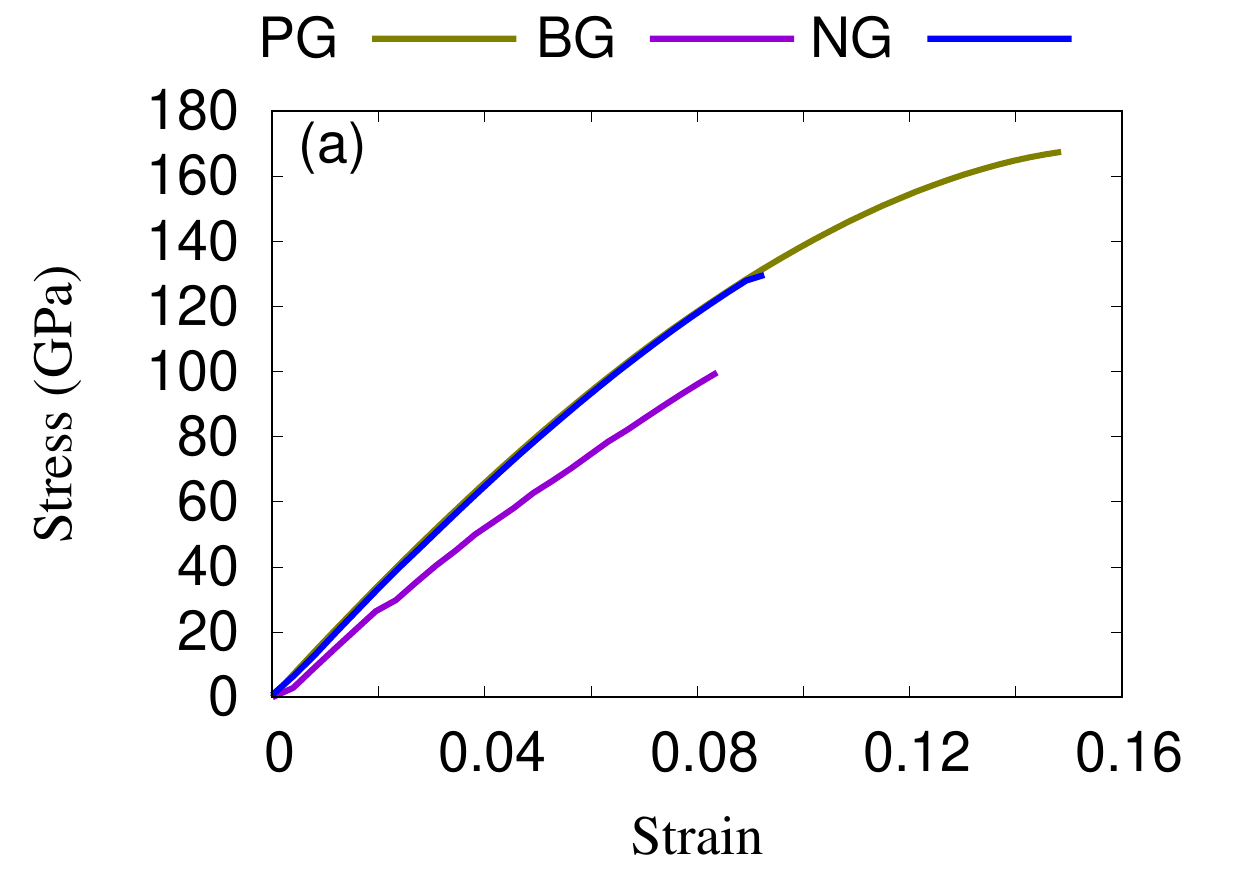}
	\includegraphics[width=0.23\textwidth]{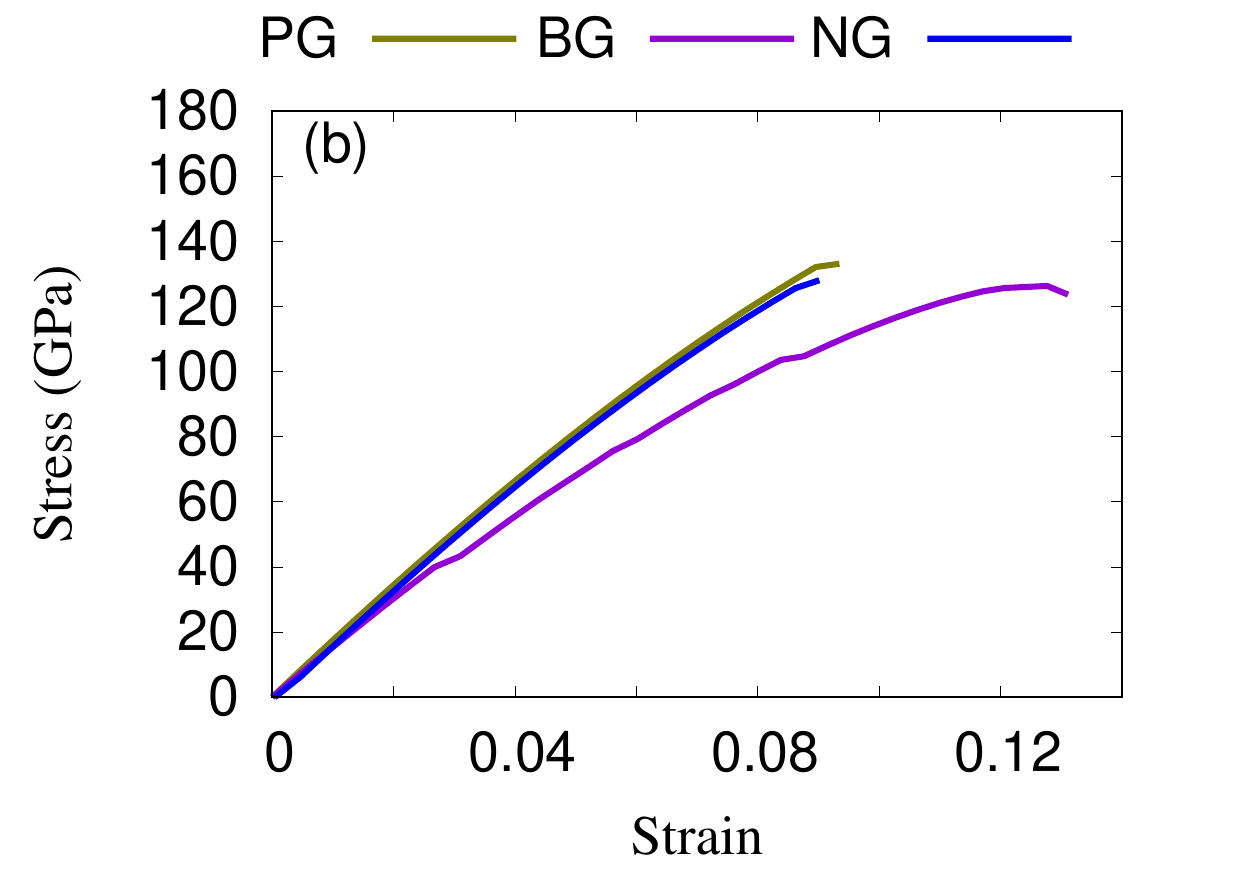}
	\caption{ Uniaxial stress-strain of PG (olive green), BG (purple), and (blue) monolayers along 
		$x$-axis (Zigzag) and $y$-axis (Armchair).}
	\label{fig02_1}
\end{figure}

The mechanical characteristic of PG, BG and NG have been studied by performing 
uniaxial tensile simulation \cite{MORTAZAVI2019733}. Results for the uniaxial stress 
found from DFT calculations of the PG, BG, and NG 
sheets along $x$- (a) and $y$-axis (b) are compared in \fig{fig02_1}. The elastic moduli of PG 
in the $x$-direction is higher than that of $y$-direction which is similar to elastic moduli 
found by Dewapriya, and et.\ al.\ \cite{Dewapriya_2013} for PG.
The linear elastic region along the $x$- and $y$-axis is similar for PG and NG indicating isotropic 
elastic response but for BG the elastic region is anisotropic character. 
Furthermore, the elastic moduli of PG is high in $x$-axis, 170 GPa, and 
$y$-axis, 140 GPa, comparing to BG and NG. This is attributed due to the stronger and larger 
binding energy of a C-C bond than a C-B or a C-N bond.

We emphasize that the position of an N- or a B-atom in the graphene structure does not influence the 
physical characteristics of the system in the case of a single B or N doped atom in a honeycomb structure. 
It means that if the N- or B-atom is substitutionally doped into the ortho or meta position, 
similar qualitative and quantitative results are obtained (not shown).

\begin{figure}[H]
\begin{table}[H]
  \captionsetup{labelformat=empty}
\noindent
\begin{tabular}[]{ >{\centering\arraybackslash}m{0cm} >{\centering\arraybackslash}m{2.4cm} >{\centering\arraybackslash}m{2.4cm} >{\centering\arraybackslash}m{2.4cm} }
	& a (nonb-BG) & b (b-BG-1) & c (b-BG-2)\\ 
	1 &
	\includegraphics[width=0.15\textwidth]{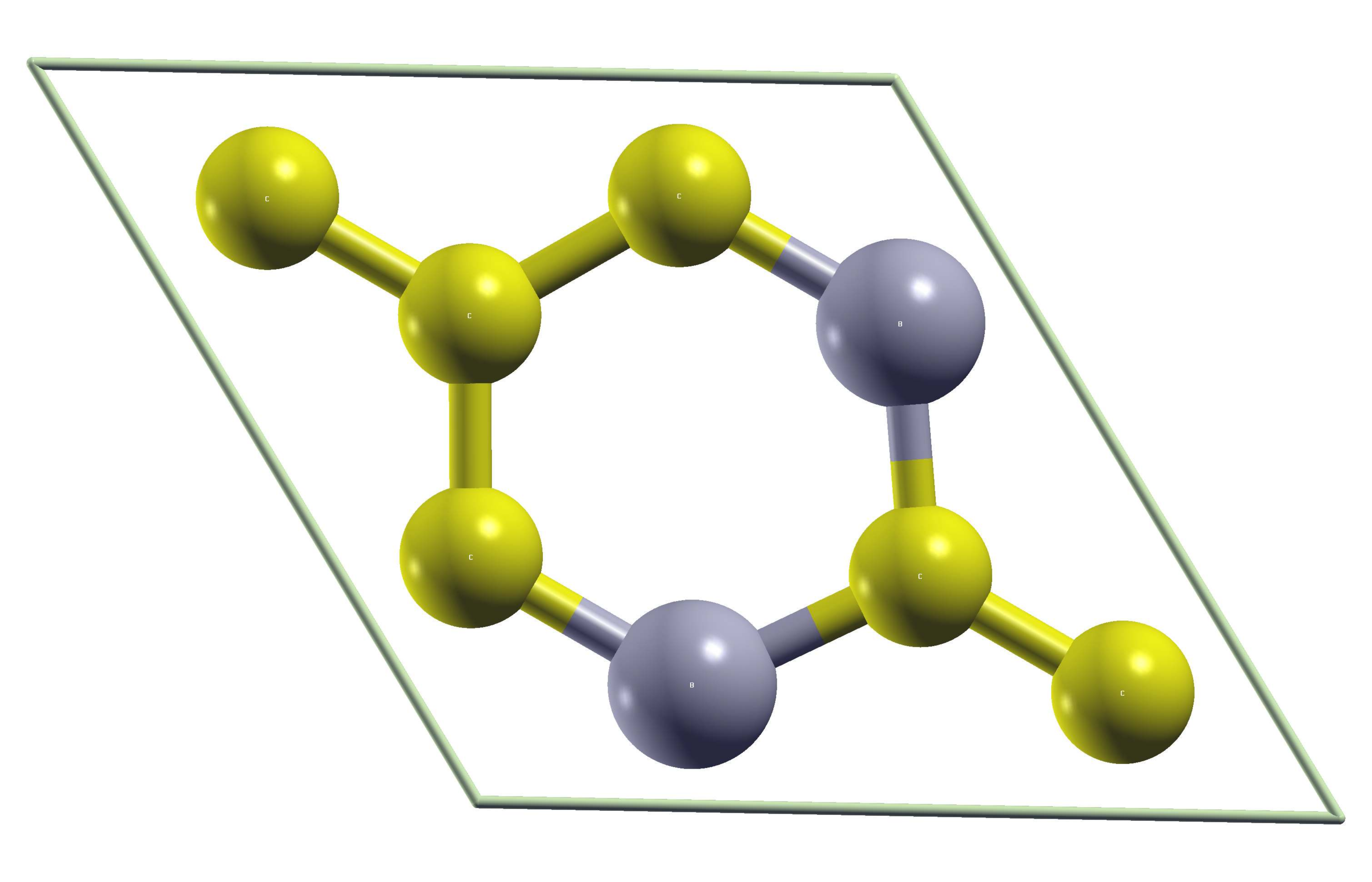} & 
	\includegraphics[width=0.15\textwidth]{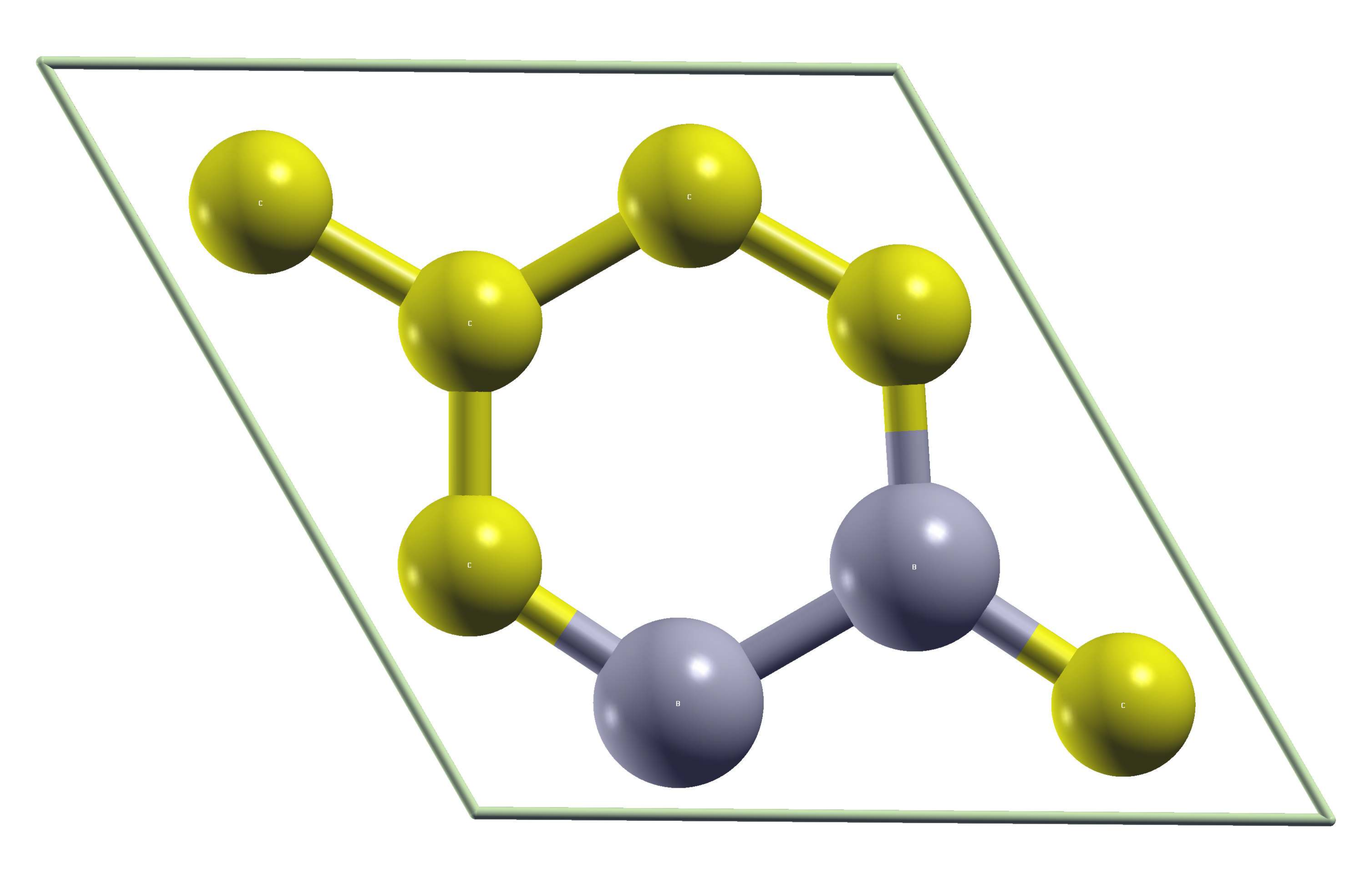} & 
	\includegraphics[width=0.15\textwidth]{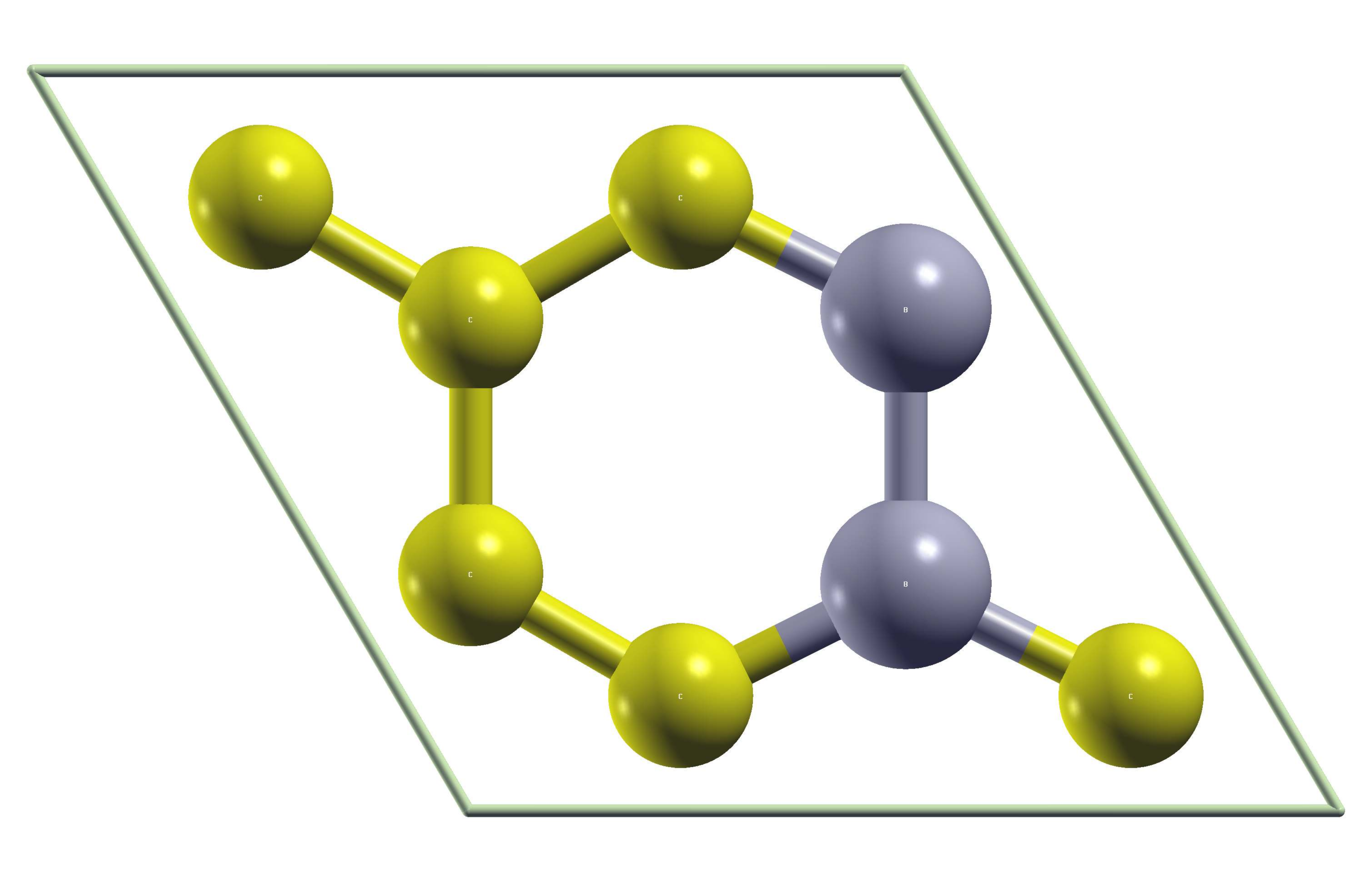} \\
	2 &
	\includegraphics[width=0.15\textwidth]{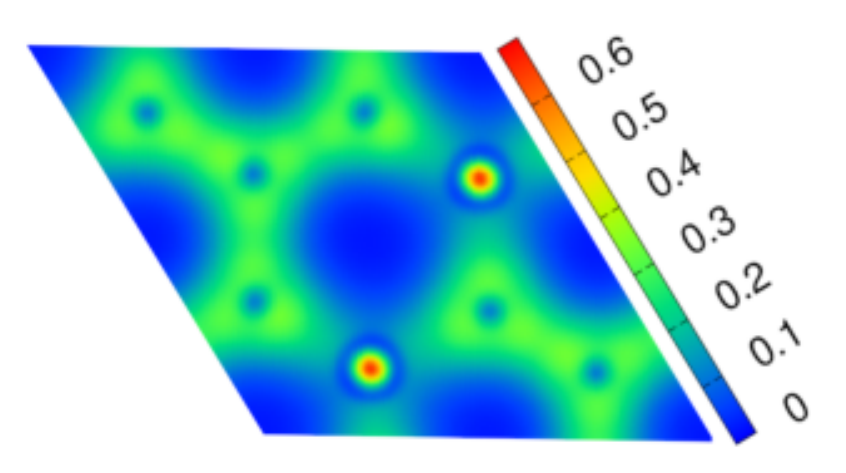}  &
	\includegraphics[width=0.15\textwidth]{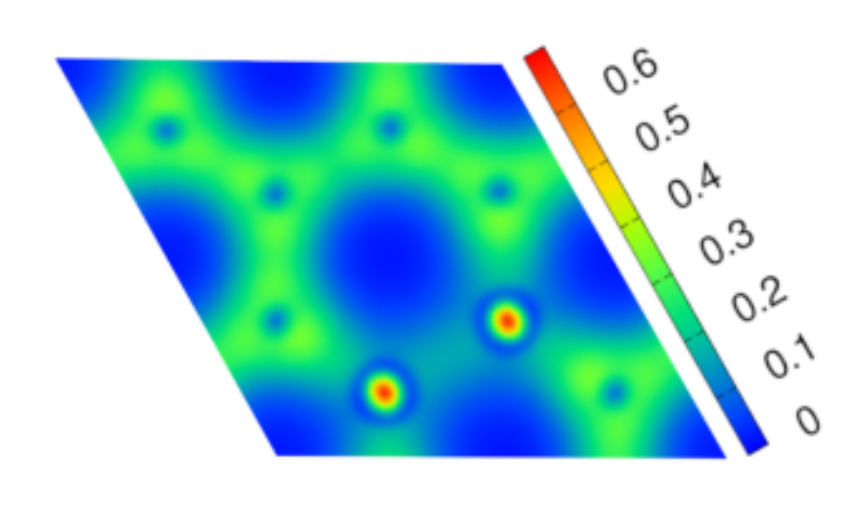} &
	\includegraphics[width=0.15\textwidth]{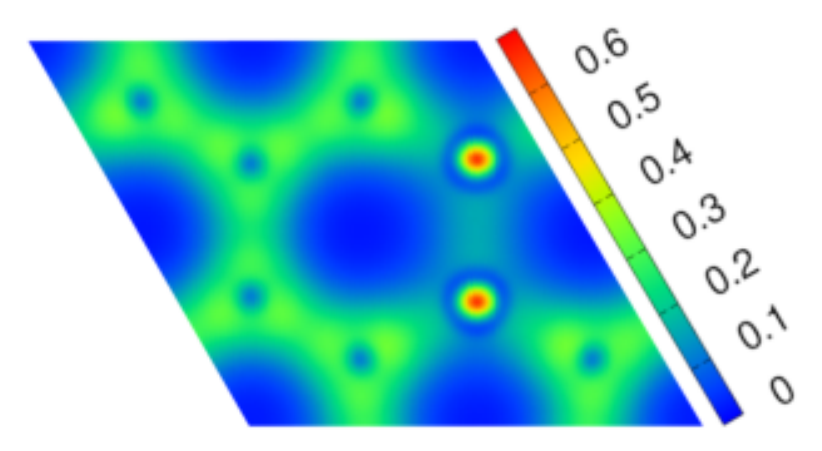} \\ 
	3 &
	\includegraphics[width=0.15\textwidth]{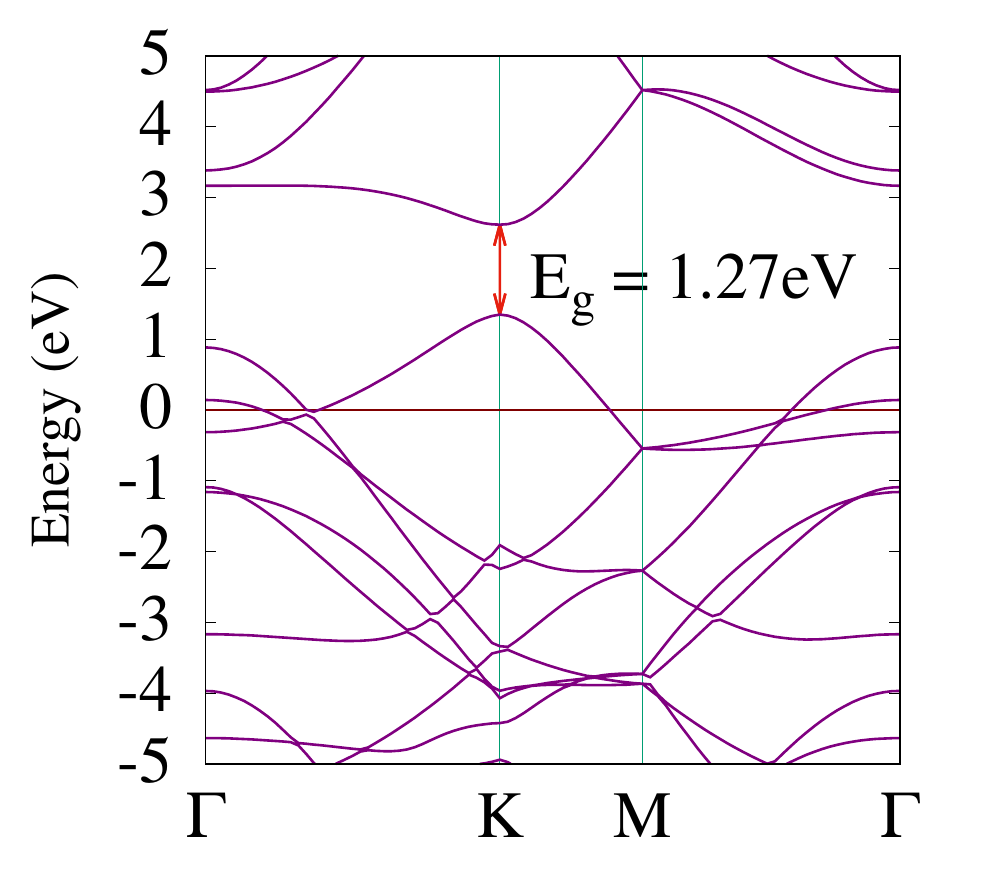} & 
	\includegraphics[width=0.15\textwidth]{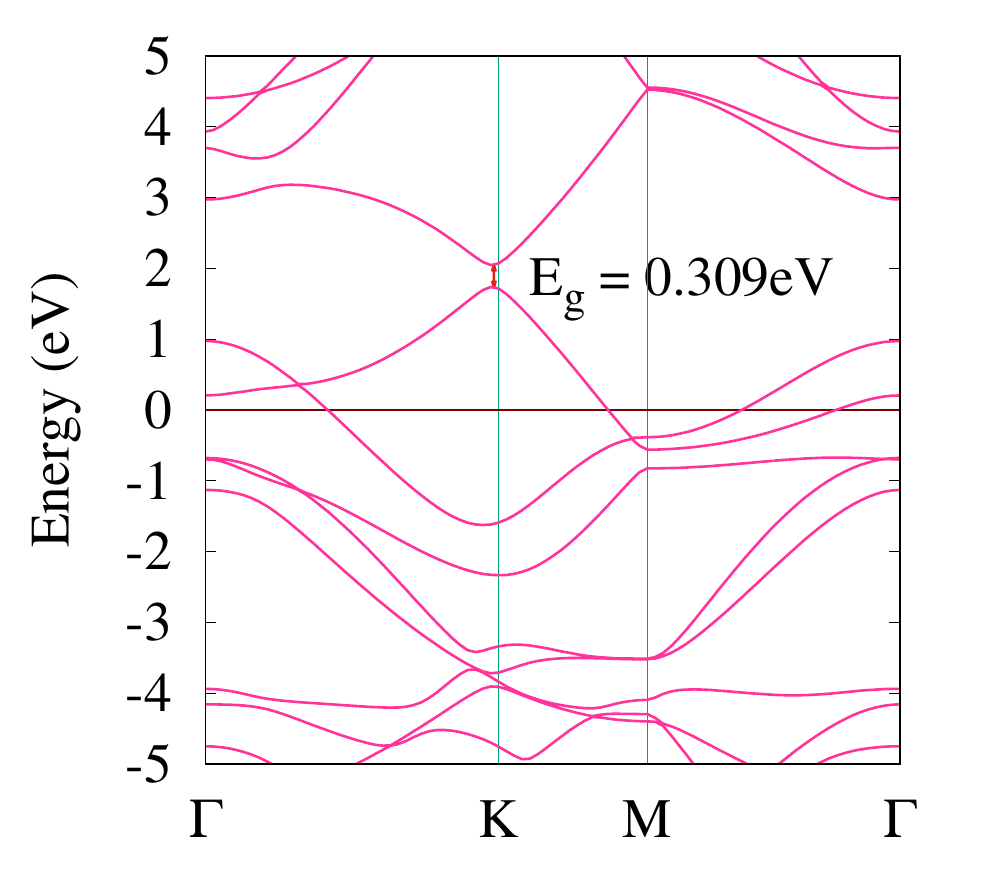} &
	\includegraphics[width=0.15\textwidth]{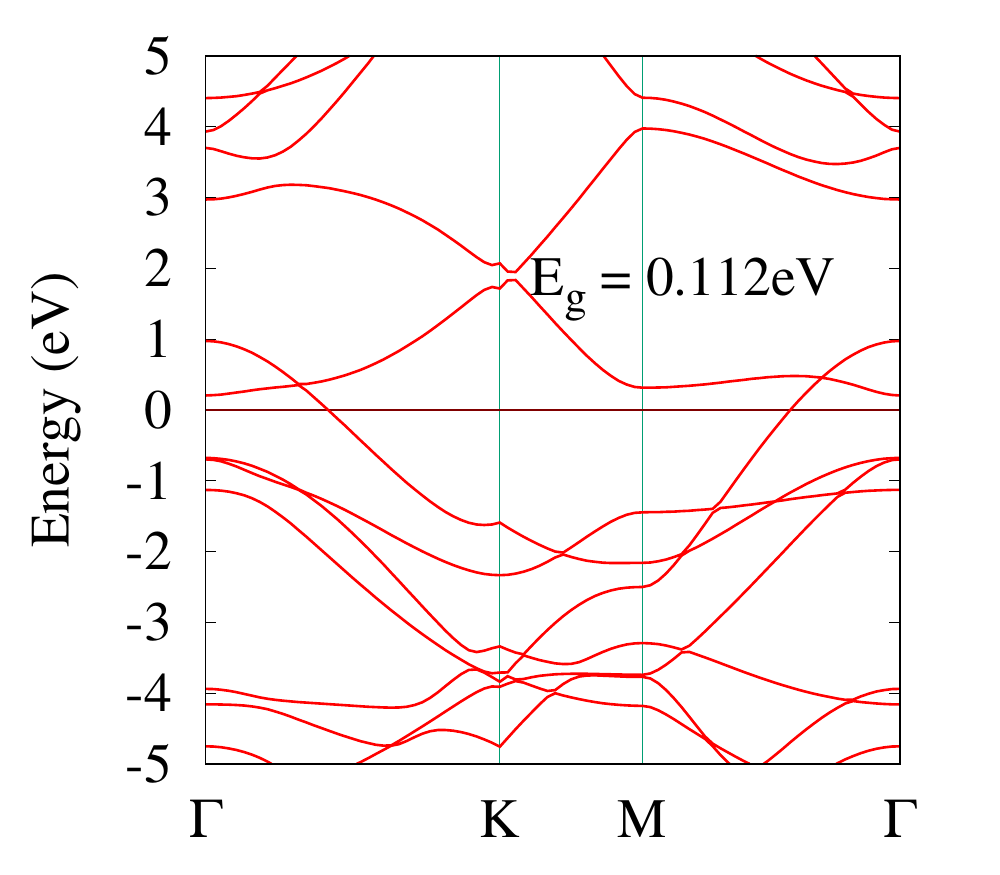} \\
	4 &
	\includegraphics[width=0.15\textwidth]{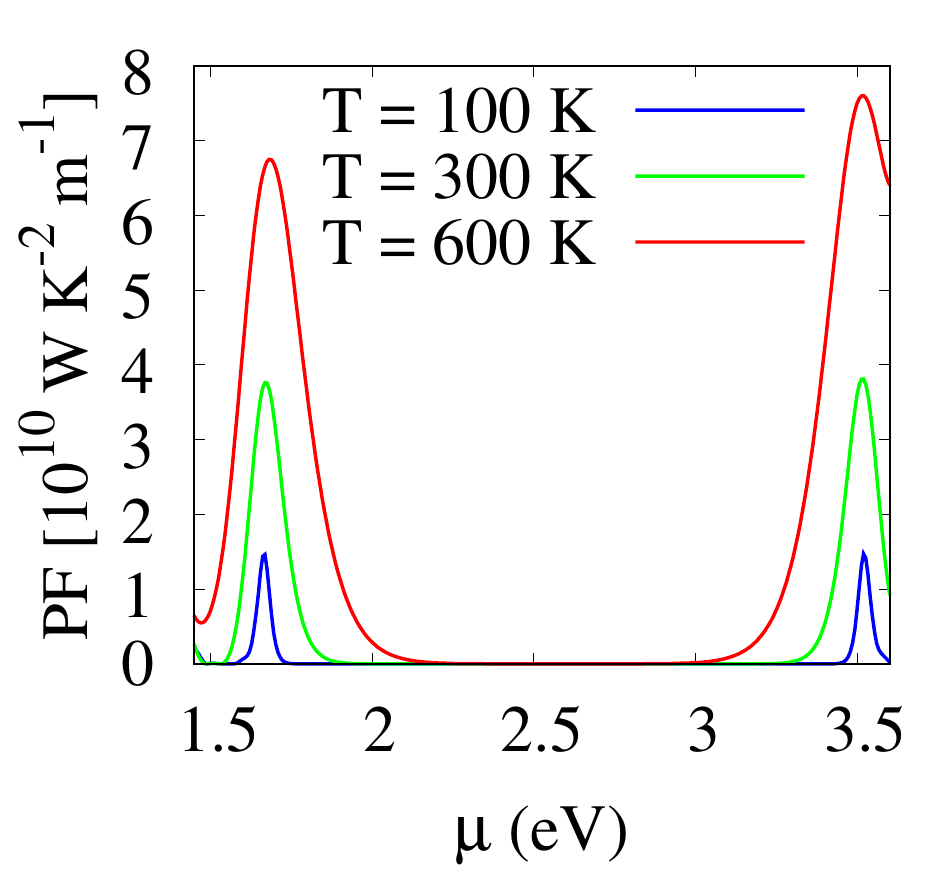} &
	\includegraphics[width=0.15\textwidth]{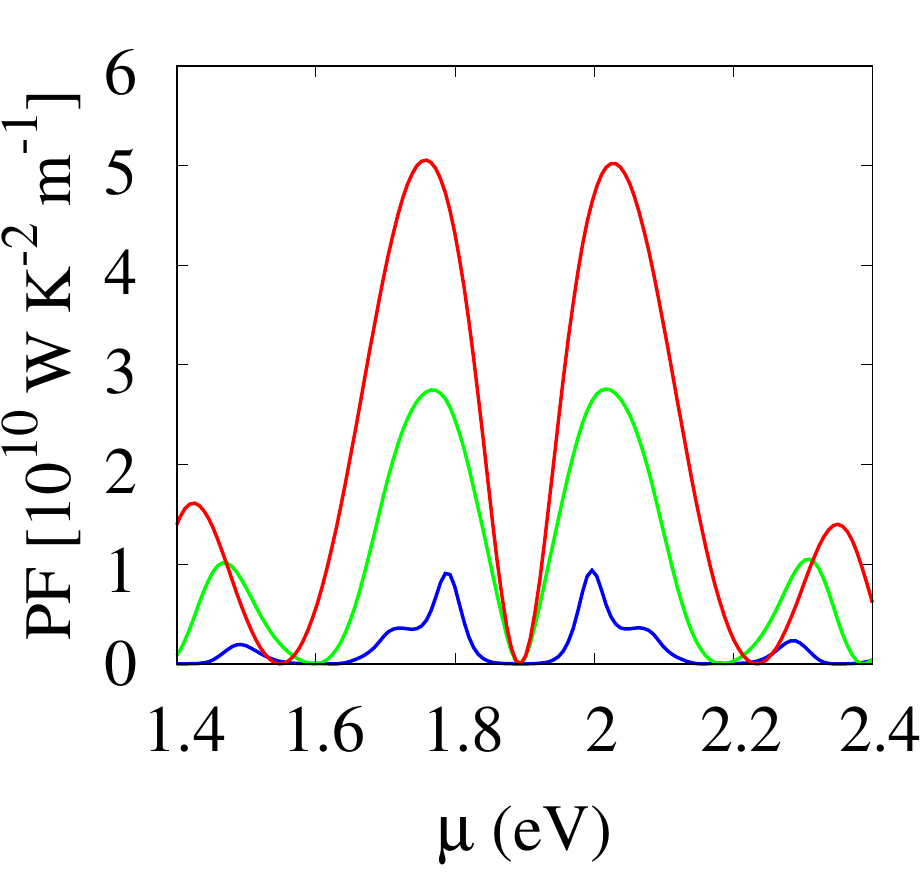} & 
	\includegraphics[width=0.15\textwidth]{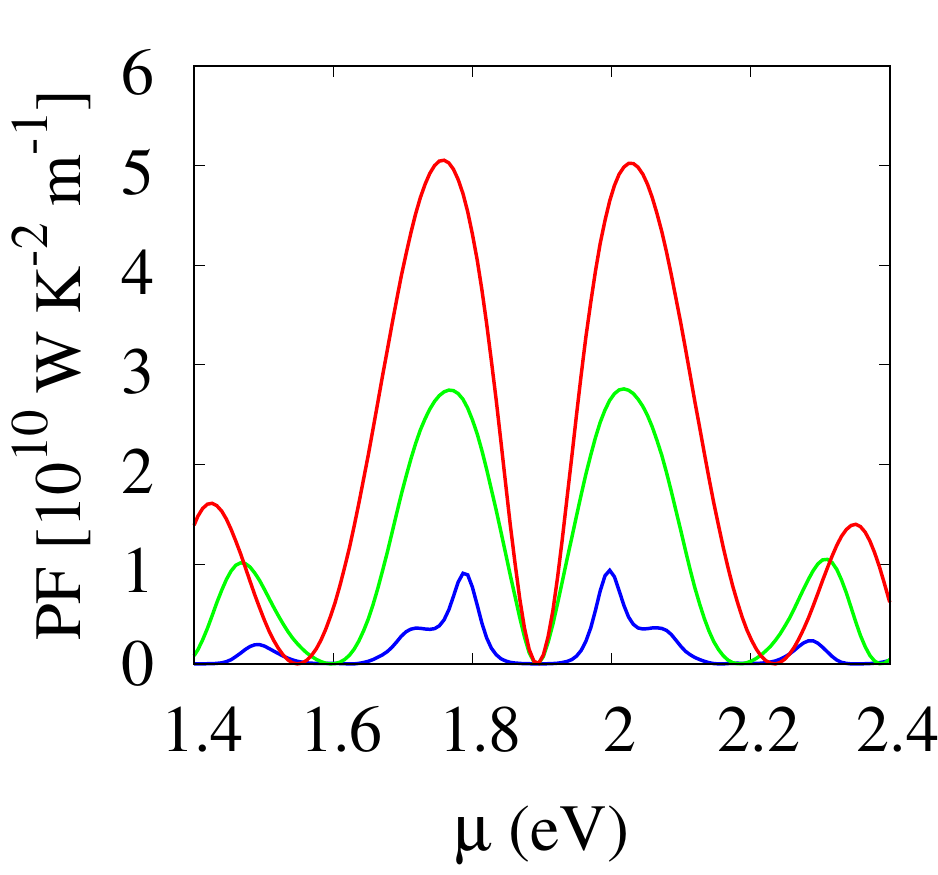} 
	\end{tabular} 
\end{table}
 \caption{First row shows the super-cell of nonb-BG (a1), b-BG-1 (b1), and  b-BG-2 (c1).
 The second row indicates the charge density distribution of the 
 ortho-para (a2), meta-para (b2), and  ortho-meta (c2) two B-atoms doped graphene structure.
 Third row demonstrates the band structure of  ortho-para (a3), meta-para (b3), and  ortho-meta (c3) 
 two B-atoms doped graphene structure.
 The fourth row displays the power factor of  ortho-para (a4), meta-para (b4), and  ortho-meta (c4) 
 two B-atoms doped graphene structure 
 for three different values of temperatures, 100 (blue), 300 (green), and 600 K (red).}
\label{fig03}
\end{figure}

The doping ratio can be further increased to $25\%$ which leads to two atoms in the $2\times2\times1$ super-cell. 
For simplicity, we show only two B-atoms doped into the graphene structure as is shown 
\fig{fig03} where the two 
B-atoms are doped into the ortho and para (a1), meta and para (b1), and ortho and meta (c1) positions.

These three structures can be classified as the bonding and the non-bonding B-atoms, where 
bonding B-doped graphene means that the B atoms make bonding between themselves, 
but in the non-bonding B doped graphene there is no B-B bonding. 
For example, the two B-atoms doped at the ortho-para positions is called non-bonding B-doped 
graphene (nonb-BG) while the other two structures are called bonding B-doped graphene (b-BG). 
We identify the b-BG as b-BG-1 and b-BG-2 which refer to the two B-atoms doped at meta-para and ortho-meta position, respectively.   

In general, a structure with a high doping ratio is less energetically and dynamically stable than one with a low doping concentration \cite{Rani2014, C5RA25239C}. 
The BG with two B-atoms is also less stable compared to the BG with one B-atom, because
the formation energy of the BG with two B-atoms is larger than that of BG with one B-atom and 
negativity is seen in phonon dispersion in the case of two B atoms doped graphene (not shown) \cite{C5RA25239C}. It shows that the two B atoms doped graphene is dynamically less stable than that of one B atom doped graphene.

In addition, we emphasize that the position of the B-doped atoms with high concentration ratio in 
the honeycomb structure of graphene influences its energetic and dynamic stability. For instance, the formation energy of 
the ortho-para BG forming nonb-BG (a1) is $-22.904$ eV which is a slightly smaller than that of 
the meta-para position in b-BG-1 (a2), $-22.328$ eV, and ortho-meta postion in b-BG-2 (a3), $-22.328$ eV.
We can thus confirm that the nonb-BG is the most energetic stable structure. Furthermore, 
the negativity of phonon dispersion for the nonb-BG is less than that of b-BG. We can thus 
say that the nonb-BG is dynamically more stable than b-BG (not shown).

The interaction energy should be taken into account in the case of nonb-BG and b-BG structures, and  
one can calculate the interaction energy between the doped atoms using \cite{doi:10.1063/1.5018065}
\begin{equation}
 \Delta E_{\rm int} =  E_{\rm T}({\rm pg})  + E_{\rm T}(2B) - 2 \times E_{\rm T}(1B),
\end{equation}
where $E_{\rm T}({\rm pg})$ is the total energy of the pristine graphene, 
and $E_{\rm T}(1B)$ and $E_{\rm T}(2B)$ are the 
total energies of the BG with one atom and two B-atoms, respectively.
The negative (positive) sign of the interaction energy displays that the two B-atoms 
attract (repel) each other, respectively \cite{doi:10.1063/1.5018065}.
The interaction energy between the two B-atoms in the nonb-BG (a1) is $-0.201$ eV indicating 
an attractive interaction, while the interaction energies of the b-BG with two B-atoms (b1) and (c1) 
are equal to $0.374$ eV demonstrating the repulsive force between the two B-atoms. 
The repulsive interaction increases the bonding length of B-B atoms, comparing 
to the C-C bonding length, which is $1.6303$ ${\angstrom}$.
The repelling of two B atoms in b-BG confirms the weaker B-B bond comparing to the B-C bond in nonb-BG. 
This effect can be seen in the electron charge distribution shown in 
the second row of \fig{fig03}, where the charge distribution between the covalent bond of B-B atoms is weaker than that of the C-B bond.

The attractive interaction between the B atoms in the nonb-BG leads to the average C-B bonding length 
of $1.528$ ${\angstrom}$ and the bandgap is thus further opened up to $1.27$ eV (a3).
It has been shown in the literature that with increasing concentration ratio, 
the bandgap is also increased \cite{C2RA22664B}. 
Our results of BG with two B-atoms confirm that not 
only the concentration ratio can increase the bandgap but also the position of doped atoms play a major 
role in the determination of the bandgap.
For example, the bandgap of the b-BG-1 and b-BG-2 is $0.309$ eV (b3), and $0.112$ eV (c3), respectively. It is decreased compared to the BG with one B-atom. This is attributed to the 
repulsive interaction between the B-atoms.
As a result, the power factor is further enhanced in the nonb-BG (a4) while 
it is decreased for the b-BG-1 and b-bG-2 shown in (b4) and (c4) compared to the BG with one B-atom.

We note that the same scenario of the BG with two B-atoms happens for 
the NG with two N-atoms except that the Fermi energy crosses the conduction band.
Therefore, we do not present the results of NG with two N-atoms.
In addition, our results of two bonded and non-bonded B atoms in graphene structure can be applied 
if the two B atoms are located at different isomer positions. It means that if the two B-atoms are located at para-para psoition,
different result should be obtained.

\begin{figure}[H]
\begin{table}[H]
  \captionsetup{labelformat=empty}
\noindent
\begin{tabular}[]{ >{\centering\arraybackslash}m{0cm} >{\centering\arraybackslash}m{2.4cm} >{\centering\arraybackslash}m{2.4cm} >{\centering\arraybackslash}m{2.4cm} }
	& a (nonb-BG) & b (b-BG-1) & c (b-BG-2) \\ 
	1 &
	\includegraphics[width=0.15\textwidth]{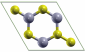} & 
	\includegraphics[width=0.15\textwidth]{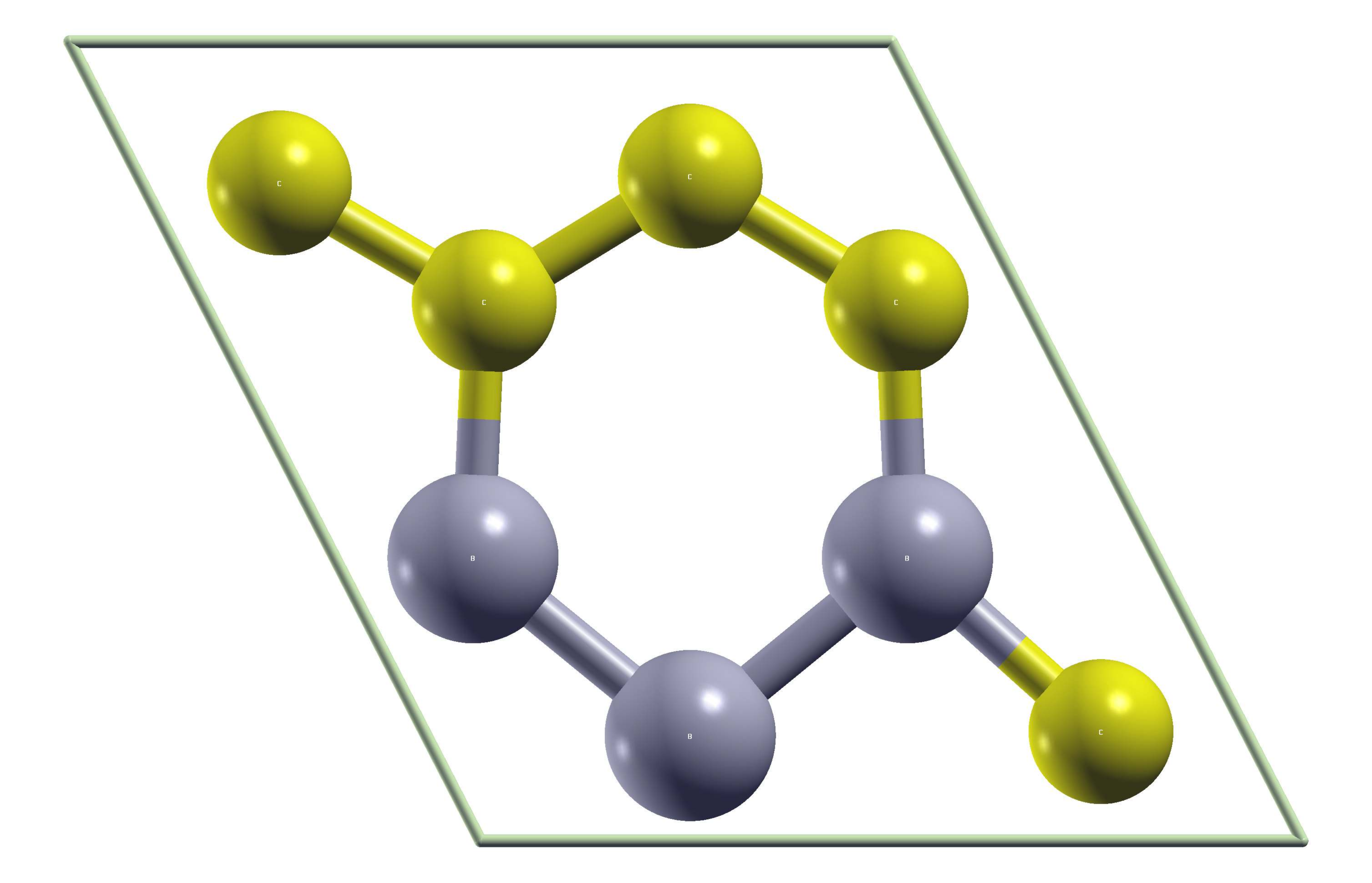} & 
	\includegraphics[width=0.15\textwidth]{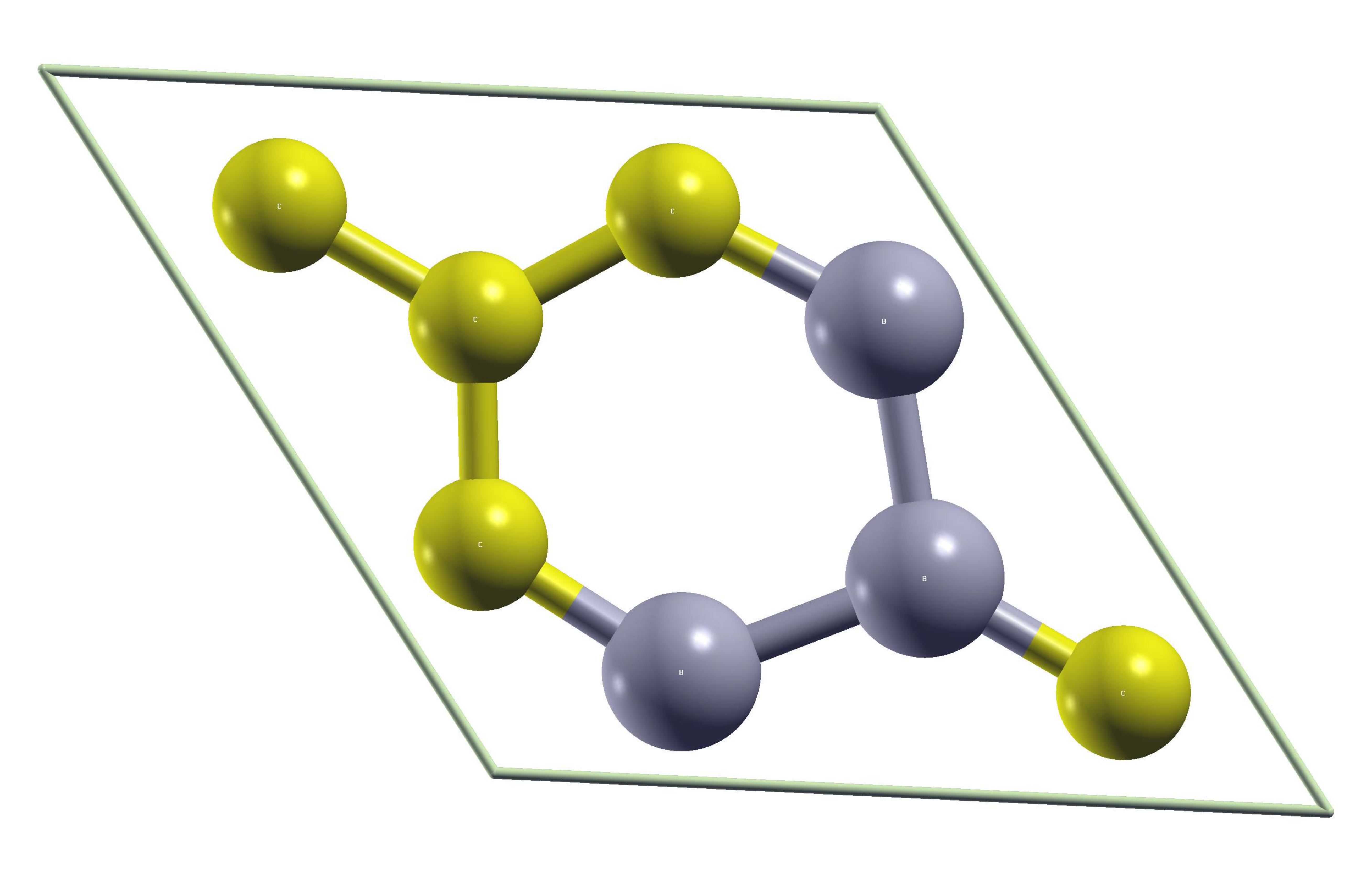} \\
	2 &
	\includegraphics[width=0.15\textwidth]{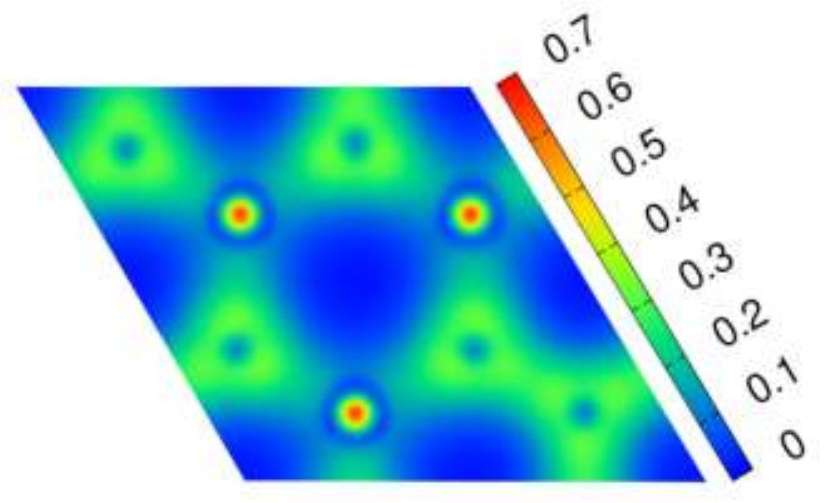}  &
	\includegraphics[width=0.15\textwidth]{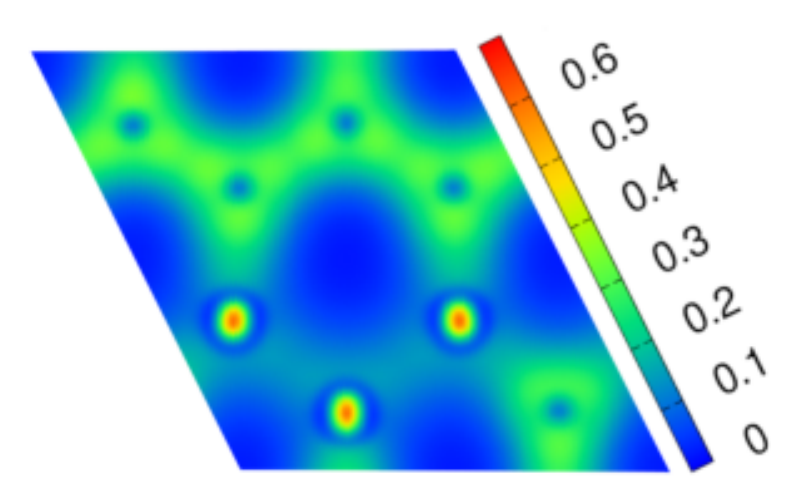} &
	\includegraphics[width=0.15\textwidth]{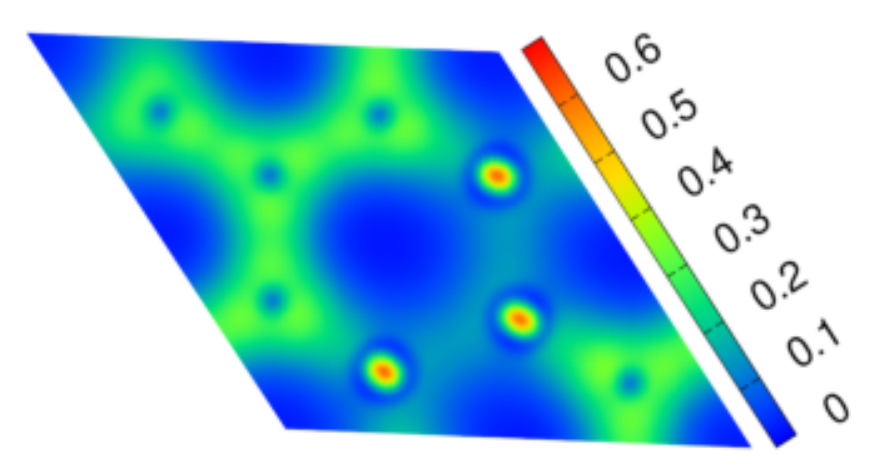} \\ 
	3 &
	\includegraphics[width=0.15\textwidth]{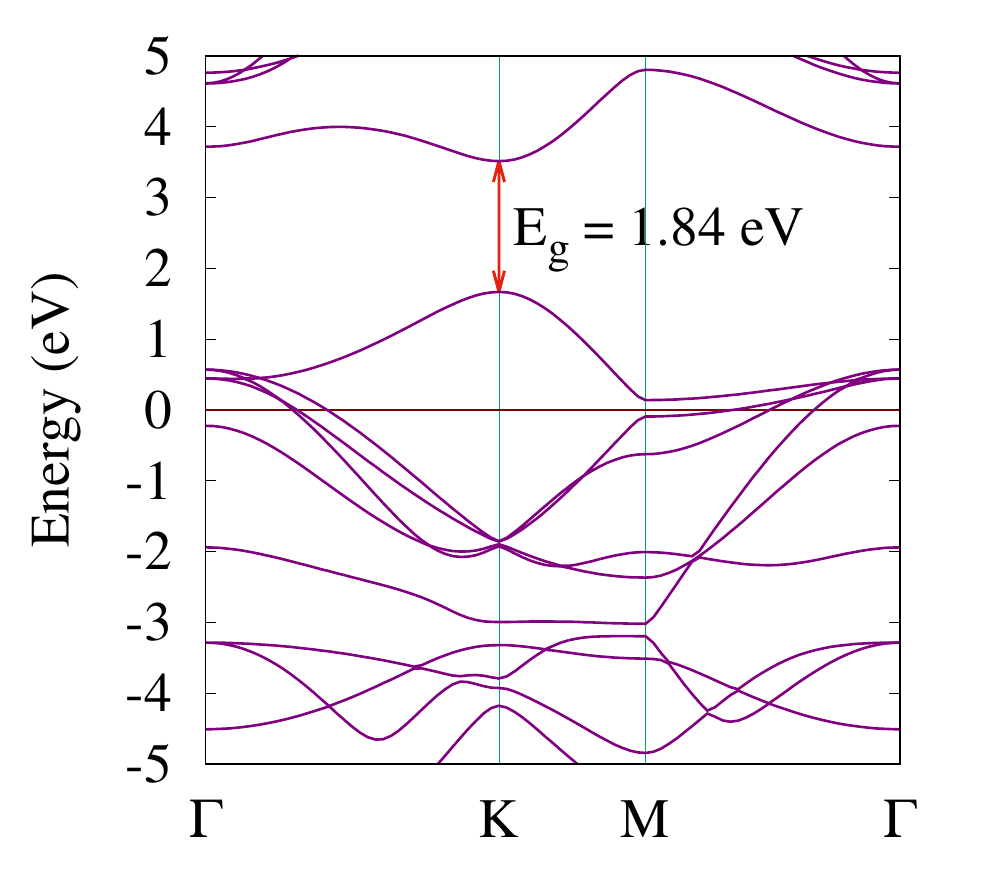} & 
	\includegraphics[width=0.15\textwidth]{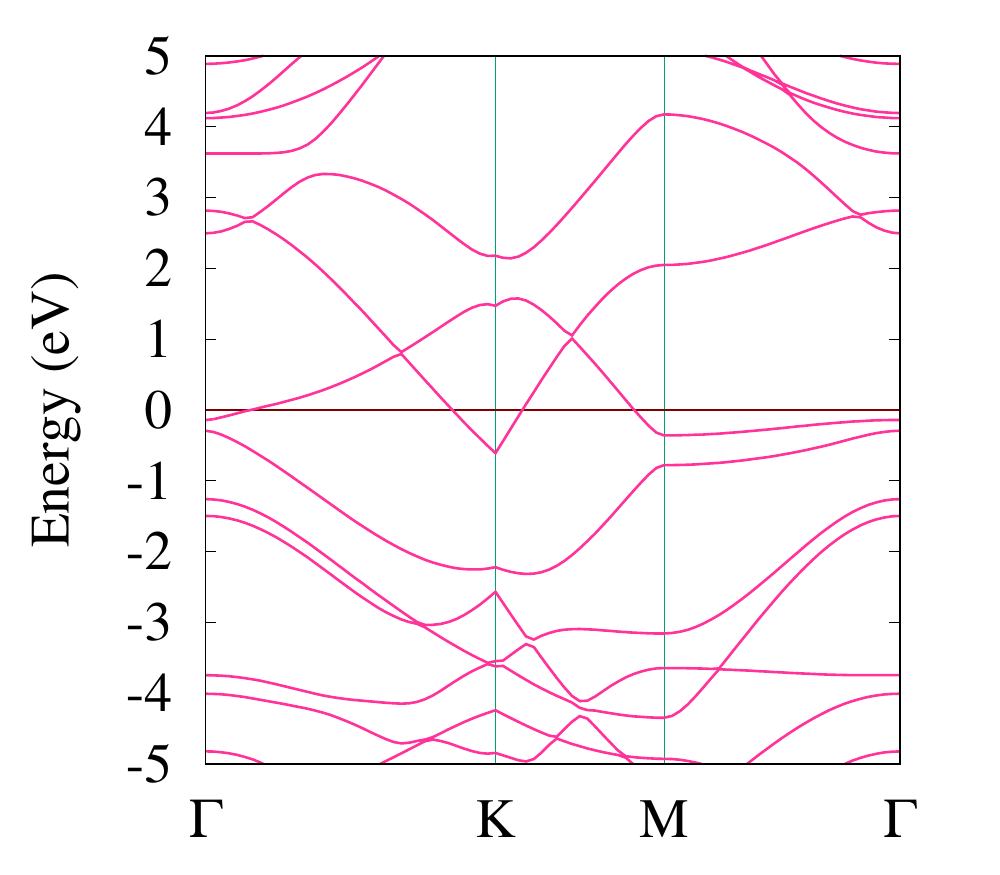} &
	\includegraphics[width=0.15\textwidth]{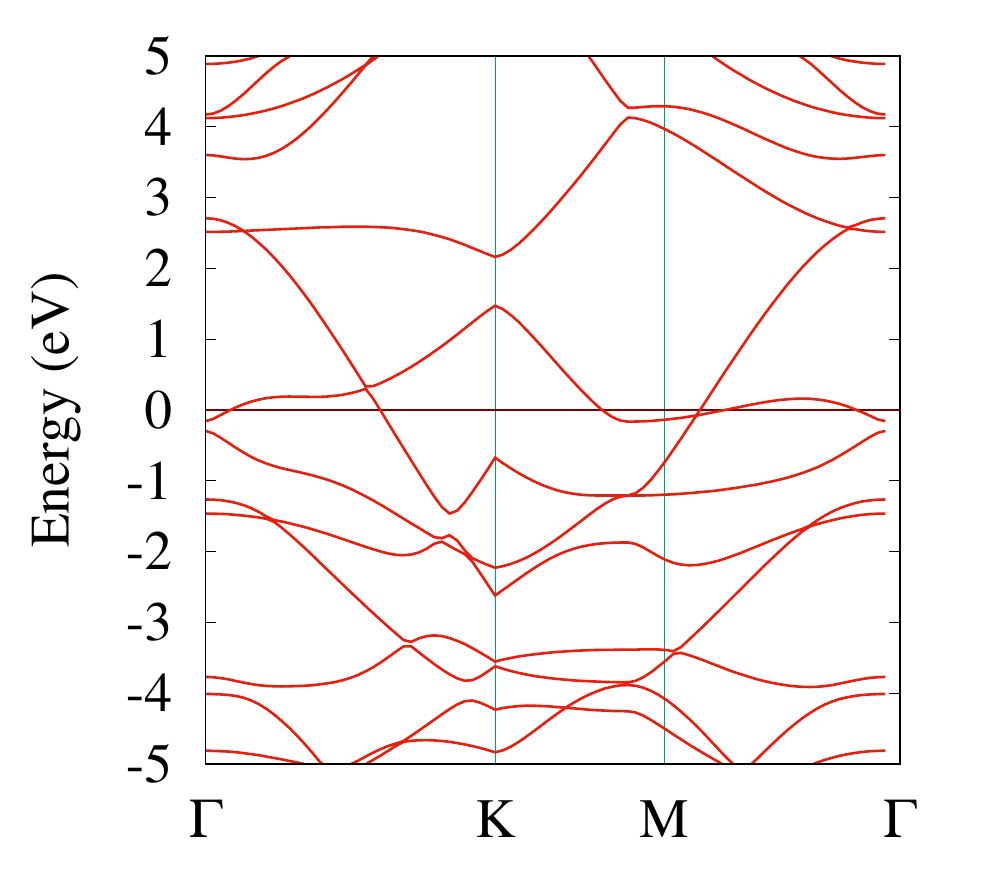} \\
	4 &
	\includegraphics[width=0.15\textwidth]{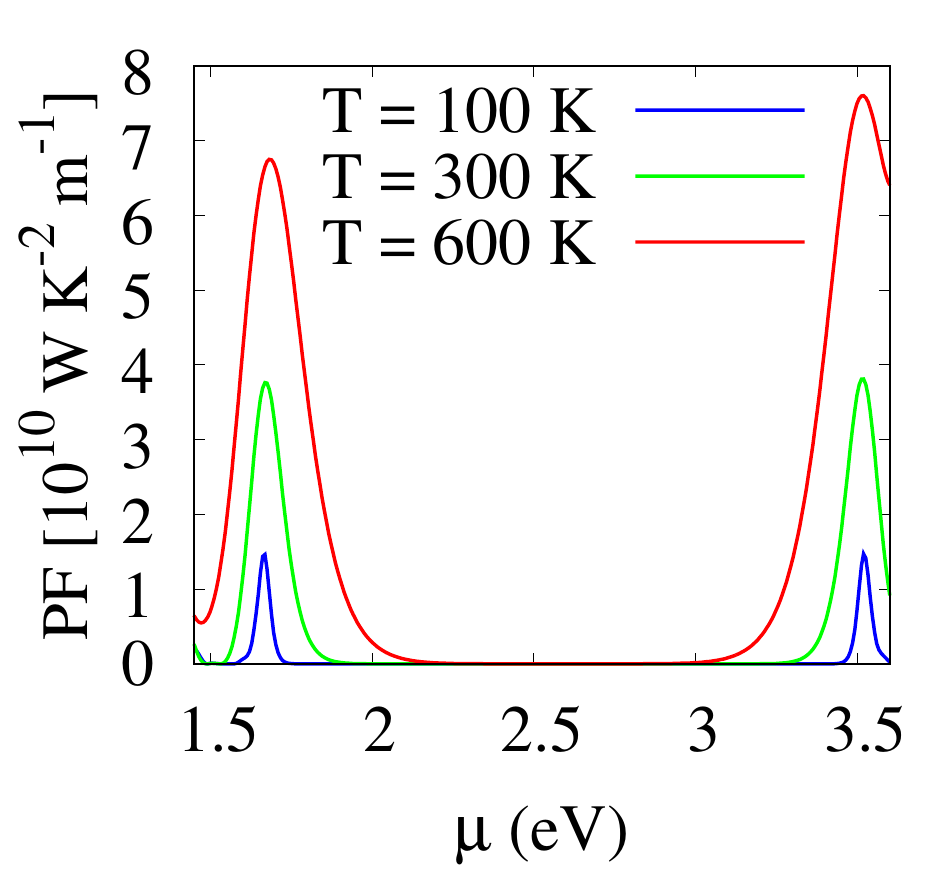} &
	\includegraphics[width=0.15\textwidth]{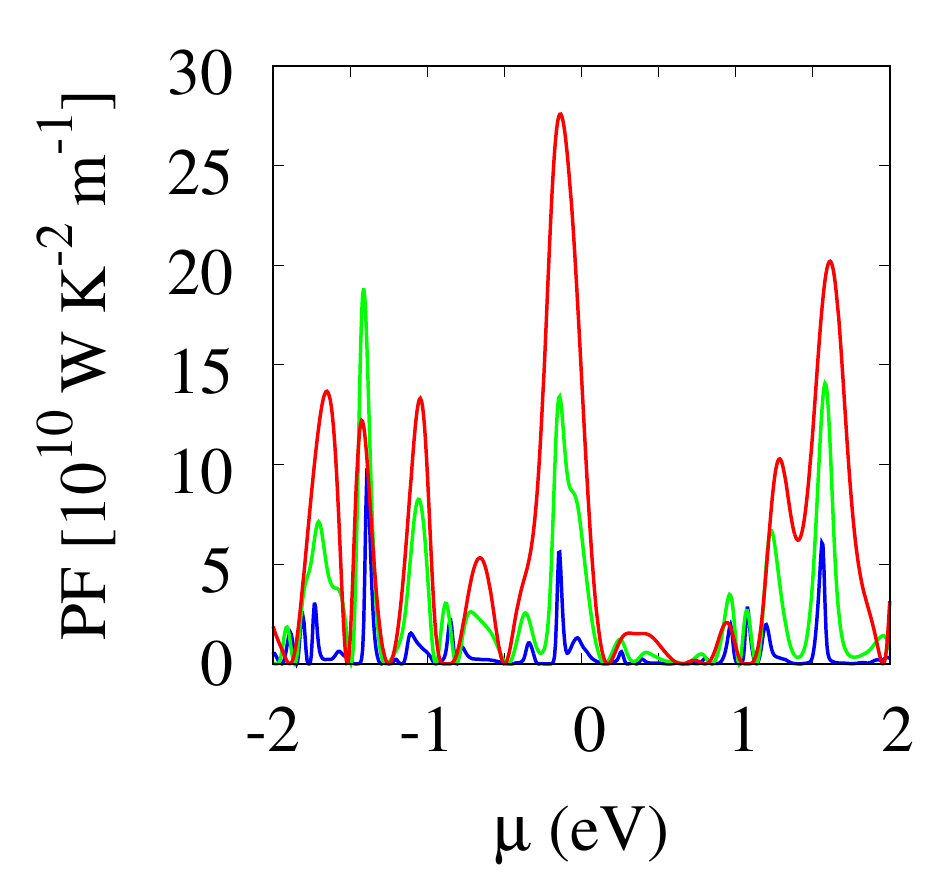} & 
	\includegraphics[width=0.15\textwidth]{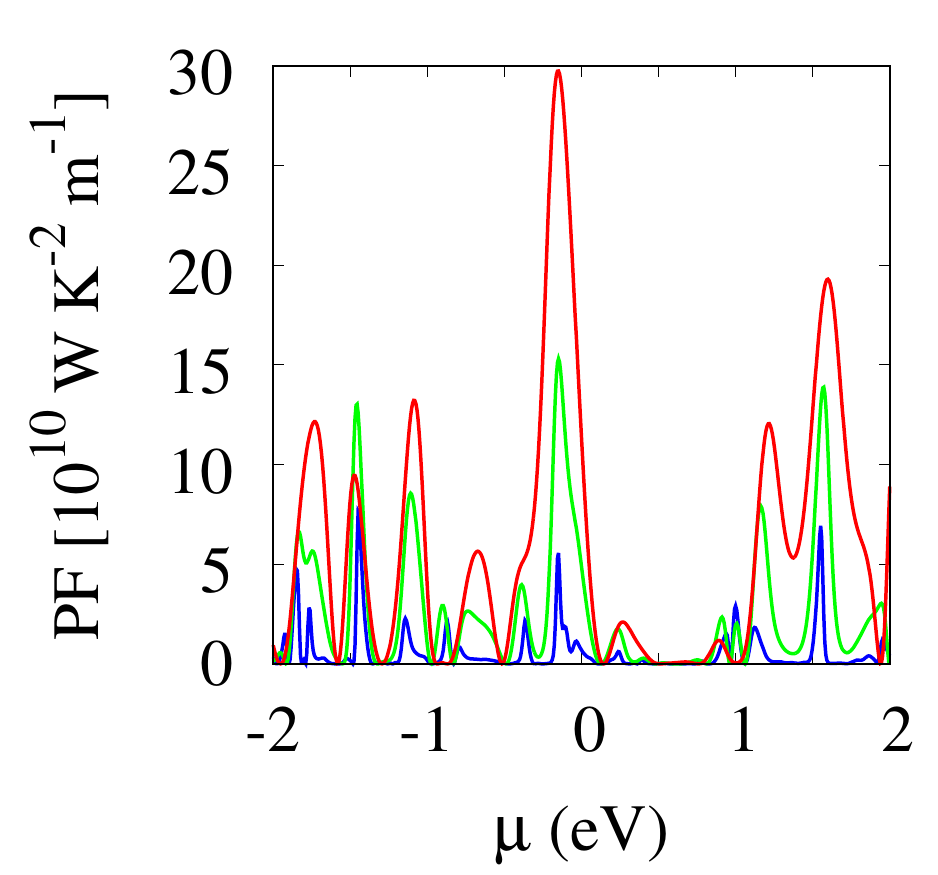} 
	\end{tabular} 
\end{table}
 \caption{First row shows the super-cell of three B-atoms doped into ortho-ortho-para (a1), nonb-BG, meta-meta-para (b1), b-BG-1, and  ortho-meta-para (c1), b-BG-2, 
 positions of the graphene structure.
 The second and third rows indicate the charge density distribution and the band structures, respectively. 
 The fourth row display the power factor for three different values of temperatures, 100 (blue), 300 (green), and 600 K (red).}
\label{fig04}
\end{figure}

The doping concentration ratio is now increased to $37.5\%$, which can be realized by 
three B- or N-atoms doped into the honeycomb structure of graphene. 
For saving space, we only present the results of three B-atoms doped into 
the ortho-ortho-para positions (a1) forming nonb-BG , and the meta-meta-para (b1), and the ortho-meta-para positions 
(c1) identifying as b-BG-1 and b-BG-2, respectively, as is shown in the first row of \fig{fig04}.

The formation energy of all three cases shown in the first row of \fig{fig04} is almost the same which is equal
to $-18.365$ eV indicating a less stable energetic structure compared to the BG with one- and two B-atoms.
But the formation energy of NG with three N-atoms 
is $-28.198$ eV which is much smaller than that of the BG with three B-atoms. 
As a result, NG is much more  energetic stable than BG. In addition, the negativity in 
the phonon dispersion is increased in BG with three B-atoms compared to BG with two B-atoms 
indicating a less dynamical stable structure (not shown).

The attractive force between the B atoms in the nonb-BG is further increased leading to a larger 
bandgap $E_{\rm g} = 1.89$ eV as is shown in \fig{fig04} (a3), while the repulsive interaction between 
the B-atoms in both types of b-BG closes the bandgap and causes an overlap of the valence and the conduction bands
[see (b3) and (c3)]. 
Consequently, the power factor for the nonb-BG is still high but 
the overlapping energies in the b-BG generate a lot of ``noise'' in the power factor [see (b4) and (c4)]
comparing to the nonb-BG (a4).

\subsection{Boron and Nitrogen codoped graphene structures}

\begin{figure}[H]
\begin{table}[H]
  \captionsetup{labelformat=empty}
\noindent
\begin{tabular}[]{ >{\centering\arraybackslash}m{0cm} >{\centering\arraybackslash}m{2.4cm}>{\centering\arraybackslash}m{2.4cm} >{\centering\arraybackslash}m{2.4cm} >{\centering\arraybackslash}m{2.4cm} }
	& a (nonb-BNG-1) & b (nonb-BNG-2) & c (b-BNG)\\ 
	1 &
	\includegraphics[width=0.15\textwidth]{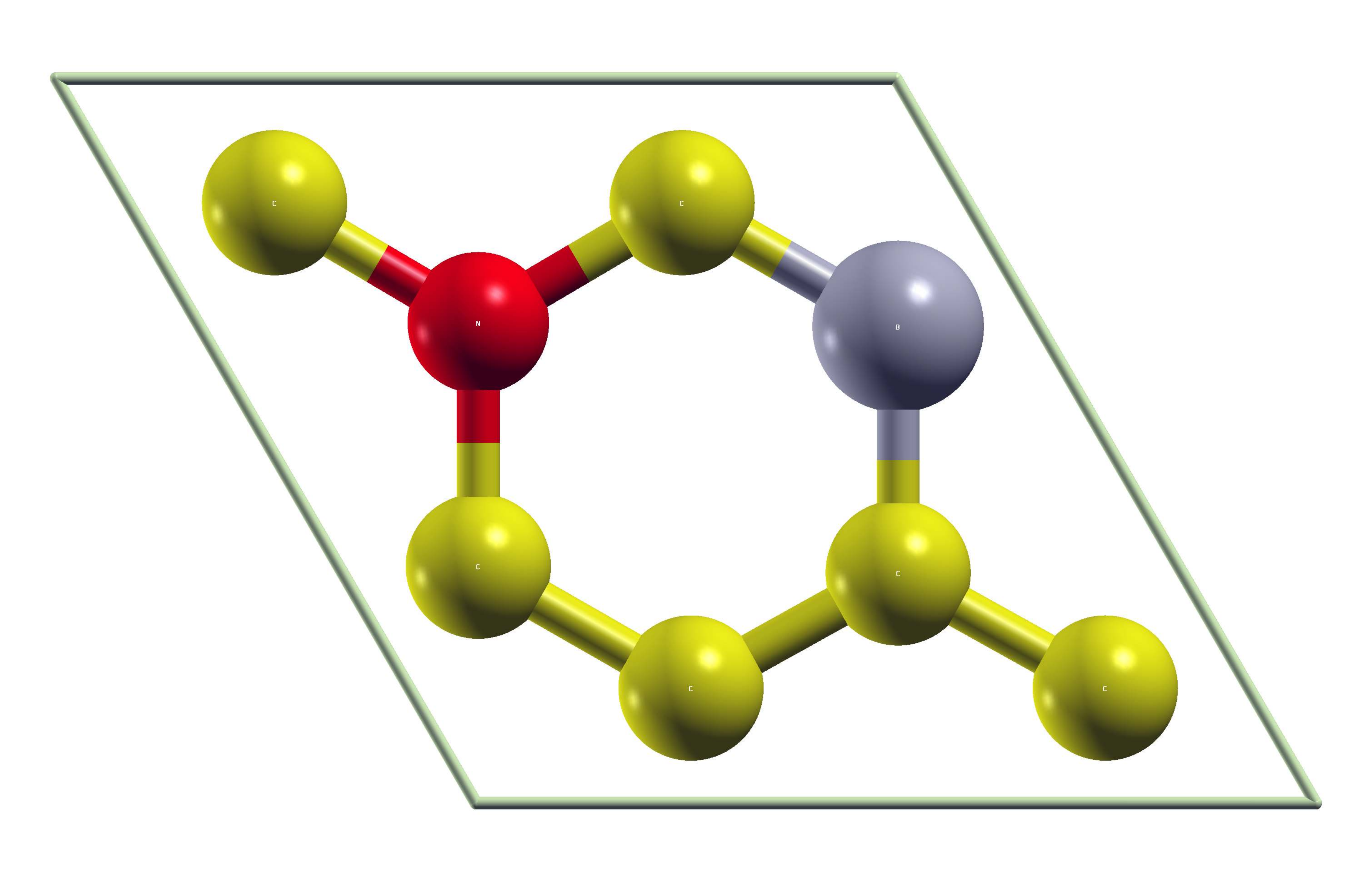} & 
	\includegraphics[width=0.15\textwidth]{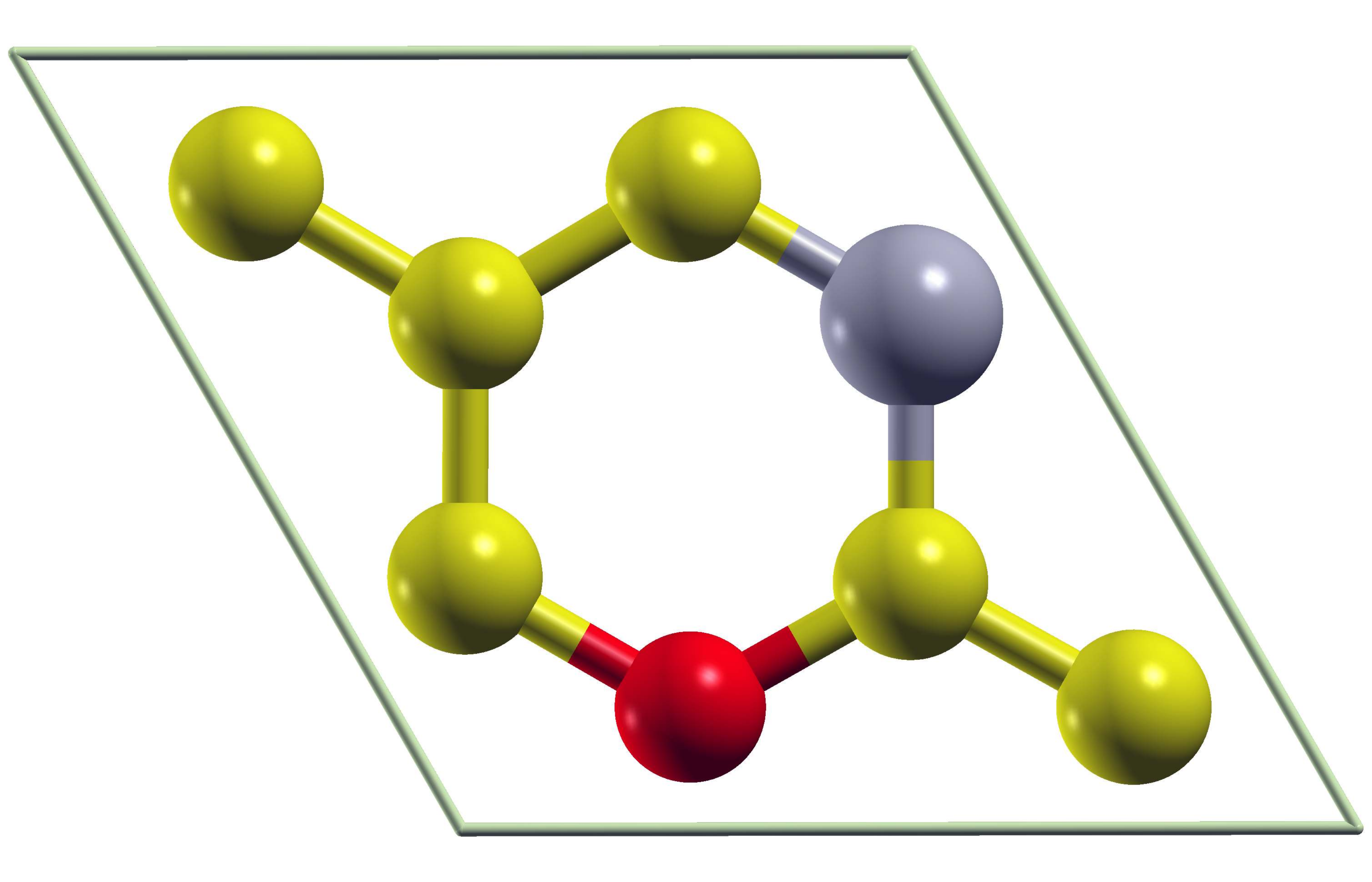} & 
        \includegraphics[width=0.15\textwidth]{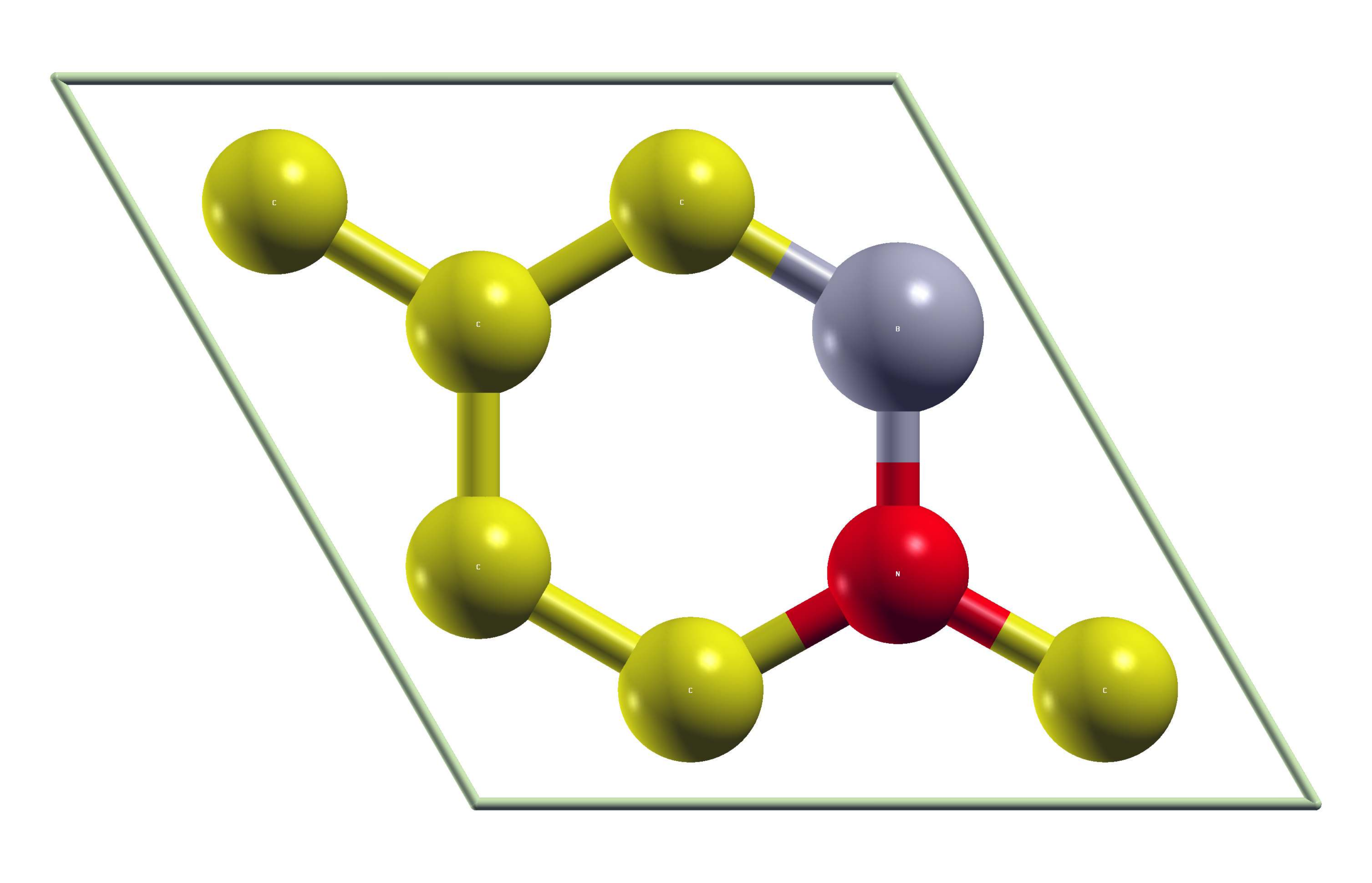} \\
	2 &
	\includegraphics[width=0.15\textwidth]{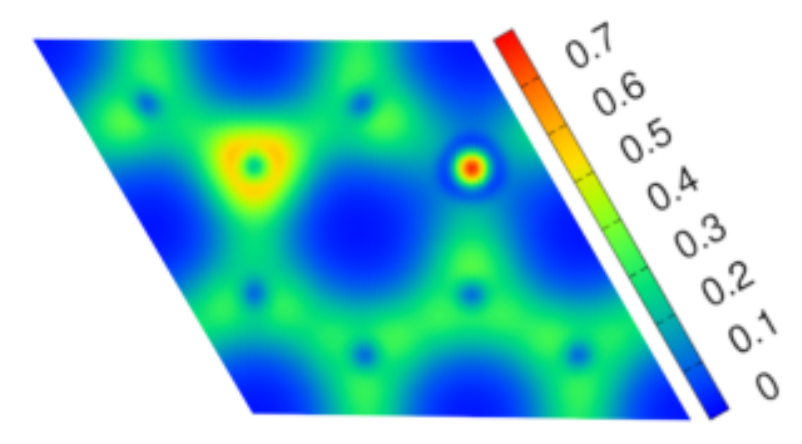}  &
    \includegraphics[width=0.15\textwidth]{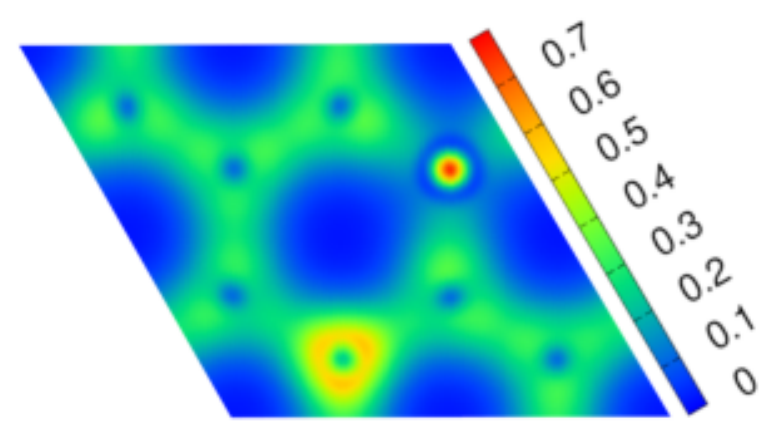} &
	\includegraphics[width=0.15\textwidth]{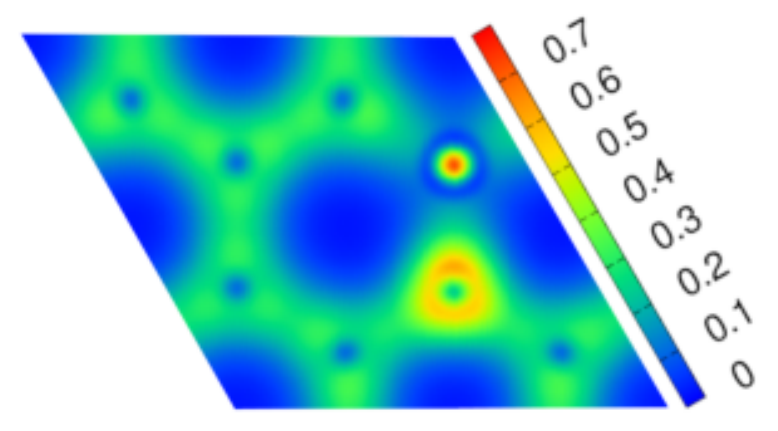} \\ 
	3 &
	\includegraphics[width=0.15\textwidth]{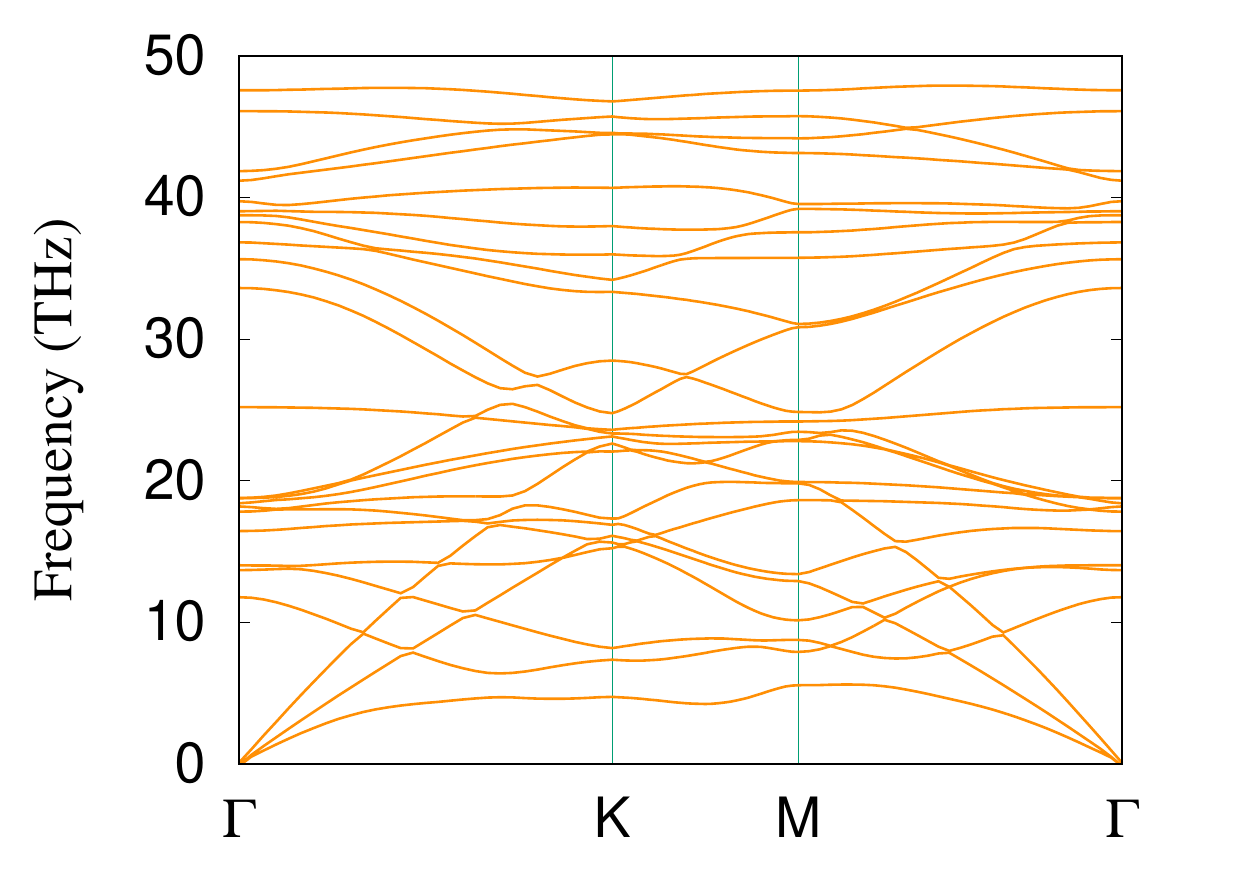} & 
	\includegraphics[width=0.15\textwidth]{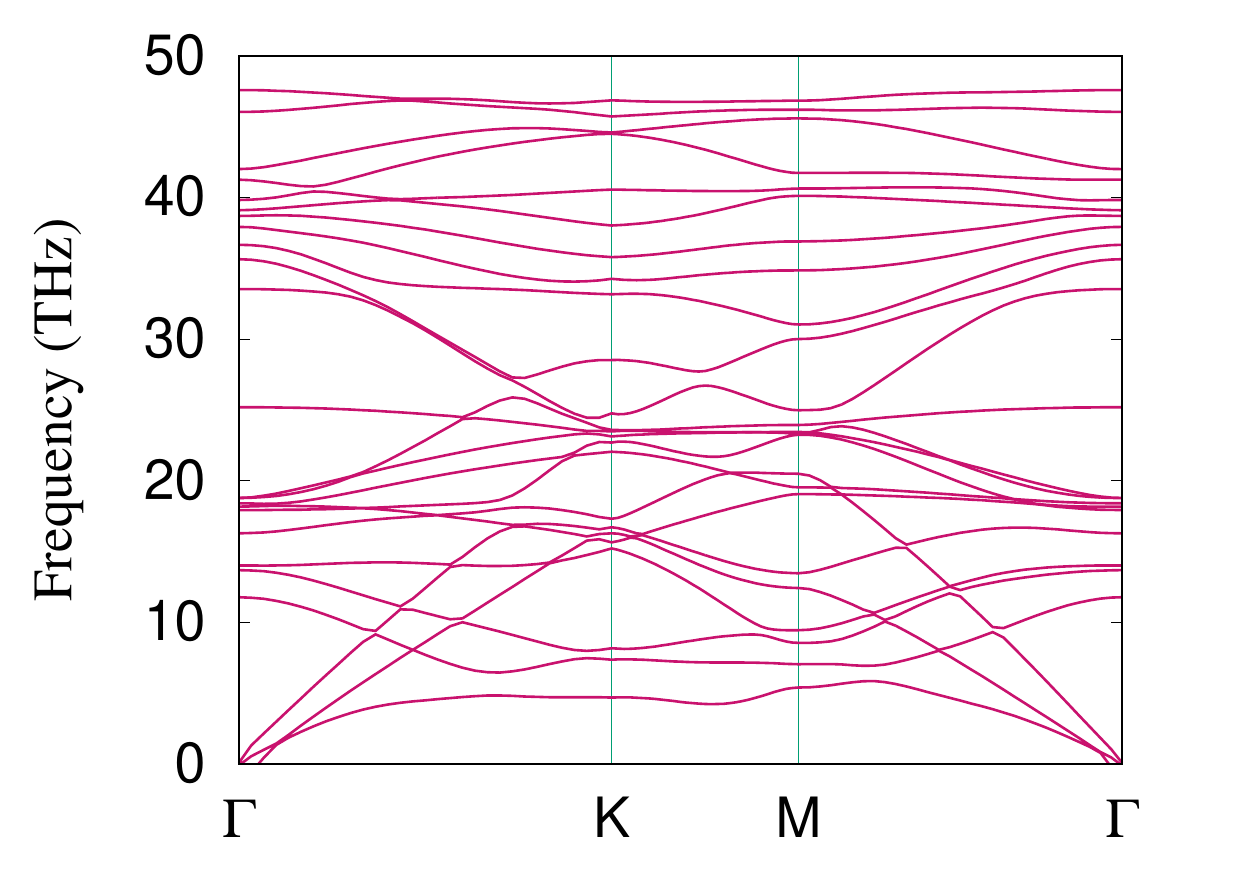} &
    \includegraphics[width=0.15\textwidth]{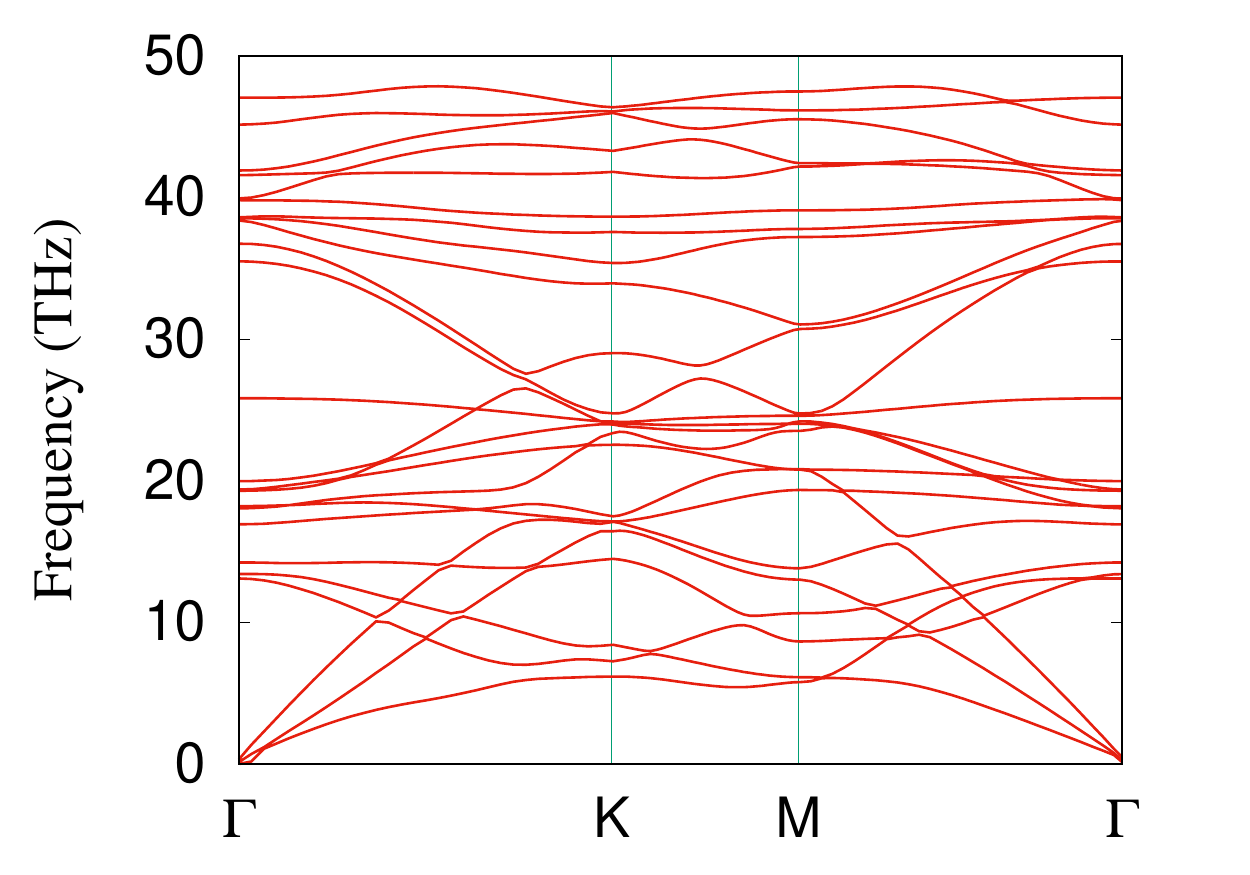} \\
    4 &
    \includegraphics[width=0.15\textwidth]{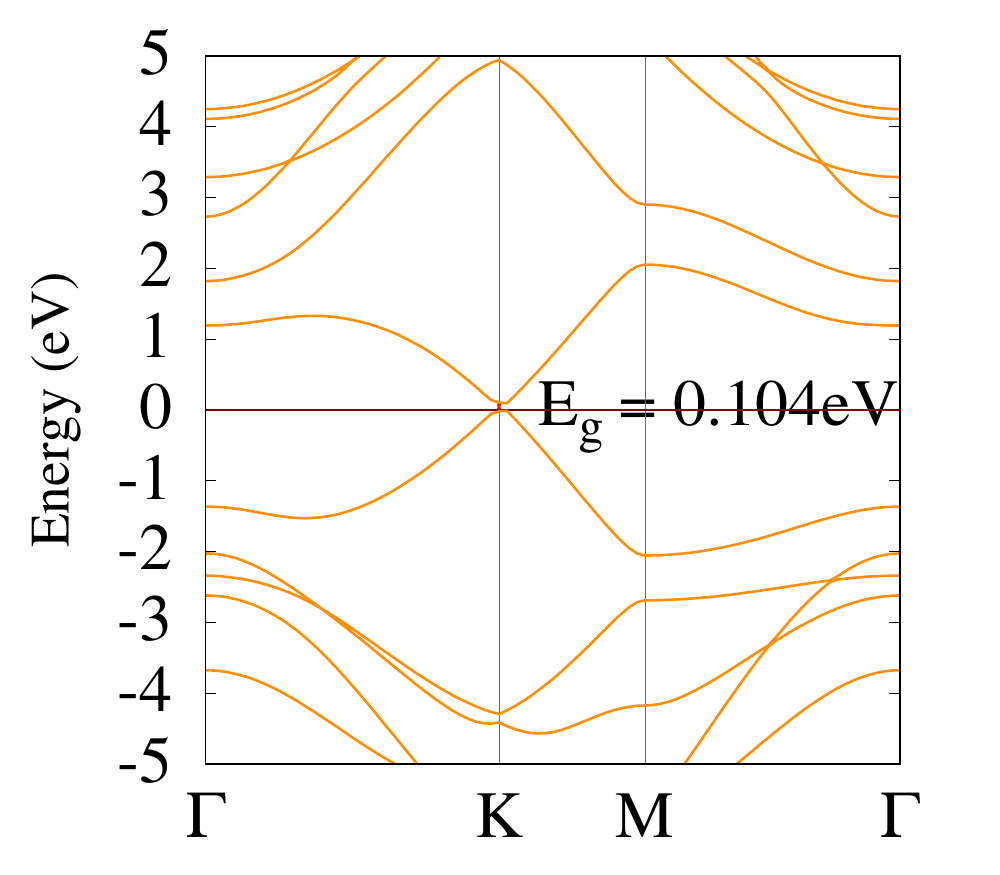} & 
    \includegraphics[width=0.15\textwidth]{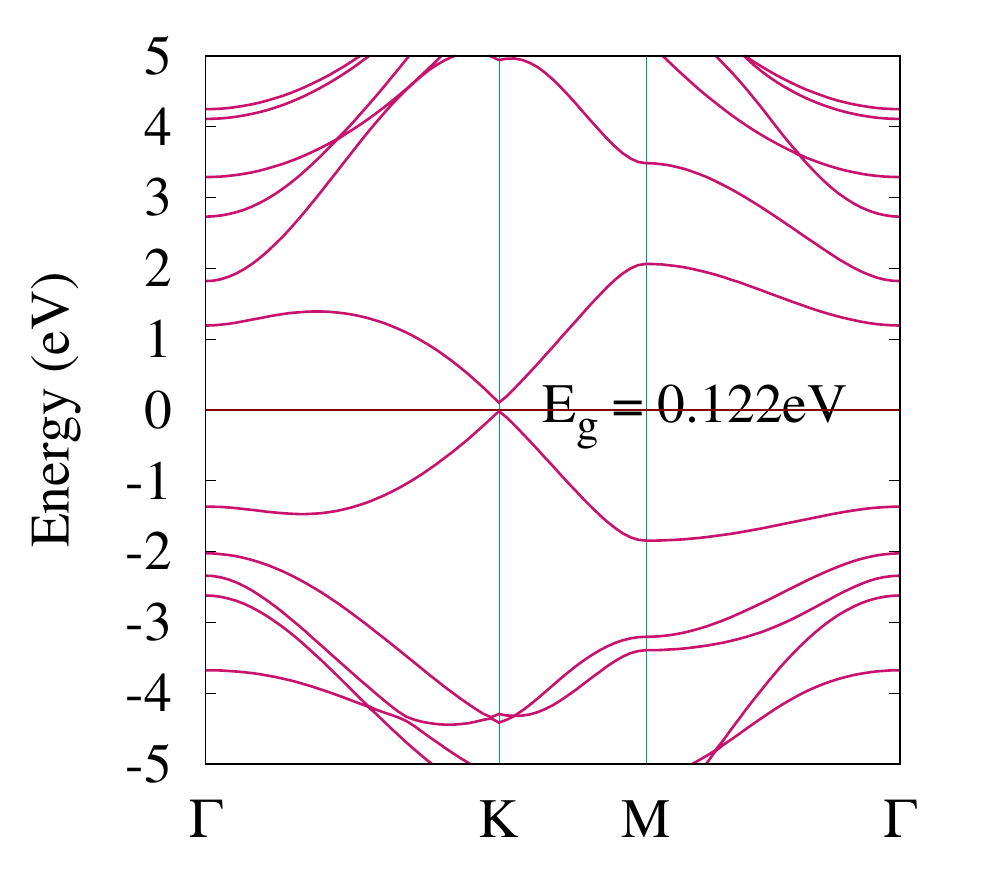} &
    \includegraphics[width=0.15\textwidth]{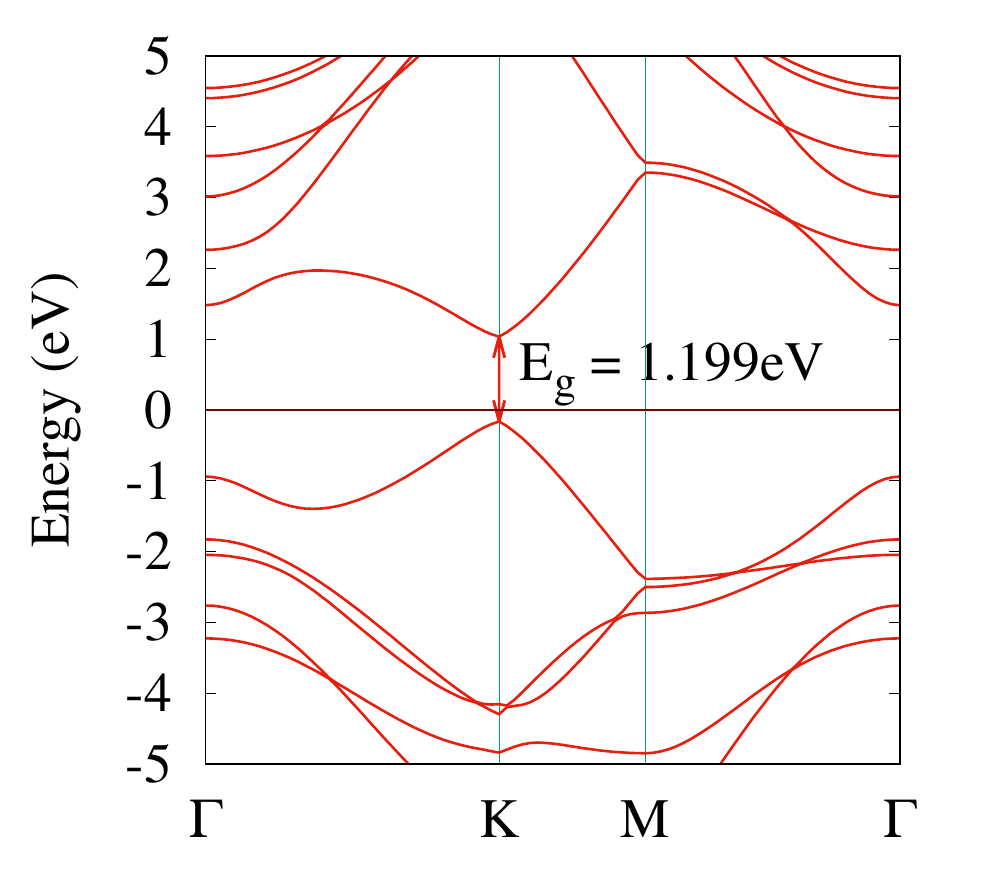} \\
	5 &
	\includegraphics[width=0.15\textwidth]{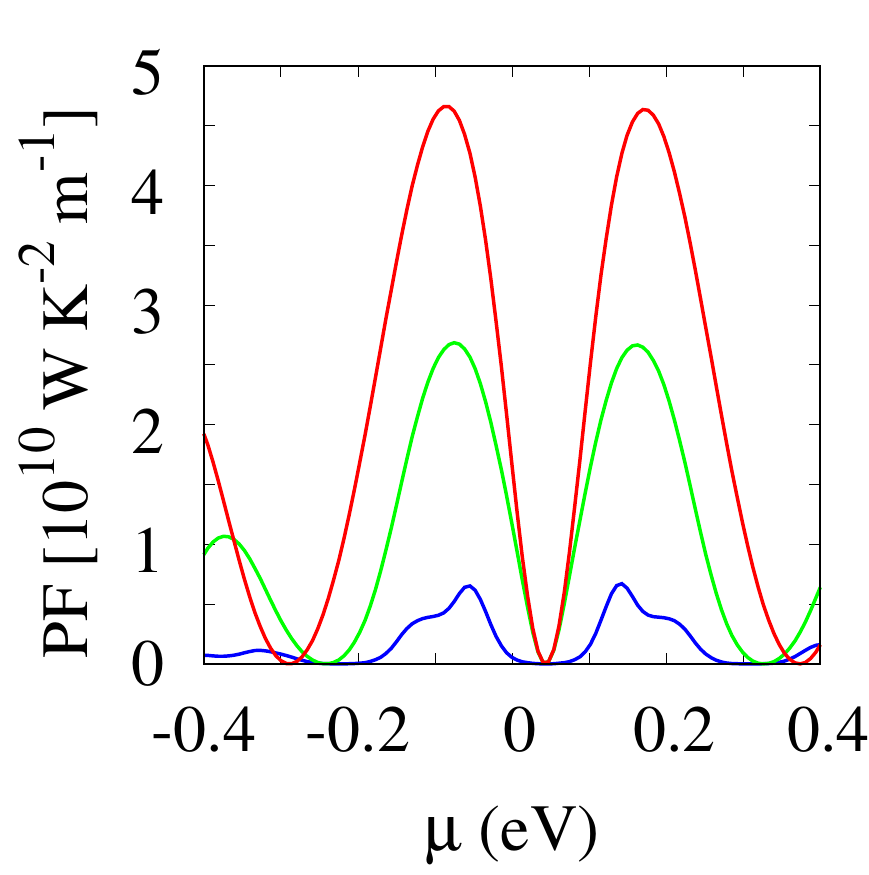} &
	\includegraphics[width=0.15\textwidth]{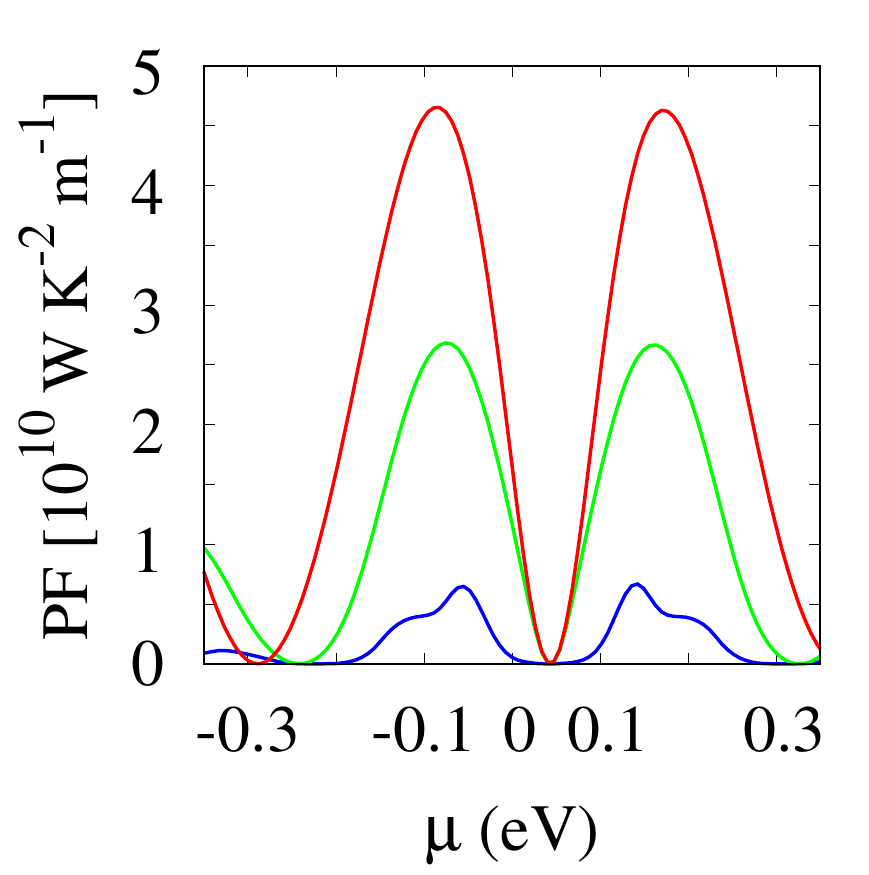} & 
        \includegraphics[width=0.15\textwidth]{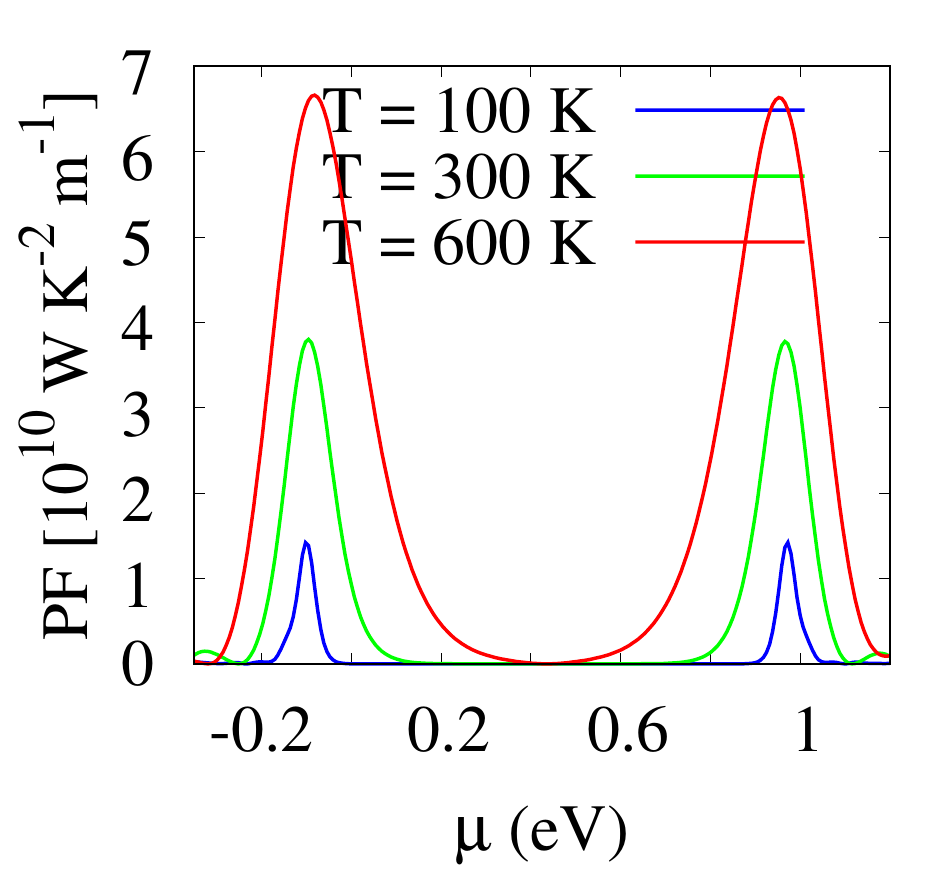} 
	\end{tabular} 
\end{table}
 \caption{First row indicates the super-cells with B and N codoped graphene at ortho positions (a1) identifying as (nonb-BNG-1),
     ortho-para positions (b1) with (nonb-BNG-2), 
     and ortho-meta positions (c1) having (b-BNG), respectively. 
     The second row indicates the corresponding charge density distribution, 
     the third row shows the phonon dispersion, and 
     the fourth row demonstrates the band structure of the systems.
     The fifth row display the power factor for three different values of temperatures, 100 (blue), 300 (green), and 600 K (red).}
\label{fig05}
\end{figure}

In this section we consider B- and N-codoped graphene
with a configuration of two or three doped atoms in the honeycomb structure.
In the case of two atoms configuration consisting of one B and one N atom, there will be three 
possible configurations which are: both B- and N-atoms are substitutionally doped at ortho positions 
(a1) showing non-bonding of B and N codoped graphene (nonb-BNG-1), ortho-para positions (b1) 
indicating the (nonb-BNG-2), and ortho-meta positions (c1) displaying bonded B and N codoped graphene (b-BNG), 
as are shown in the first row of \fig{fig05}. 
There are more configurations of the  B and N atoms in the graphene structure. For example: 
if we exchange the position of B and N atoms in the structure shown in column (a), the physical properties should be unchanged.
Furthermore, if both B and N atoms are put at meta-meta position, the physical characteristics will be similar to the 
ortho-ortho position structure shown in column (a).
It has been shown that the nonb-BNG is an effective catalyst compared to b-BNG for an oxygen reducing 
reaction in batteries \cite{doi:10.1021/ja310566z, doi:10.1002/cssc.201900060}.
Here, we show that this may not hold true for the application of thermal devices.

The nonb-BNG structures shown in (a1) and (b1) of \fig{fig05} have almost the same formation energy, 
$E_f = -28.024$ eV, while the formation energy of the b-BNG presented in (c1) is slightly 
lower, $-29.238$ eV. It indicates that the b-BNG is the most energetically stable system among these structures.

In both nonb-BNG and b-BNG structures, the N atom attracts electrons and builds a high-electron region; while the B atom loses electrons and generates a low-electron region [see second row of \fig{fig05}]. 
But in the b-BNG, the lone-pair electrons from the N atom are neutralized by the vacant orbital of B atom, causing a high electron density between the B-N bond (c2).
This indicates a strong and stiff covalent bond between the B and the N atoms compared 
to the B-C and the B-N bonds.

The phonon dispersion of all three configurations is shown in the third row. 
The positive value of the phonon dispersion indicates a dynamical stability of the structures. 
The interaction of the B and the N atoms with graphene structure results in pronounced
changes in the phonon spectrum compared to pristine graphene shown in \fig{fig02}(a3).
By introducing the BN domain in a graphene nanosheet, it was found that the electron 
states in graphene referring to the $\pi$-electrons are sensitive for the redistribution of charge 
in graphene due to B and N atoms. This redistribution of electron charge breaks the local symmetry of 
graphene which is the main reason for the opening of a bandgap. Furthermore, the Coulomb dipole from 
the B and the N atoms breaks the symmetry of the graphene structure leading to the 
bandgap opening \cite{C2NR11728B}.

The strong attractive interaction between the B and N in the b-BNG
opens up a bandgap to $1.199$ eV (c4) while the attractive interaction in nonb-BNG is weak leading 
to a small bandgap [see (a4) and (b4)].
The localized electron between the B-N bonds and the large bandgap in the b-BNG cause a weak 
contribution of the $\pi$ electrons to the transport leading to decrease in the electric conductivity 
and an increase in the Seebeck coefficient. 
As a result, the power factor for b-BNG is increased compared to the nonb-BNG structure shown in 
the fifth row of \fig{fig05}.

We notice that if the position of dopant atoms is exchanged but the bonding and non-bonding between 
the doped atoms are kept, the same results should be obtained.
For example, exchanging the position of N and B atoms at ortho positions (a1), meta-position (b1), and ortho-meta position (c1) in \fig{fig05} should not change the results of band structure and power factor.

\begin{figure}[htb]
	\centering
	\includegraphics[width=0.23\textwidth]{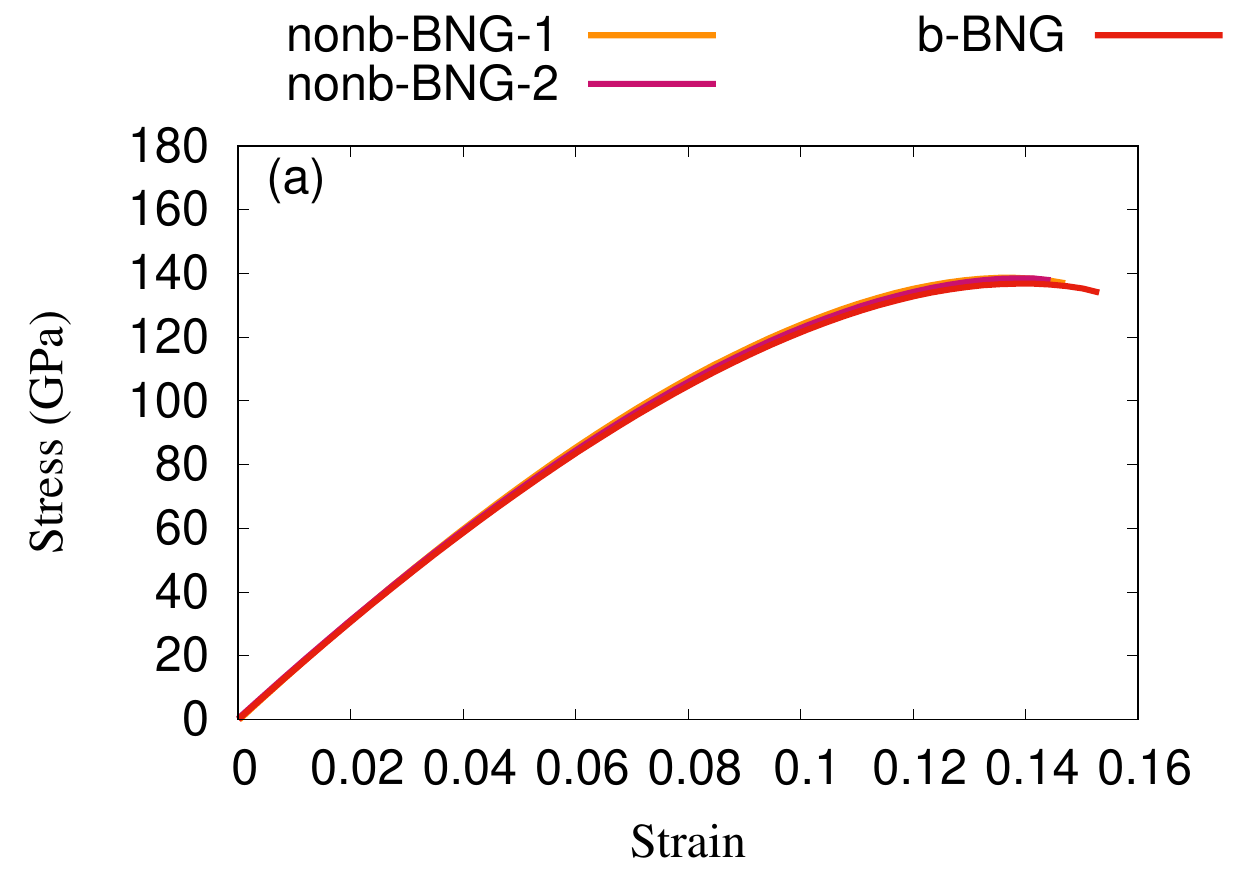}
	\includegraphics[width=0.23\textwidth]{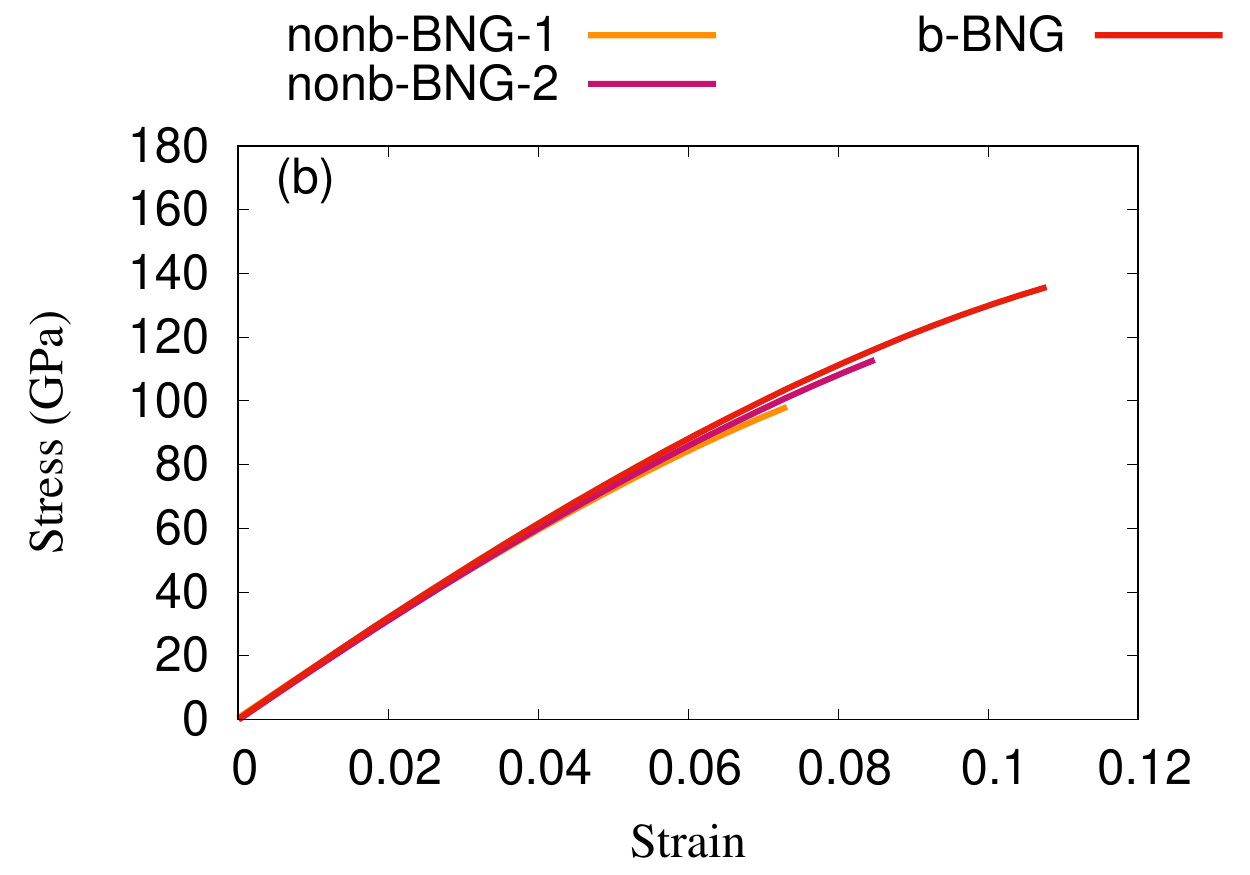}
	\caption{Uniaxial stress-strain of nonb-BNG-1 (orange), nonb-BNG-2 (purple), and b-BNG (red) monolayers along $x$-axis, Zigzag, (a) and $y$-axis, Armchair, (b).}
	\label{fig05_1}
\end{figure}

The stress-strain curves of nonb-BNG and b-BNG are shown in \fig{fig05_1}, where 
the stress is assumed to be along the $x$-axis (a) and $y$-axis (b) of the structures.
The comparison of the stress-strain relations along the $x$-axis, zigzag, and $y$-axis, armchair, shows
that the stress of the armchair sheet is $6\%$ larger than that of the zigzag sheet for the b-BNG (red color). 
Their failure strains are also different with $0.15$ for the zigzag, and 
$0.11$ for the armchair. This is attributed to the B-N bond in the armchair direction which is 
stronger than the B-C and the N-C bonds along the zigzag direction.
On the other hand, in the 
nonb-BNG structures (orange and purple color), the zigzag sheet is stronger than the armchair sheet.
In the nonb-BNG-1, the zigzag sheet has stress $129.58$~GPa at strain $~0.148$ which is larger than that 
of the armchair sheet of $98$~GPa at strain 0.074 (orange color), while in nonb-BNG-2 the stress of 
zigzag sheet is $15\%$ larger than that of armchair sheet but with different frailer strain 
(purple color) \cite{Dewapriya_2013}. We can conclude that the stress of the nonb-BNG and the b-BNG is 
almost the same in the zigzag sheet, while the stress of the b-BNG in armchair sheet is higher than 
that of the nonb-BNG.

\begin{figure}[H]
\begin{table}[H]
  \captionsetup{labelformat=empty}
\noindent
\begin{tabular}[]{ >{\centering\arraybackslash}m{0cm} >{\centering\arraybackslash}m{2.4cm} >{\centering\arraybackslash}m{2.4cm} >{\centering\arraybackslash}m{2.4cm} }
	& a (nonb-BNG) & b (b-BNG-1)  & c (b-BNG-2)\\ 
	1 &
        \includegraphics[width=0.15\textwidth]{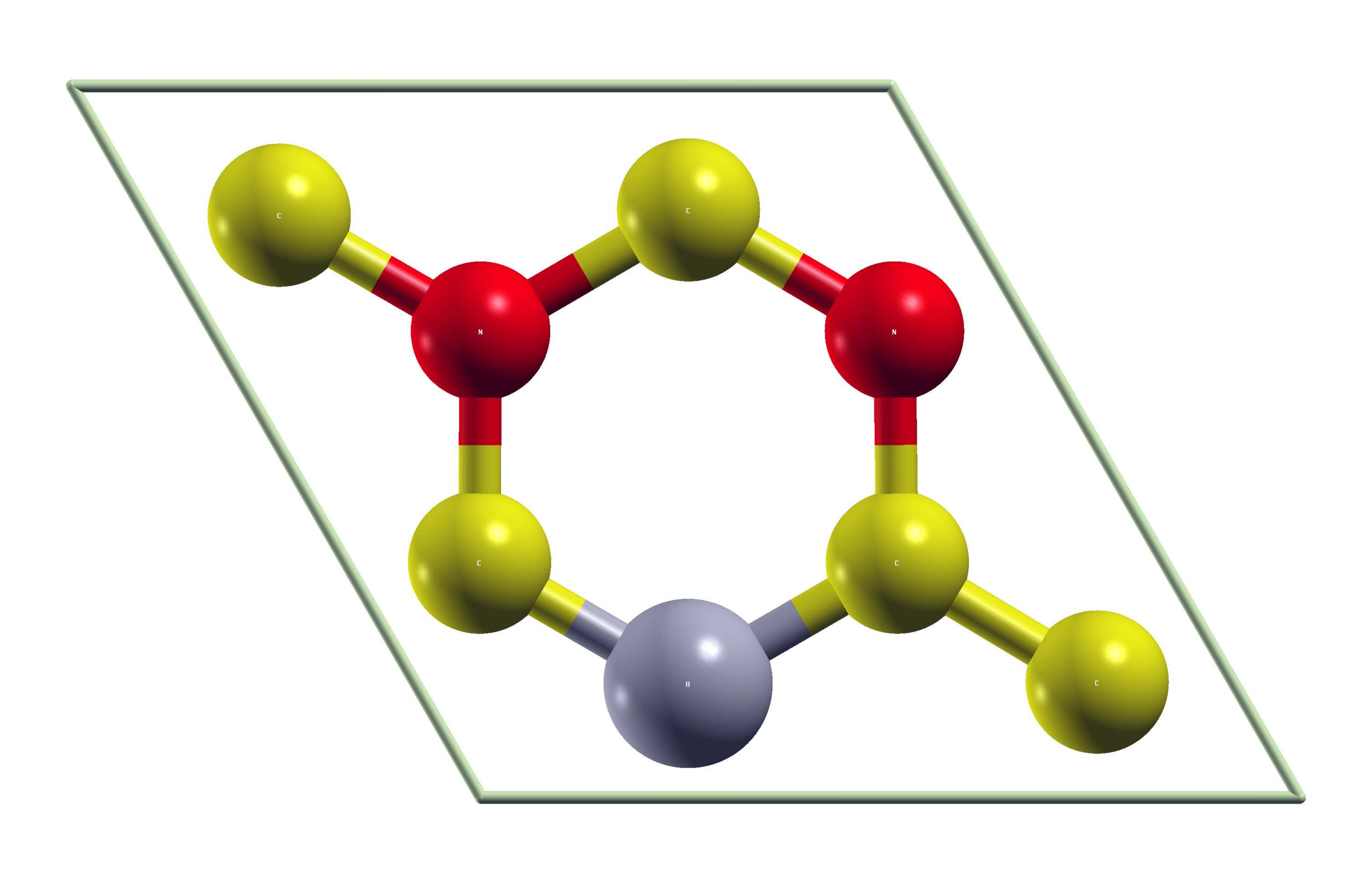} & 
	\includegraphics[width=0.15\textwidth]{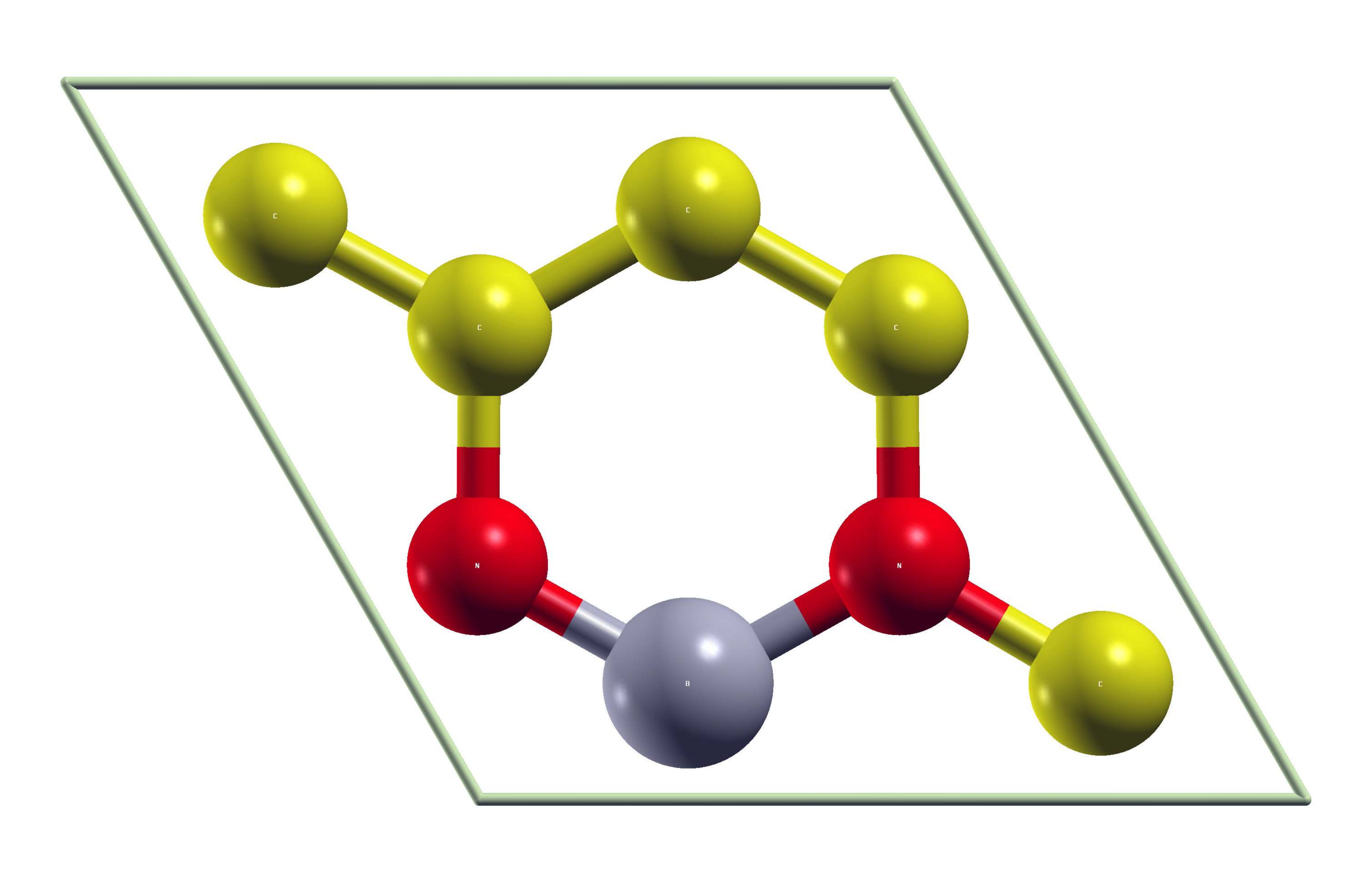} & 
	\includegraphics[width=0.15\textwidth]{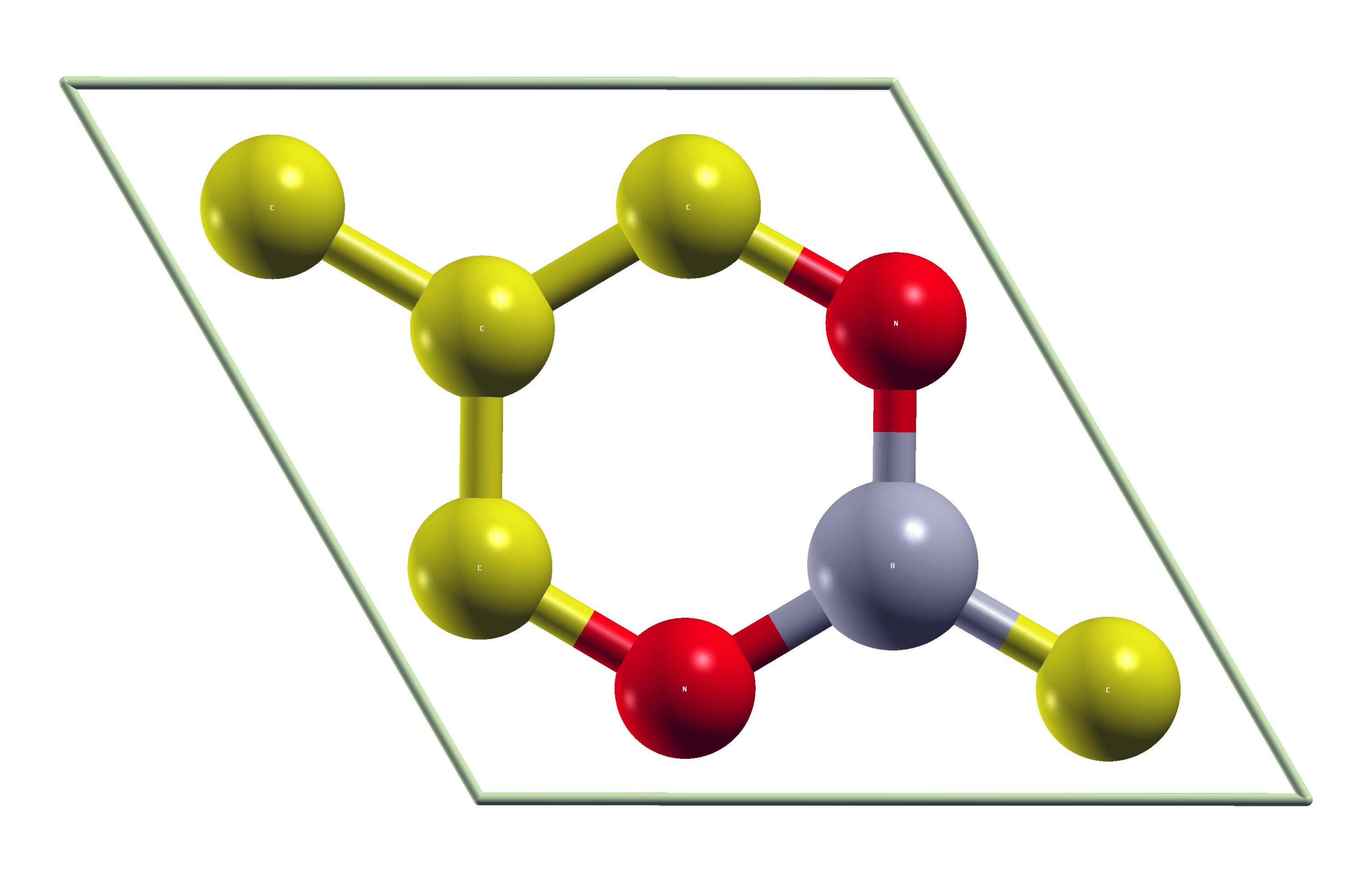} \\
	2 &
        \includegraphics[width=0.15\textwidth]{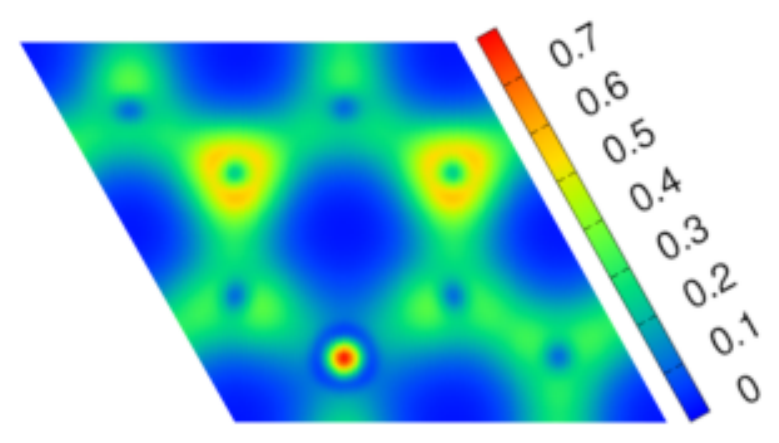} & 
	\includegraphics[width=0.15\textwidth]{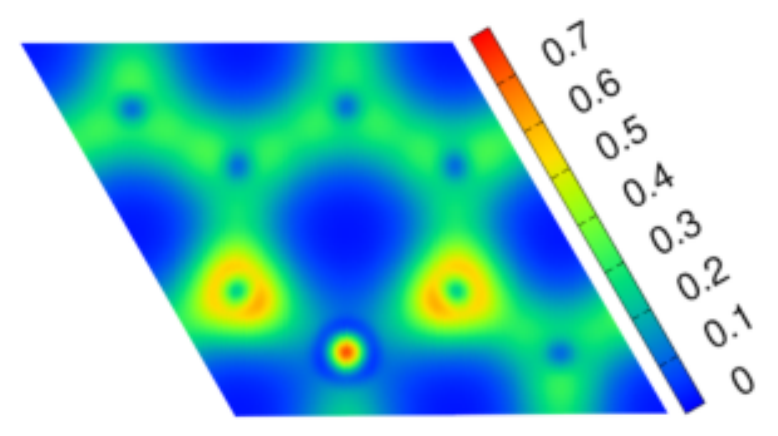}  &
	\includegraphics[width=0.15\textwidth]{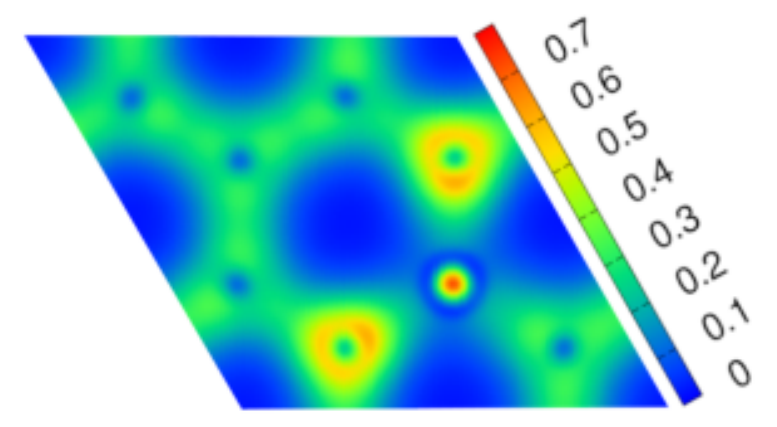} \\
	3 &
	\includegraphics[width=0.15\textwidth]{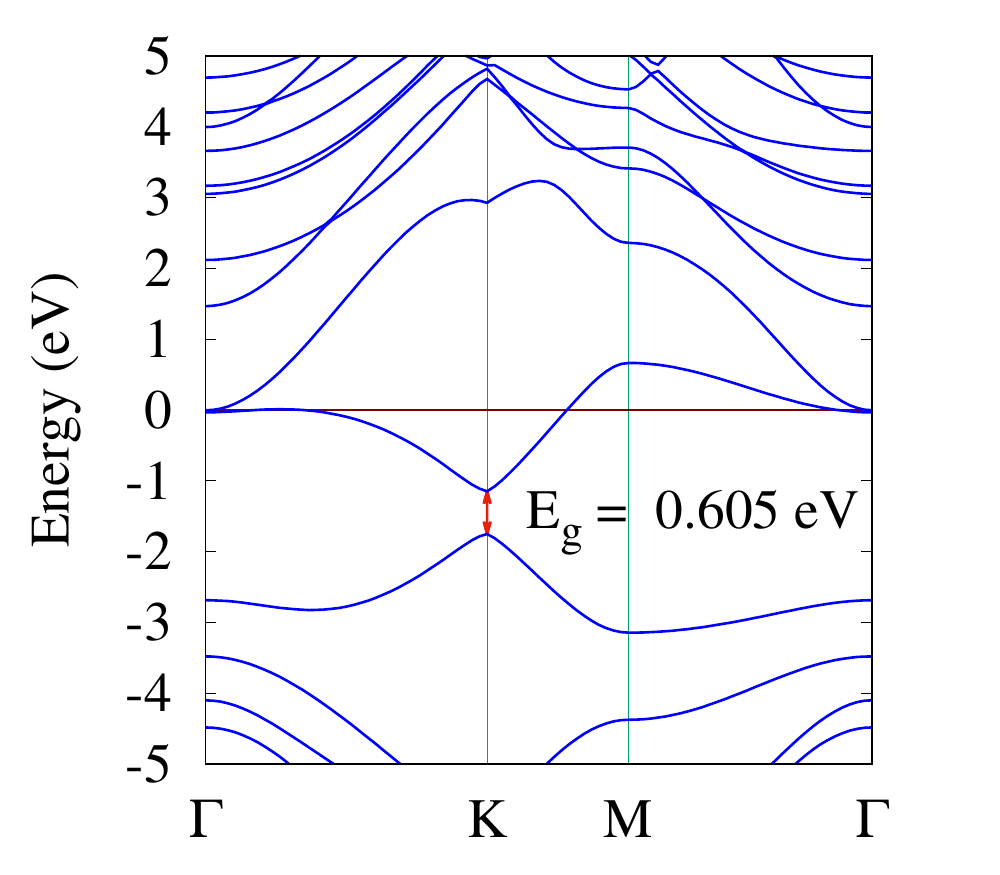} & 
	\includegraphics[width=0.15\textwidth]{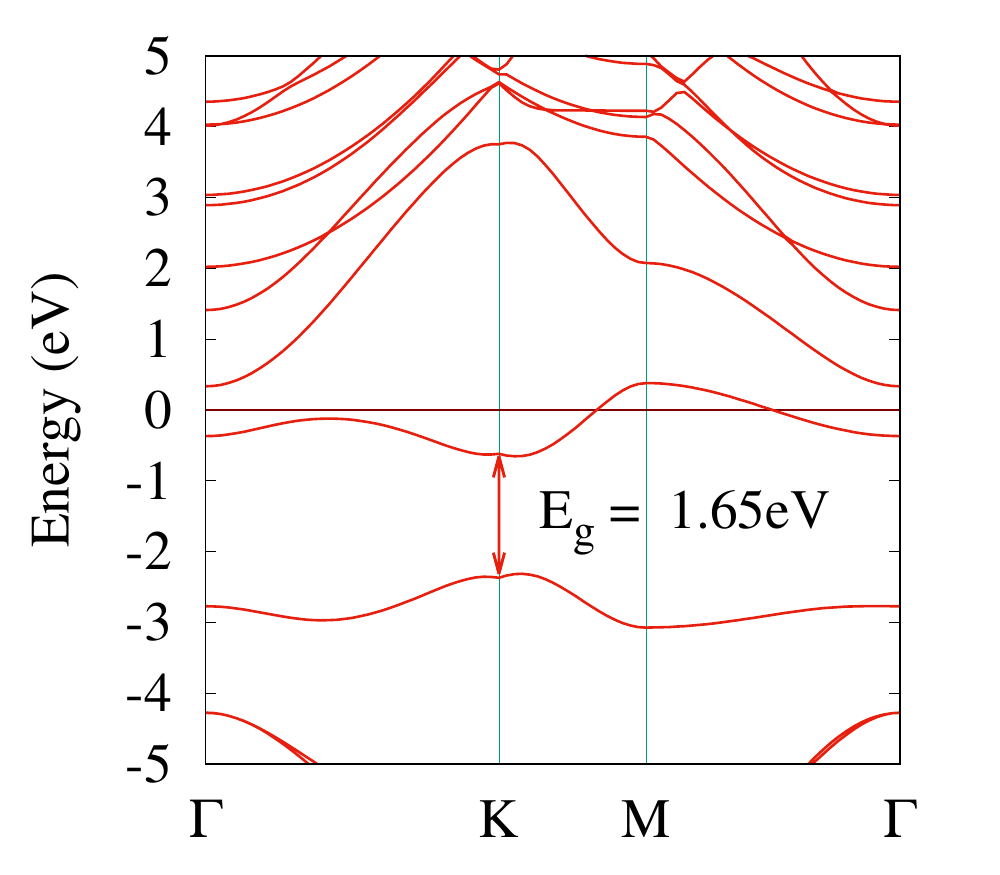} & 
        \includegraphics[width=0.15\textwidth]{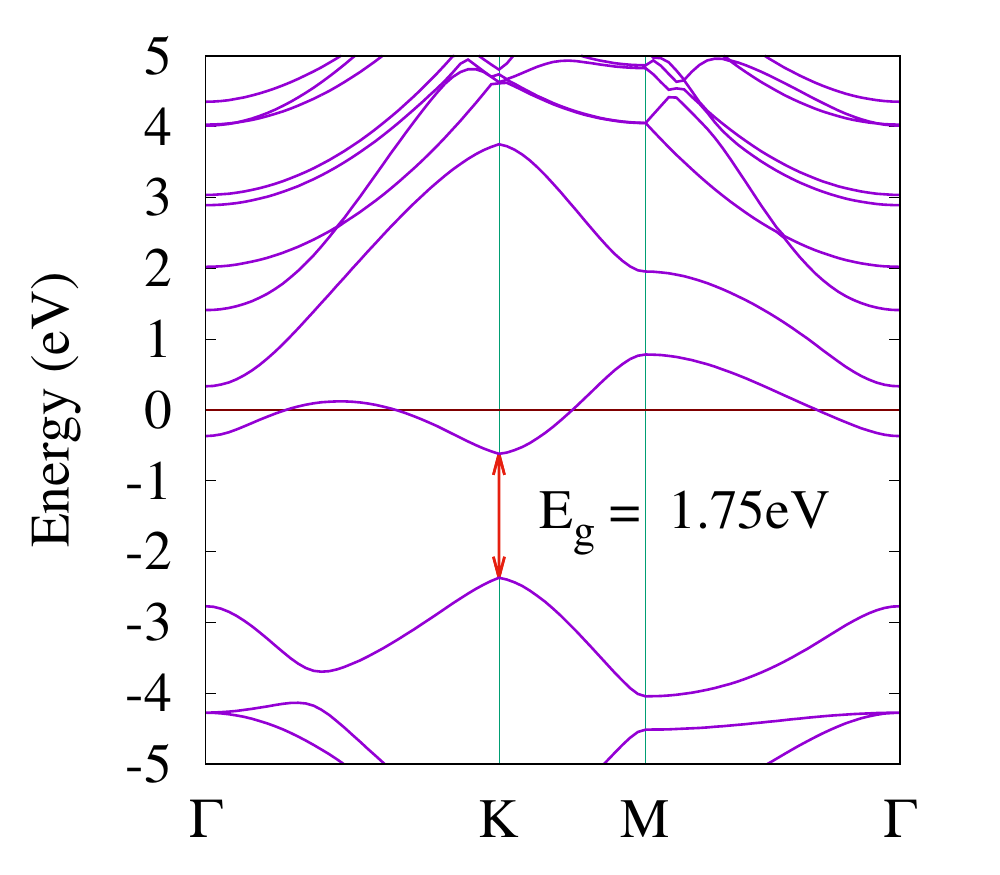} \\
	4 &
	\includegraphics[width=0.15\textwidth]{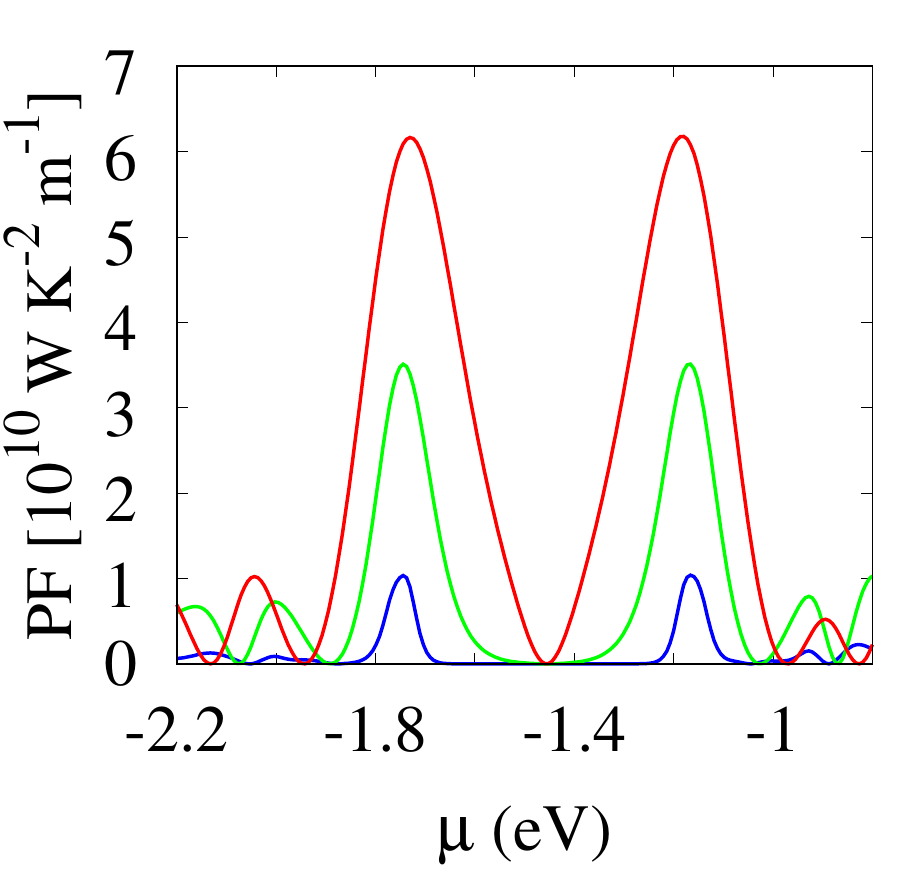} & 
	\includegraphics[width=0.15\textwidth]{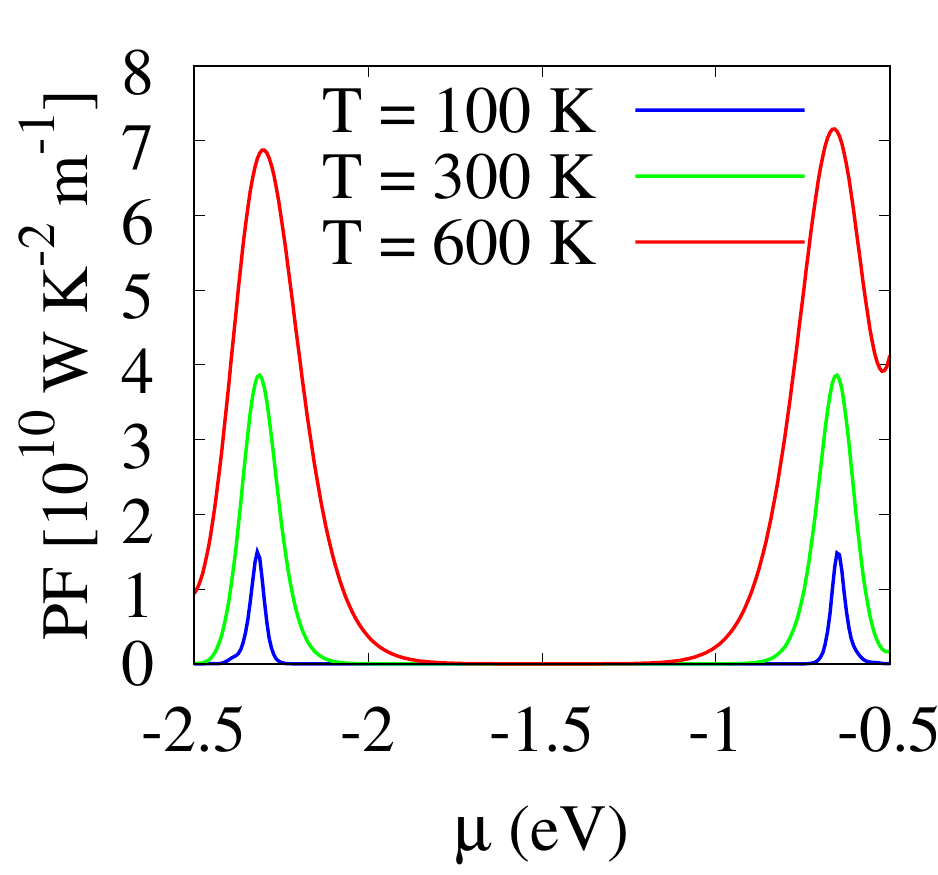} &
        \includegraphics[width=0.15\textwidth]{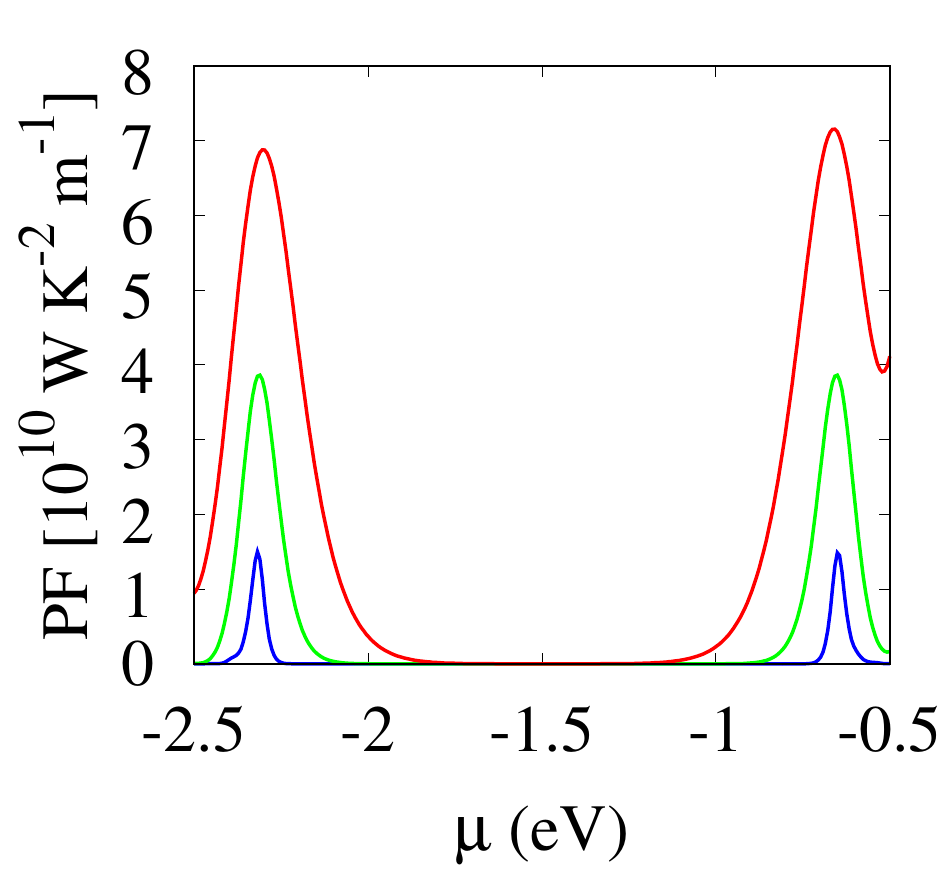} \\
	\end{tabular} 
\end{table}
 \caption{
     First row demonstrates the super-cell structure of one B-atom at para position and two N-atoms at ortho positions (a1) having 
     nonb-BNG, one B-atom at para position and two N-atoms at meta
     meta positions (b1) demonstrating as b-BNG-1, and one B atom at meta position and two N atoms at ortho-para position (c1), respectively, indicating 
     b-BNG-2.
     The second row indicates the corresponding charge density distribution and
     the third row demonstrates the band structure of the systems.
     The fourth row display the power factor for three different values of temperatures, 100 (blue), 300 (green), and 600 K (red).}
\label{fig06}
\end{figure}

We further increase the doping ratio to $37.5\%$ in which the structures consist of a codoped graphene nanosheet with one B and two N atoms
shown in \fig{fig06}. The configuration of doped atoms are: two N atoms at ortho-positions 
and a single B atom at a para-position (a1) indicating an (nonb-BNG), two N atoms at meta-positions and a single B atom at a para-position (b1) identifying as  (nonb-BNG-1), and two N atoms at ortho-para positions and one B atom at a meta position 
(c1), showing as (b-BNG-2). 
The formation energy of the nonb-BNG structure is found to be $-26.949$ eV which is smaller 
than that of the b-BNG, $-29.27$ eV, demonstrating that the nonb-BNG is more energetically 
stable than the b-BNG. The phonon dispersion of these three structures shows 
that the b-BNG structures is dynamically stable but the the non-BNG structure is less stable (not shown).

In the second row of \fig{fig06}, the corresponding electron density distribution of the first row are shown, where a high electron density between N-B bonds are found in the b-BNG-1 and b-BNG-2 [see (b2) and (c2)]. 
The ``strong'' attractive interaction between the B and N atoms in the b-BNG structures causes wide bandgaps which are  $1.65$ and $1.75$ eV for the b-BNG with the meta-para and ortho-meta-para configurations 
of doped atoms shown in (b3) and (c3), respectively. 
But the bandgap of nonb-BNG is much smaller, $0.605$ eV, due to a weak attractive interaction (a3).
In addition, since the number of N atoms is twice that of B atoms in the graphene sheets, 
the downward-shift in band structure is seen due to the predominant ratio of N atoms.
Consequently, the power factor of the b-BNG is higher than that of the nonb-BNG, as is presented in the fourth row of \fig{fig06}.

\begin{figure}[H]
	\begin{table}[H]
		\captionsetup{labelformat=empty}
		\noindent
		\begin{tabular}[]{ >{\centering\arraybackslash}m{0cm} >{\centering\arraybackslash}m{2.4cm} >{\centering\arraybackslash}m{2.4cm} >{\centering\arraybackslash}m{2.4cm} }
			& a & b & c \\ 
			1 &
			\includegraphics[width=0.15\textwidth]{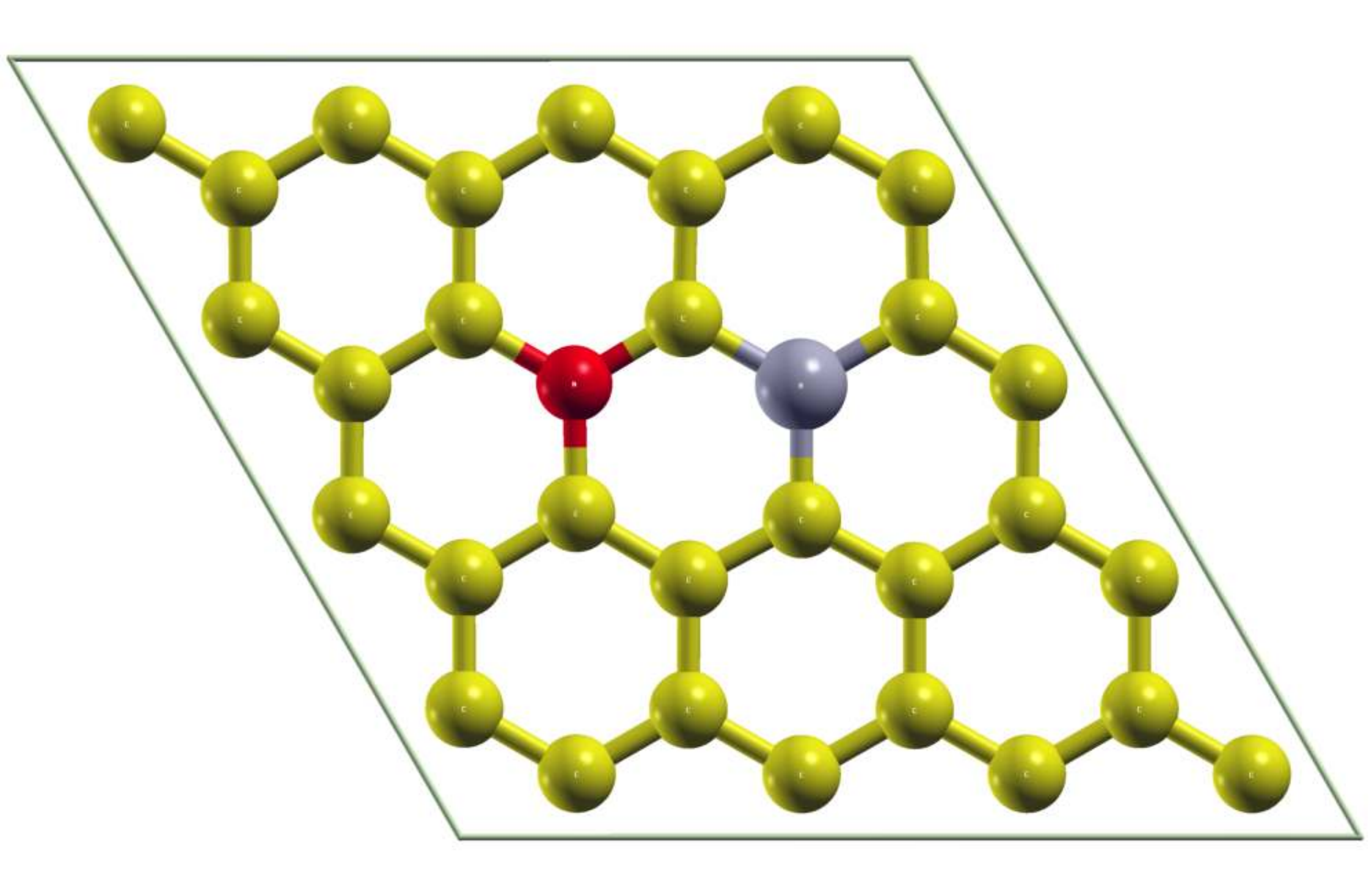} & 
			\includegraphics[width=0.15\textwidth]{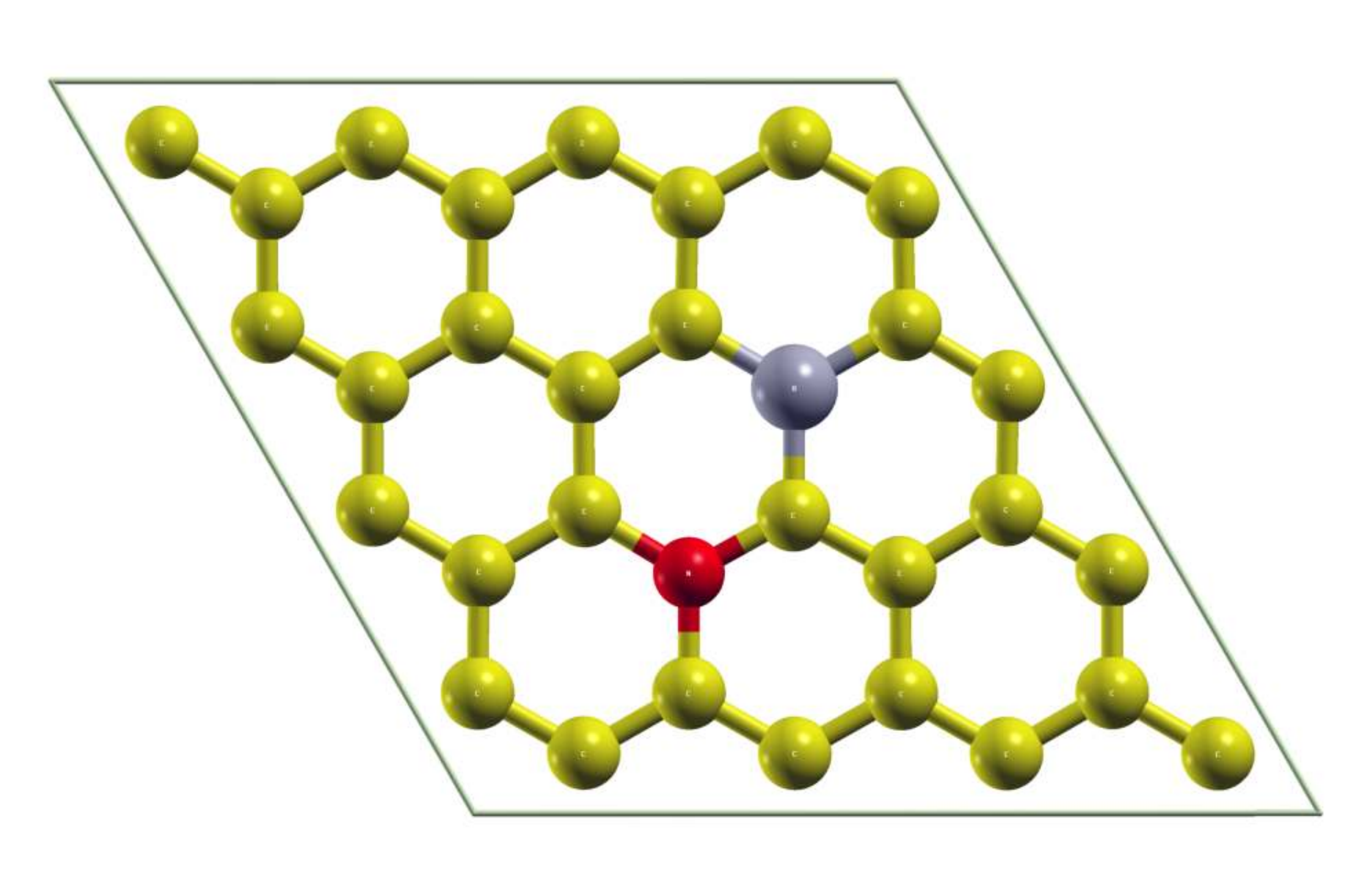} & 
			\includegraphics[width=0.15\textwidth]{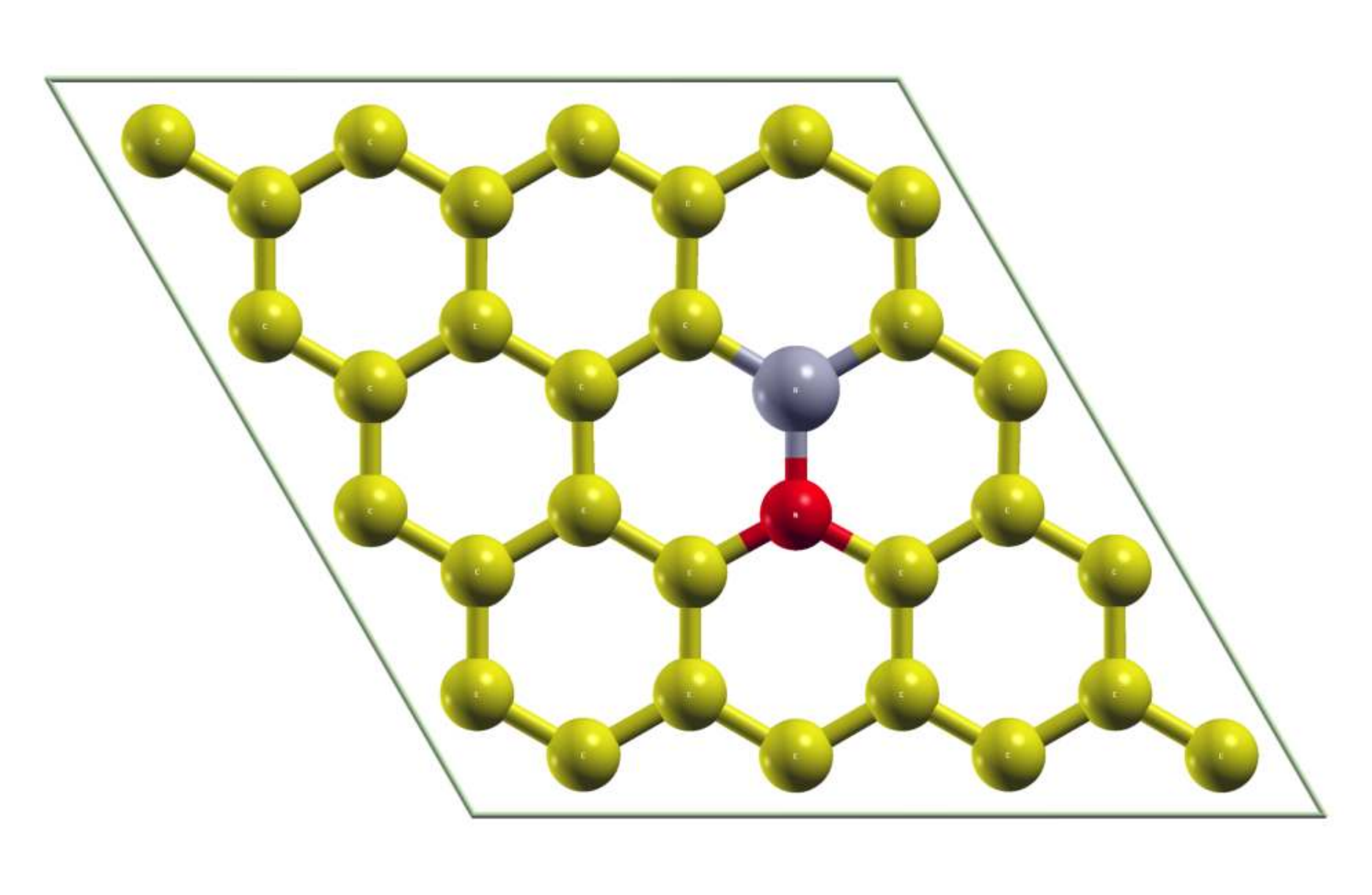} \\
			2 &
			\includegraphics[width=0.15\textwidth]{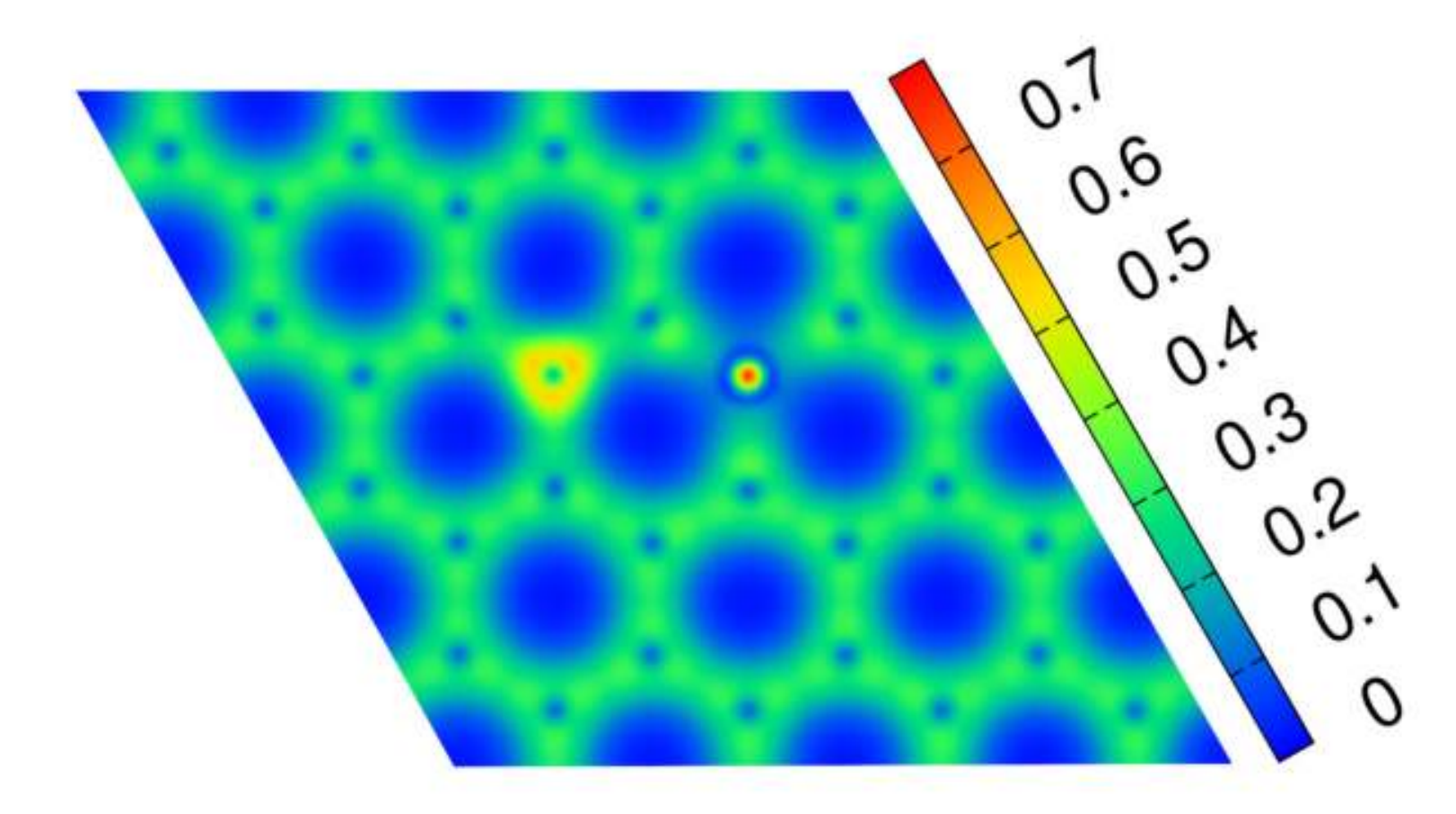}  &
			\includegraphics[width=0.15\textwidth]{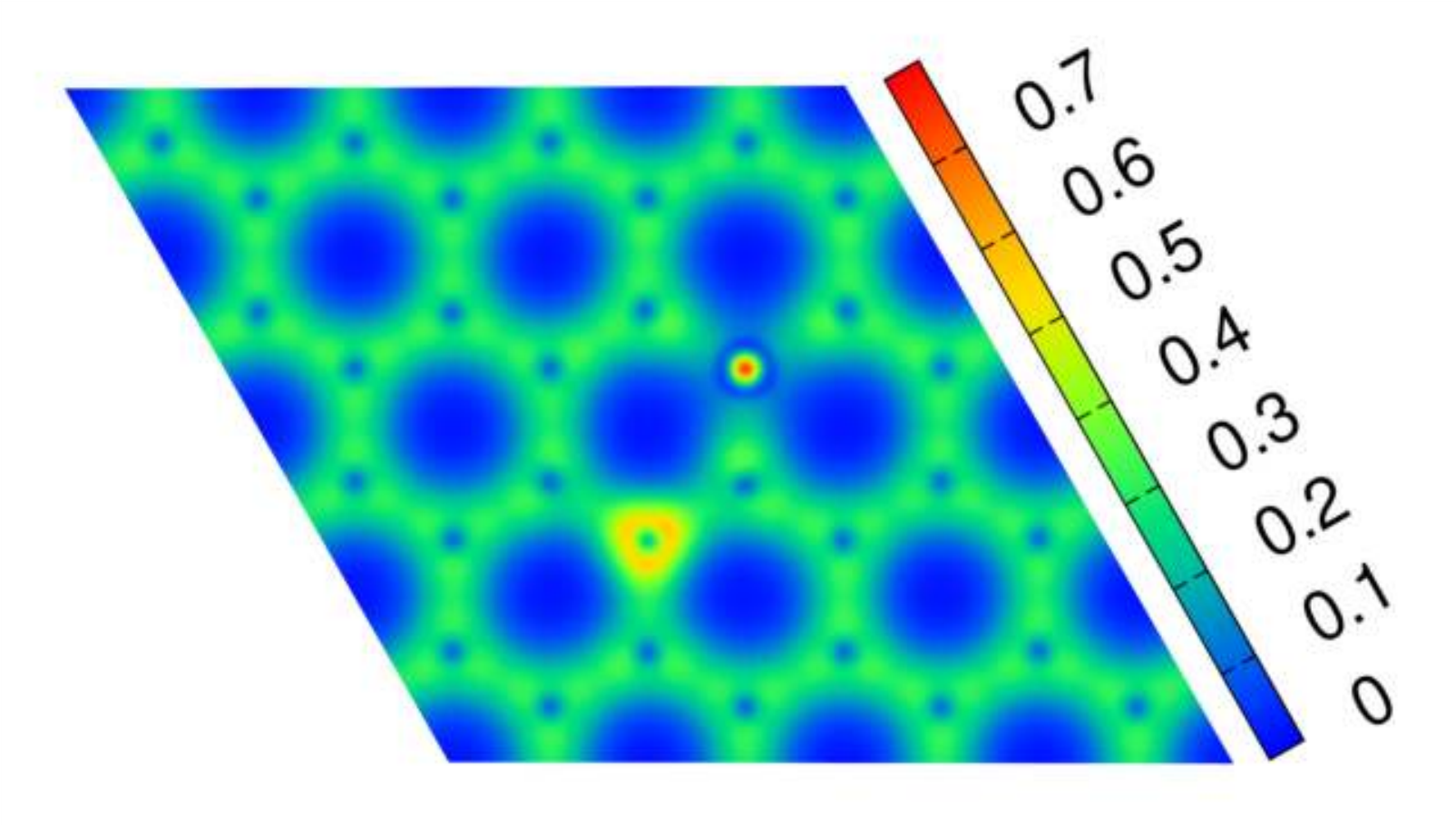} &
			\includegraphics[width=0.15\textwidth]{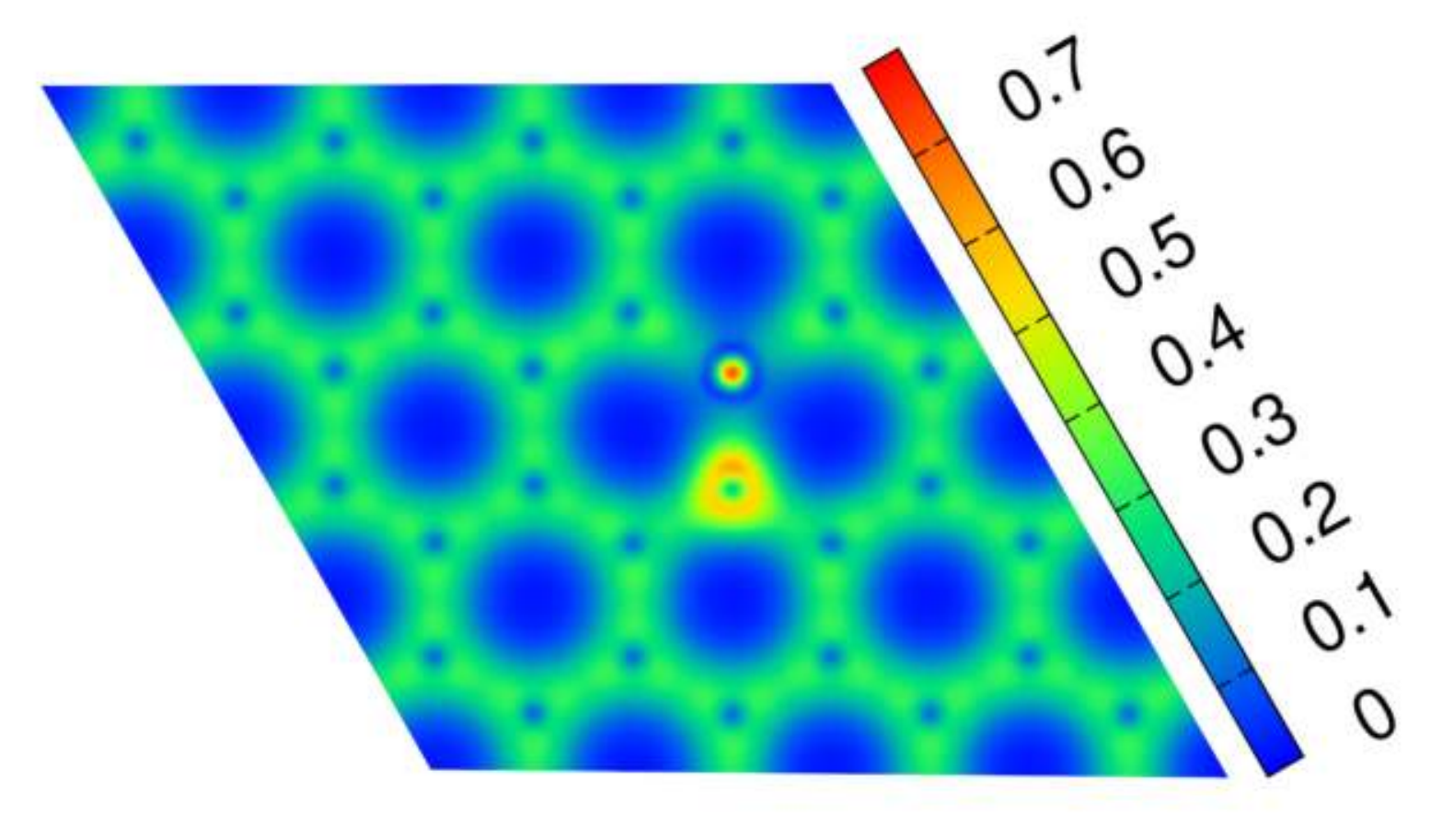} \\ 
			3 &
			\includegraphics[width=0.15\textwidth]{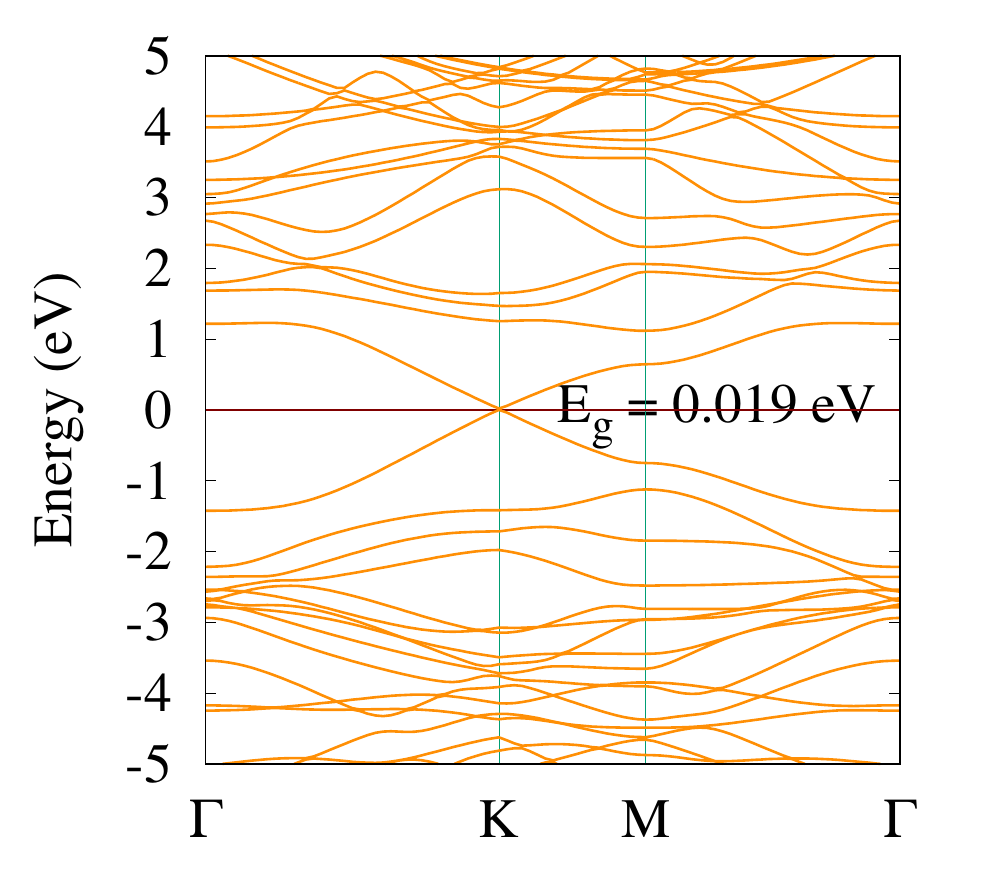} & 
			\includegraphics[width=0.15\textwidth]{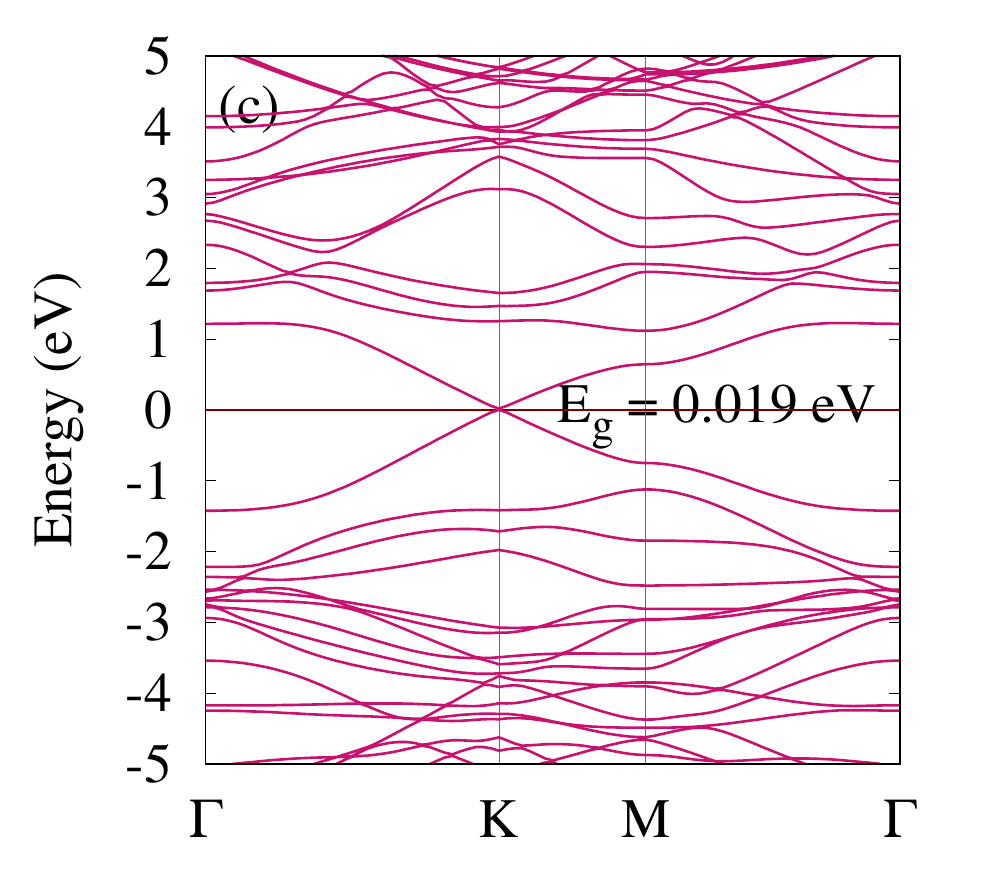} &
			\includegraphics[width=0.15\textwidth]{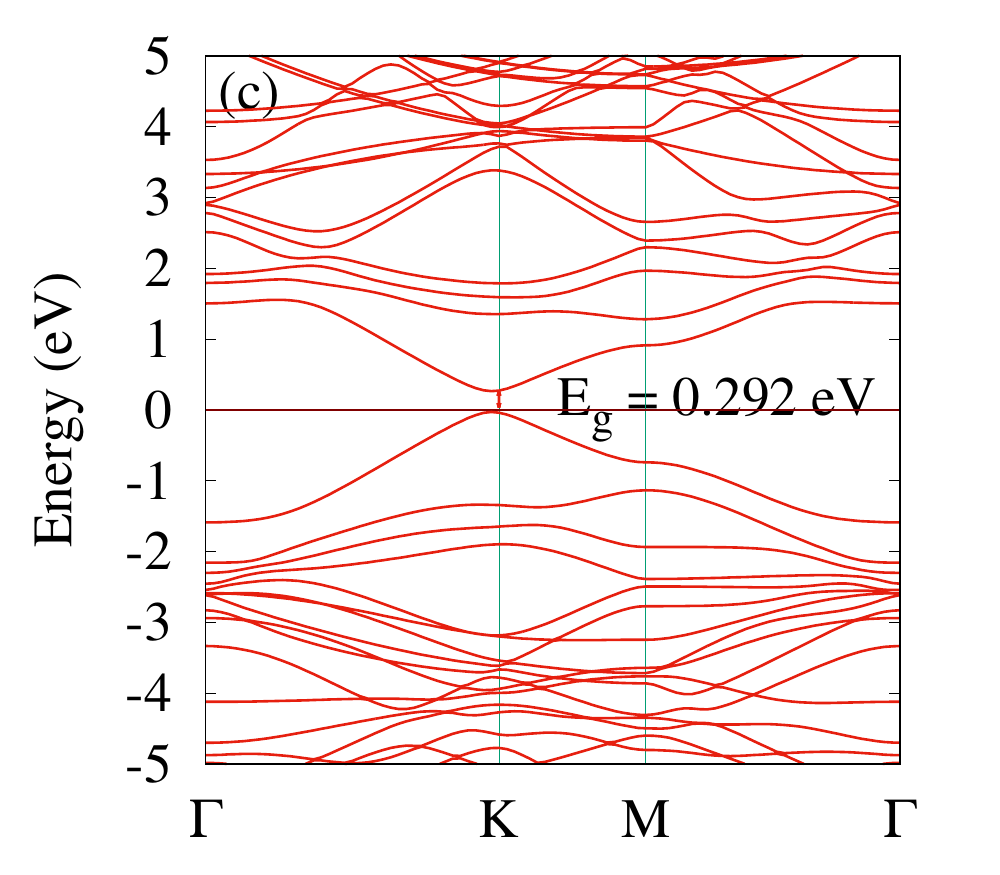} \\
			4 &
			\includegraphics[width=0.15\textwidth]{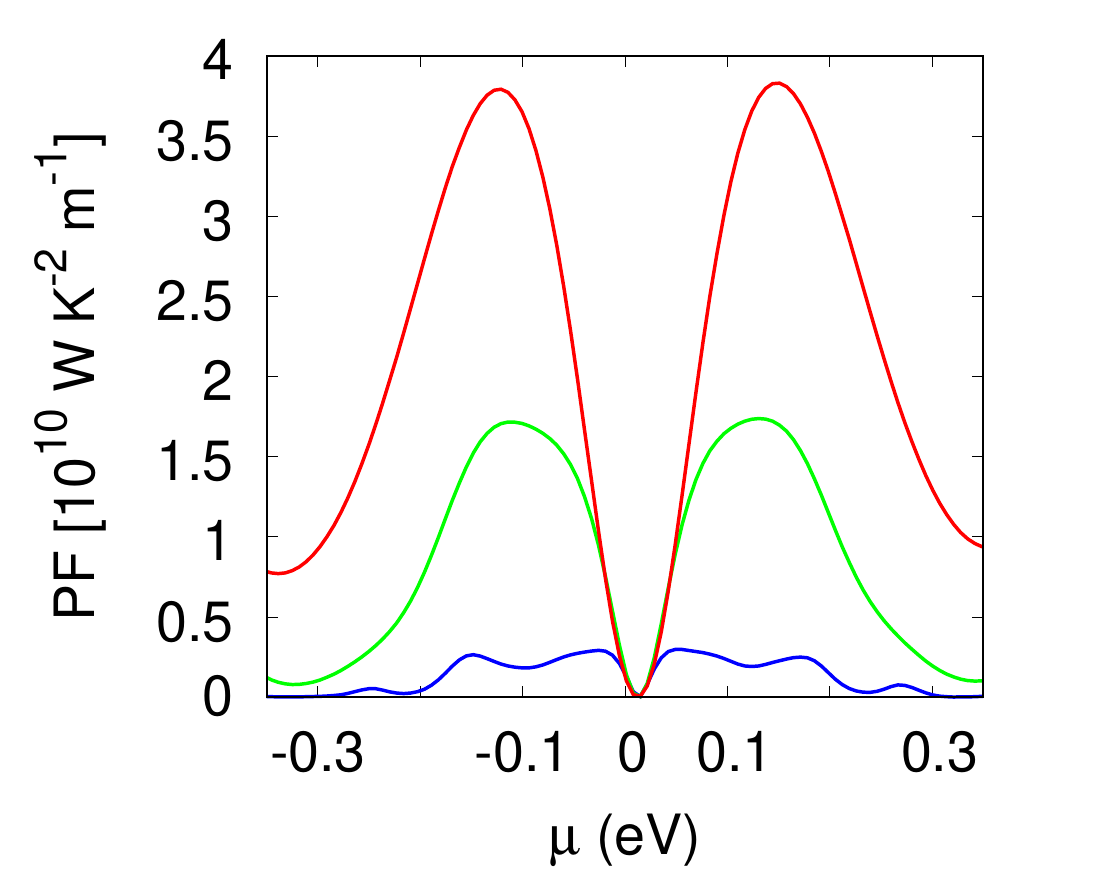} &
			\includegraphics[width=0.15\textwidth]{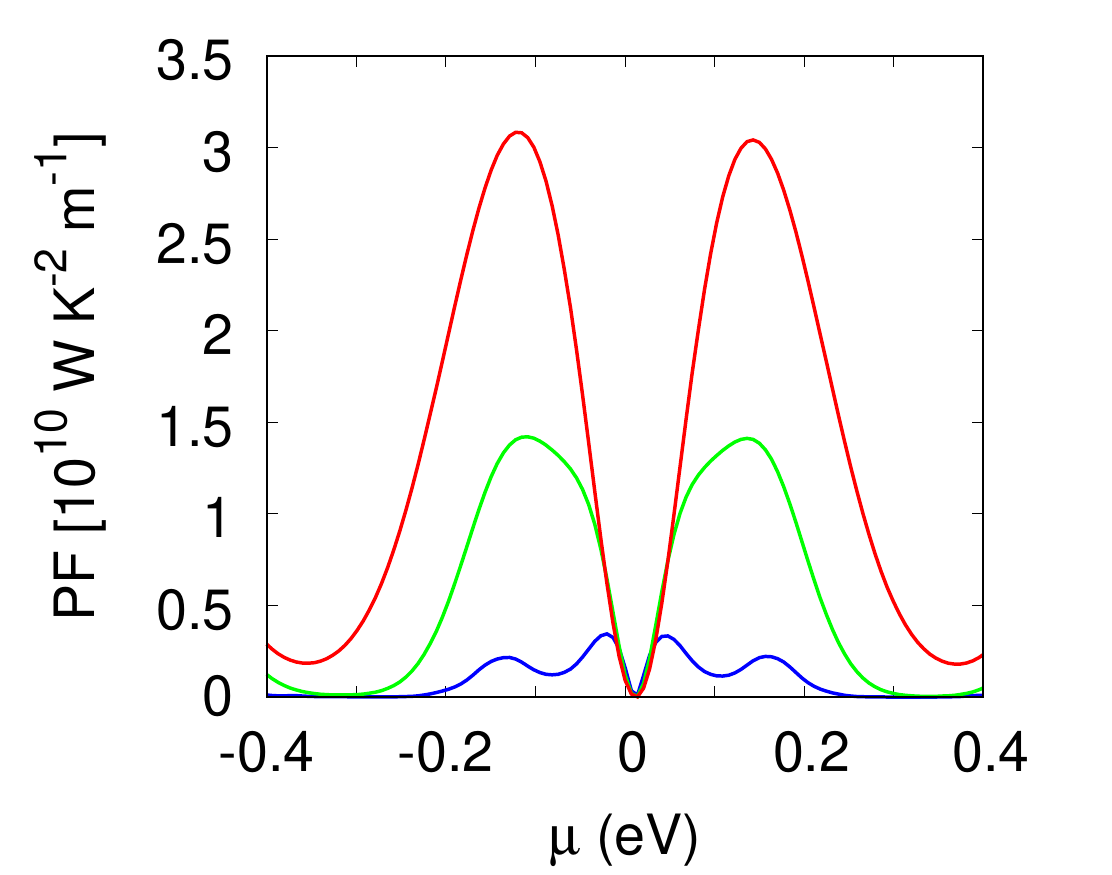} & 
			\includegraphics[width=0.15\textwidth]{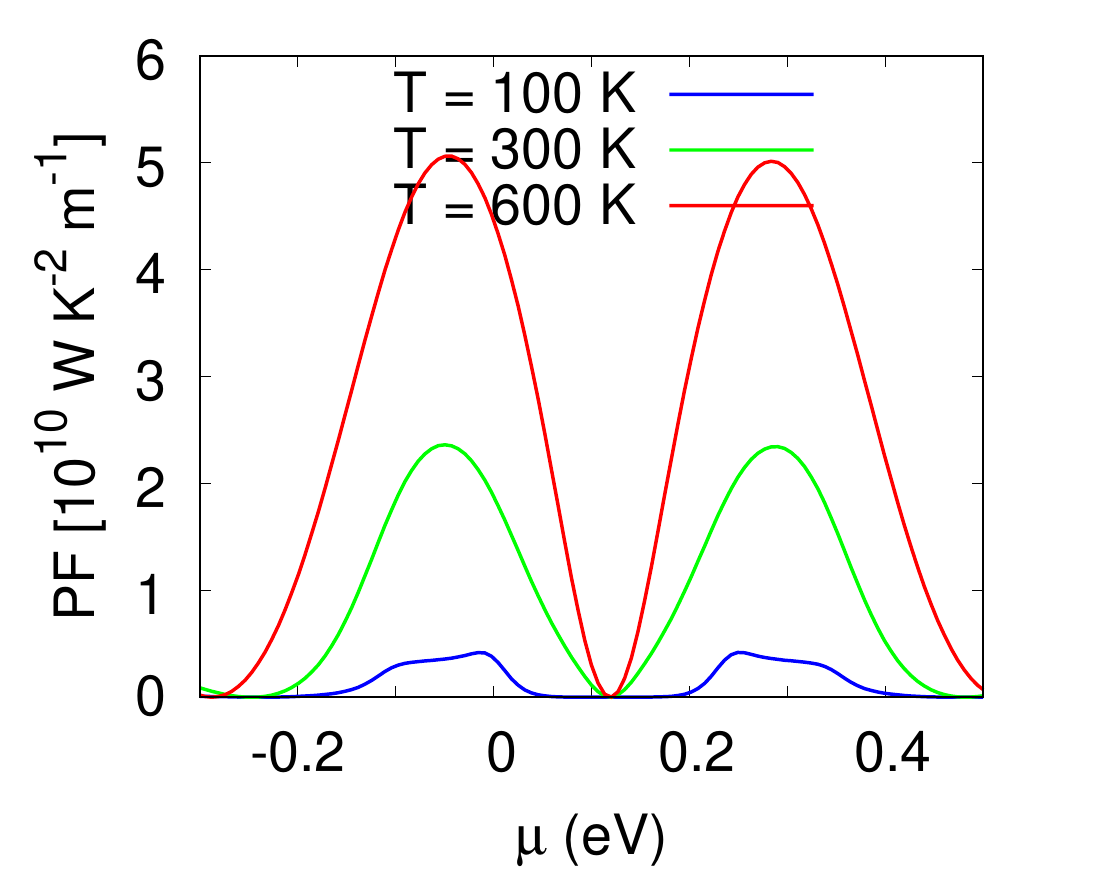} 
		\end{tabular} 
	\end{table}
	\caption{First row indicates the $4\times4\times1$ super-cells with B and N codoped graphene at 
		the same configuration of doped atoms shown in \fig{fig05} which are ortho positions (a1) showing nonb-BNG, ortho-para positions (b1) with nonb-BNG, 
		and ortho-meta positions (c1) having b-BNG, respectively. 
		The second row indicates the corresponding charge density distribution and
		the third row demonstrates the band structure of the systems.
		The fourth row display the power factor for three different values of temperatures, 100 (blue), 300 (green), and 600 K (red).}
	\label{fig07}
\end{figure}

Finally, we present the case of two atoms of B and N doped in a 4$\times$4$\times$1 
super-cell consisting of 32 atoms (see \fig{fig07}) with the same B and N atom configuration as 
presented in \fig{fig05}. Our aim here is to show that our previous results in a 2$\times$2$\times$1 
super-cell can be generalized to a larger super-cell.
It can be clearly seen that the band structure of all three atomic configurations are qualitatively 
similar to \fig{fig05} except the band gap here is smaller which is due to the low concentration 
ratio ($6.25\%$) of doped atoms. Since the band gap of the b-BNG structure (\fig{fig07}(c3)) is larger 
than that of the nonb-BNG, we thus expect to have higher power factor in a b-BNG. The same qualitative 
result was obtained for 2$\times$2$\times$1.

\section{Conclusions}
\label{Sec:Conclusion}

We have studied the stability, the electronic structure, and the thermal properties of the B and N doped 
and codoped graphene nanosheets using DFT calculations. Using the the PAW potentials and 
the Perdew-Burke-Ernzerhof exchange and correlation functionals of the 
generalized gradient approximation, one can calculate the electronic properties of the system by a SCF 
Kohn-Sham equation. We observe that a bandgap is opened up for the Boron- or Nitrogen-doped graphene due 
to the symmetry breaking of the doping atoms. 
As a result, the power factor representing the thermal efficiency of the system is enhanced. 
Increasing the doping ratio, the attractive interactions between the non-bonding atoms 
increase the bandgap, and the power factor is thus further enhanced. On the other hand, 
the repulsive interaction between bonded atoms decreases the bandgap and the power factor, likewise.

In the B and the N codoped graphene, the non-bonding of the doped atoms forms a weak 
attractive interaction between the atoms leading to 
a small bandgap, and correspondingly the power factor is decreased. 
But in the case of bonded B and N atoms, the attractive interaction is strong and the power factor 
is enhanced. 

\section{Acknowledgment}
This work was financially supported by the University of Sulaimani and 
the Komar University of Science and Technology.
The computations were performed on resources provided by the Division of Computational Nanoscience 
at the University of Sulaimani. VG acknowledges the financial support from the Research Fund of 
the University of Iceland, and the Icelandic Research Fund, grant no.\ 163082-051,

%\bibliographystyle{elsarticle-num} 
%\bibliography{Ref_2.bib}

\end{document}